\documentclass[11pt,draft]{IEEEtran}
\usepackage{color}     
\usepackage{epsf,psfrag,amssymb,amsfonts,latexsym,cite,verbatim,enumerate,subfigure}
\usepackage{srcltx,amsmath,cases, graphicx}
\usepackage{times, multirow, array}
\usepackage[mathscr]{eucal}

\def\psfancypar#1#2{\begingroup\def\par{\endgraf\endgroup\lineskiplimit=0pt}
               \setbox2=\hbox{\large\sc #2}
               \newdimen\tmpht \tmpht \ht2 \advance\tmpht by \baselineskip
               \font\hhuge=Times-Bold at \tmpht
               \setbox1=\hbox{{\hhuge #1}}
               \count7=\tmpht \count8=\ht1
               \divide\count8 by 1000 \divide\count7 by \count8 
               \tmpht=.001\tmpht\multiply\tmpht by \count7 
               \font\hhuge=Times-Bold at \tmpht
               \setbox1=\hbox{{\hhuge #1}}
               \noindent
                \hangindent1.05\wd1
               \hangafter=-2 {\hskip-\hangindent
               \lower1\ht1\hbox{\raise1.0\ht2\copy1}%
                \kern-0\wd1}\copy2\lineskiplimit=-1000pt}

\newcommand{\E}{\mbox{{\rm E}}}

\newcommand{\abf}{\mbox{${\bf a}$}}

\def\boxit#1{\vbox{\hrule\hbox{\vrule\kern3pt
        \vbox{\kern3pt#1\kern3pt}\kern3pt\vrule}\hrule}}

\def\reals{ { {\rm  I \kern-0.15em R }  } }
\def\complex{ {\,{{\rm C} \kern-0.50em \raise0.20ex {  |}}\, }}

\def\mubf{\hbox{\boldmath$\mu$\unboldmath}}

\def\Sigmabf{\hbox{$\bf \Sigma$}}

\def\abf{{\bf a}}

\def\gbf{{\bf g}}
\def\hbf{{\bf h}}

\def\tbf{{\bf t}}
\def\ubf{{\bf u}}
\def\vbf{{\bf v}}
\def\wbf{{\bf w}}
\def\xbf{{\bf x}}
\def\ybf{{\bf y}}

\def\xbf{{\bf x}}
\def\ybf{{\bf y}}
\def\Abf{{\bf A}}

\def\Dbf{{\bf D}}

\def\Hbf{{\bf H}}
\def\Ibf{{\bf I}}

\def\Rbf{{\bf R}}

\def\Wbf{{\bf W}}
\def\Xbf{{\bf X}}

\def\Hc{{\cal H}}

\def\Nc{{\cal N}}

\def\be{\vskip .3cm \begin{equation}}
\def\ee{\end{equation} \vskip .4cm \noindent}
\newcommand{\R}{\mbox{$\hat {\bf R}_{N}$}}

\def\Rxx{\Rbf_{\ssstyle X\kern-.1em X}}

\let\ssstyle=\scriptscriptstyle

\def\Kout{\setbox1=\hbox{\Huge\bf K}\hbox to
1.05\wd1{\hspace{.05\wd1}
}}
\def\Sout{\setbox1=\hbox{\Huge\bf S}\hbox to 1.05\wd1{\hspace{.05\wd1}
}}

  \ifx\LabelFigloaded\MYundefined\relax
  \else
   \fi

  \def\LabelFigloaded{\relax}

  \chardef\LabelFigCatAt\the\catcode`\@
  \catcode`\@=11

 \let\LabelFigwlog@ld\wlog
 \def\wlog#1{\relax}

 \ifx\\\MYundefined@
    \let\\\relax
 \fi

  \def\ms@g{\immediate\write16}

 \def\N@wif{\csname newif\endcsname }
 \def\Temp@ {\N@wif\ifIN@}
 \ifx\INN@\MYundefined@
    \else \let\Temp@\relax
 \fi
 \Temp@

  \def\IN@{\expandafter\INN@\expandafter}
  \long\def\INN@0#1@#2@{\long\def\NI@##1#1##2##3\ENDNI@
    {\ifx\m@rker##2\IN@false\else\IN@true\fi}%
     \expandafter\NI@#2@@#1\m@rker\ENDNI@}
  \def\m@rker{\m@@rker}
 
  \newtoks\Initialtoks@  \newtoks\Terminaltoks@
  \def\SPLIT@{\expandafter\SPLITT@\expandafter}
  \def\SPLITT@0#1@#2@{\def\TTILPS@##1#1##2@{%
     \Initialtoks@{##1}\Terminaltoks@{##2}}\expandafter\TTILPS@#2@}

 \def\Shifted@@#1#2#3{\setbox0=\hbox{#3}%
   \raise -\dp0\vbox {\kern-#2%
       \hbox {\kern#1\unhbox0\kern-#1}%
           \kern#2}}

 \newcount\gridcount
 \newbox\auxGridbox@ \newbox\hGridbox@ \newbox\vGridbox@
 \newbox\Labelbox@ \newbox\auxLabelbox@
 \newbox\Coordinatebox@
 \newtoks\Labeltoks@
 \newdimen\Wdd@ \newdimen\Htt@
 \newdimen\Wddd@ \newdimen\Httt@
 
 \def\Wr@{\immediate\write16}

 \newdimen\GL@wd
 \def\GridLineWidth#1{\GL@wd=#1}

 \def\gobble#1{}
 \def\EdgeErr@{\Wr@{}%
      \Wr@{\string\Edges\space argument
      1, 10, 100 or 1000 please\string!}%
      }

 \newcount\Edgect@

 \def\Sweepup#1\endSweepup{}

 \def\SetEdges@{%
    \edef\Zr@@s{\expandafter\gobble\number\Edgect@\empty}%
        \count255=0\Zr@@s\relax
        \ifnum\count255=\z@\else\EdgeErr@\show\tailtest\fi
        \count255=1\Zr@@s\relax
        \ifnum\count255=\Edgect@\relax\else\EdgeErr@\show\leadtest\fi
    \EdgGl@b\edef\Zr@s{\expandafter\gobble\Zr@@s\empty}
    \ifnum\Edgect@>\@ne\relax\EdgGl@b\let\L@Dc\empty
        \else\EdgGl@b\edef\L@Dc{\string.}\fi
    \ifnum\Edgect@>\@ne\relax
        \EdgGl@b\edef\Edgescale@##1{\divide##1 by \Edgect@}%
        \else\EdgGl@b\edef\Edgescale@##1{}\fi
    }

 \def\Edges#1{\Edgect@=#1\relax
     \let\EdgGl@b\global \SetEdges@}

 \Edges{1}

 \def\hhrule{\hrule height \GL@wd\vskip-.\GL@wd}

 \def\hRule@{%
   \advance\gridcount -2%
   \vfil\hhrule\vfil
   \llap{\smash{\raise -2.5pt
     \hbox{\L@Dc\number\gridcount\Zr@s\kern2pt}}}%
   \hhrule
   }

\def\vvrule{\vrule width \GL@wd \kern-\GL@wd}

 \def\vRule@{\advance\gridcount 2%
   \hfil\vvrule\hfil
   \setbox\auxGridbox@=\vbox to 0pt
      {\vskip \Htt@\vskip 2pt
        \hbox to 0pt{\hss\L@Dc\number\gridcount\Zr@s\hss}\vss}%
      \wd\auxGridbox@=0pt \box\auxGridbox@
   \vvrule
   }

 \def\PlaceGrid@@{\gridcount=10 
  \setbox\hGridbox@=\hbox{%
        \hbox{%
             \hskip-.4pt\vrule
             \vbox to \Htt@{%
               \offinterlineskip\parindent=\z@\relax
               \hbox to \Wdd@{\hfil}
               \hRule@\hRule@\hRule@\hRule@
               \vfil\hhrule\vfil}%
             \vrule\hskip-.4pt}
    }%
  \gridcount=0%
  \setbox\vGridbox@=\hbox{%
      \vbox{\offinterlineskip\parindent=0pt\hsize=0pt
         \vskip-.4pt\hrule%
         \hbox to \Wdd@{%
                 \vtop to \Htt@{\vfil}%
                 \vRule@\vRule@\vRule@\vRule@
                 \hfil\vvrule\hfil}%
         \hrule\vskip-.4pt}}%
  \wd\hGridbox@=0pt\ht\hGridbox@=0pt
  \wd\vGridbox@=0pt\ht\vGridbox@=0pt
  \hbox{\box\hGridbox@\box\vGridbox@}%
  }

 \def\LabelsGlobal{\def\LabGl@b{\global}}
 \def\LabelsLocal{\def\LabGl@b{}}
 \LabelsGlobal 

 \def\SetLabels#1\endSetLabels{%
   \LabGl@b\Labeltoks@={#1()\\}%
   }

 \LabGl@b\Labeltoks@={()\\}

 \def\ShowGrid{\LabGl@b\let\PlaceGrid@\PlaceGrid@@}
 \def\HideGrid{\LabGl@b\let\PlaceGrid@\relax}
 \def\Grids{\ShowGrid\LabGl@b\let\GridSwitch@\ShowGrid}
 \def\noGrids{\HideGrid\LabGl@b\let\GridSwitch@\HideGrid}

 \noGrids

 \def\bAdjust@@{%
     \setbox\auxLabelbox@=\hbox{\raise \dp\auxLabelbox@
            \box\auxLabelbox@}}
 \def\bAdjust@{\let\vAdjust@\bAdjust@@}

 \def\eAdjust@@{\dimen0=-.5\ht\auxLabelbox@
     \advance\dimen0 by .5\dp\auxLabelbox@
     \setbox\auxLabelbox@=
            \hbox{\raise\dimen0\box\auxLabelbox@}}
 \def\eAdjust@{\let\vAdjust@\eAdjust@@}

 \def\tAdjust@@{%
     \setbox\auxLabelbox@=\hbox{\raise-\ht\auxLabelbox@
            \box\auxLabelbox@}}
 \def\tAdjust@{\let\vAdjust@\tAdjust@@}

 \let\vAdjust@\relax

 \def\lAdjust@{\let\hAdjust@\rlap}
 \def\rAdjust@{\let\hAdjust@\llap}

 \let\hAdjust@\relax\let\vAdjust@\relax

 \def\FetchLabel@#1(#2)#3\\{%
     \IN@0#2@@\ifIN@
        \setbox0=\hbox{\ignorespaces#1#3\unskip}%
        \ifdim\wd0>0pt
           \ms@g{}%
           \ms@g{ !!! Bad label(s)? !!!}%
           \message{ #1(#2)#3}%
        \fi
        \def\LabelMole@##1\endFetchLabel@{%
            \IN@0()\\@##1@%
            \ifIN@\def\Temp@{\FetchLabel@##1\endFetchLabel@}%
            \else\def\Temp@{}%
            \fi
            \Temp@
           }%
     \else
       \ignorespaces#1\unskip
       \setbox\auxLabelbox@=%
         \hbox to 0pt{\hss\ignorespaces\hAdjust@
          {\ignorespaces#3\unskip}\hss}%
       \vAdjust@
       \let\hAdjust@\relax\let\vAdjust@\relax
       \AugmentLabelBox@@{#2}%
       \ht\Labelbox@=0pt\dp\Labelbox@=0pt
       \let\LabelMole@\FetchLabel@%
     \fi\LabelMole@}

 \newtoks\XYSep@ 
 \def\SetXYSeparator#1{%
     \IN@0#1@@\ifIN@\XYSep@{*}%
     \else
     \XYSep@{#1}%
     \fi
     }

 \SetXYSeparator*

 \def\AugmentLabelBox@@#1{%
     \IN@0\the\XYSep@ @#1@\ifIN@
       \SPLIT@0\the\XYSep@ @#1@%
       \setbox\Labelbox@=\hbox to 0pt{%
         \unhbox\Labelbox@
         \Shifted@@{\the\Initialtoks@\Wddd@}%
         {\the\Terminaltoks@\Httt@}%
         {\box\auxLabelbox@}}%
     \else
         \ms@g{}%
         \ms@g{ !!! Bad insertion point. !!!}%
         \message{ (#1\ this point was rejected.)}%
     \fi
    }

 \def\FetchOption@#1[#2]#3\endFetchOption@{%
    \def\temp{#1}
    \ifx\temp\empty
       \Edgect@=#2\relax
       \let\EdgGl@b\relax
       \SetEdges@
       \Cleaner@#3%
    \fi}

 \def\Cleaner@#1[@]{\Labeltoks@{#1}}
     
 \def\PlaceLabels@@{\mathsurround=0pt
     \def\Cr@{\\}%
     \let\L\lAdjust@\let\R\rAdjust@
     \let\B\bAdjust@\let\E\eAdjust@\let\T\tAdjust@
     \expandafter\FetchOption@\the\Labeltoks@[@]\endFetchOption@
     \Wddd@=\Wdd@ \Edgescale@\Wddd@ 
     \Httt@=\Htt@ \Edgescale@\Httt@
     \expandafter\FetchLabel@\the\Labeltoks@\endFetchLabel@
     \box\Labelbox@
     }%

 \let \PlaceLabels@\PlaceLabels@@

 \def\AffixLabels#1{\setbox\Coordinatebox@=\hbox{#1}%
      \Wdd@=\wd\Coordinatebox@ \Htt@=\ht\Coordinatebox@
      \advance\Htt@ \dp\Coordinatebox@
      \hbox{\copy\Coordinatebox@\kern-\Wdd@ 
           \Shifted@@{0pt}{-\dp\Coordinatebox@}%
           {\PlaceLabels@\PlaceGrid@}%
           \kern\Wdd@}%
      \GridSwitch@ 
      \LabGl@b\Labeltoks@{()\\}%
      }
 
   \let\wlog\LabelFigwlog@ld   
   \catcode`\@=\LabelFigCatAt  

                                By

              Raymond S\'eroul <A18645@FRCCSC21.BITNET>
                                and 
              Laurent Siebenmann <lcs@topo.math.u-psud.fr>
    
              VERSIONS: July 1991, Oct 1991, Jan 1992, July 1992

INTRODUCTION

      This labelling package is intended for TeX users who
rely on non-TeX sources for for their graphics inserts.  It
provides means for adding TeX labels to such inserts with a
minimum of fuss. 

       For most labels, TeX users have in the past found it
reasonably convenient to rely on non-TeX sources. Typical
occasions when an inescapable need for TeX labels seemed to
arise are

 (a) when the graphics program lacks certain exotic or complex
mathematical symbols

 (b) when the very highest typographical quality is wanted for the
labels

 (c) when labels included with the graphics fail to print, 
 and you cannot figure out why (cf. boxedeps.doc).  The labels
 provided by labelfig.tex are 100

       Since this package first appeared, many users, who in the
past scarcely dreamed of using TeX labels, have come to use
nothing but.  So it is now appropriate to add

Intoxication Warning:  TeX labels may be addictive and expensive. 

     If you have a fast preview you may disagree, and even find
that this package provides an agreeable paste-up environment; see
extra applications at end.

     Note to publishers: It is possible and convenient to ultimately
export the TeX labels produced by labelfig.tex to become an integral
part of the EPS file. This is often desired by a publisher who typically
uses an "upmarket" graphics or page layout program, with which the
staff is skilled in perfecting figures.  See Appendix I for
a recipe.

     The authors are grateful to Patrick Ion of Math Reviews for
helpful comments and encouragement.

BASIC INSTRUCTIONS

    After reading in the macro file using

preview or proof your figure with a coordinate grid printed on
top, by typing the following:

    \ShowGrid  
    \AffixLabels{<the graphics insertion>}

Here <the graphics insertion> is what you would type to insert
the graphics object alone without the grid.  This must provide
for the space around it. For example <the graphics insertion>
might well be \BoxedEPSF{MyFigure scaled 700} using the
boxedeps.tex macro package (from same source); this provides a
TeX box containing the encapsulated PostScript insert specified by
the file MyFigure. \AffixLabels{...} provides the grid (supposing
\ShowGrid is present) and later, once you have specified labels
using the grid, it will "tack on" the labels.

     The grid is a sort of (usually elongated) checkerboard of
ten rows and ten columns and its (internal) partitions are by
default numbered  .1, ... ,.9  both horizontally (X-coordinate
running left to right) and vertically (Y-coordinate running bottom
to top).  Thus the points enclosed by the grid correspond to the
points of the unit square in the cartesian "X-Y" plane, the lower
left corner corresponding to the origin (0,0).  By extrapolation,
the full page corresponds to a larger rectangle in the plane.

     These coordinates serve to position labels as follows.
Before the \AffixLabels{...} command type label specifications:

  \SetLabels
   (<X-coordinate>*<Y-coordinate>) <first label> \\
   .
   .
   .
   (<X-coordinate>*<Y-coordinate>)  <last label> \\
  \endSetLabels

Each row specifies one label and is terminated by \\.  In each
row, the position indicator comes first; it is written as a
standard cartesian point except that the X- and Y- coordinates
are separated by * rather than a comma because TeX allows a
comma as decimal point. There are no dimension units to specify
as the unit is the grid itself.

     By default, this cartesian point specifies where the middle
of the baseline of the label will be located.  However if you precede
the point by \L [or \R] the left [or right] edge of the baseline will
be located there. Similarly you may also precede the point by \T, \E,
or \B to vertically align the top equator or bottom of the label box
at the specified point.  This gives nine standard positions of
the label with respect to the insertion point --- corresponding to
the eight principle points of the compas and the center

                     \L\T     \T      \R\T

                     \L\E     \E      \R\E

                     \L\B     \B      \R\B

But this neglects the default "baseline" level of TeX,
giving potentially three more positions

                     \L    <no tag>   \R

For text, the baseline level is often the preferred. Its relation to
the others is variable. It will often coincide with the bottom level,
as happens for "X".  But it is often distinct, as for "g", in which
case you have in all 12 distinct positions rather than 9.

     It is convenient to think of this specification of label
position as attaching the label by a thumb-tack to the coordinate
grid. There are up to twelve positions of the thumb-tack on the
label, while the position of the thumb-tack on the coordinate grid is
arbitrary.  Normally, one choses the position of the thumb-tack on
the label to be the one that is the closest to the item being
labeled.  There are good reasons for this "rule of thumb":

   (a)  It facilitates correct positioning at first try.

   (b)  If the scale of the figure must be altered after labels
have been affixed, the labels have a good chance of remaining well
positioned.

   (c)  The visible grid need not extend beyond the "bounding box"
for the figure, because the best preferred position is always
(at least almost) within the bounding box .

The second reason is particularly important. Indeed it often
happens that scale has to be altered after labelling begins, in
order to either provide space for the labels, or to adjust
proportions between the labels and the figure.  (The size of labels
is unaffected by scaling.)

     Here is an artificial but self-contained test which uses
TeX rules to make a graphics object.

TEST

    Do not skip this!

 \def\FrameIt#1{\hbox{\vrule$\vcenter {\hrule\kern3pt%
             \hbox {\kern3pt #1\kern3pt}%
               \kern3pt\hrule}$\relax\vrule}}

 \def\Caption#1#2{\FrameIt{%
       \vtop {\hsize=#1\relax \parindent=0pt
         \leftskip=0pt \rightskip=0pt plus15pt
         \parfillskip=0pt
         \lineskip=1pt\baselineskip=0pt
         #2}}}

 \def\FirstQuadrant{\hbox to 100pt{\vrule\vbox to 100pt{%
        \hbox to 100pt{\hfil}\vfil\hrule}\hss}}

  \SetLabels
    \R(.5*.2) $\zeta\,\cdot$\\
    (.9*-.10) $\xi$\\
    \R(-.03*.9) $\eta$\\
    \T(.5*.9) \Caption{70pt}{%
          \it The norm of
          $g(\xi+i\eta)$ is indicated on
          contours of this invisible surface.}\\
  \endSetLabels

  \AffixLabels{\FirstQuadrant}

  \end

  Note that the coordinates to use for labels are indicated on the
edges of the grid (when visible) corresponding to the conventional
x- and y- axes of the Cartesian plane. By default the grid is
1-by-1. However, by the command \Edges{100}, you can change this
to 100-by-100 and many users find this alternative most
convenient. Place the command \Edges{...} in your style file (or
header) since its effect is is global. Other possible edge values
are 10 and 1000.

  If you use the command \Edges{...} at all, do so with care.  For
if you accidentally delete an \Edges{...} command your labels will
abruptly be badly misplaced and may logically but mysteriously
generate "dimension too big" errors under TeX and "off page" errors
under your driver.  

  You can dictate the edgescale for an individual figure by giving
the scale in brackets immediately after \SetLabels.  Thus, to
import into an article using say \Edge{100} a figure labelled using
another edgescale, say the original 1-by-1 default, you can use
\SetLabels[1]...\endSetLabels.

GETTING IT DOWN PAT

     Complicated labeling deserves the same respect as
complicated mathematics.  Do not expect it to come out perfect the
first time!  What is needed in either case is a mechanism to
repeatedly typeset troublesome pieces.

     One mechanism is always available.  One does complicated
labelling in a separate "test" file involving just the figure being
labelled;  a texpert will know how to \dump TeX's current state as
a temporary format that restarts rapidly at each retry.  Usually,
one then pastes the completed labelled figure back into the main
TeX file, but, of course, one can also \input it as an auxiliary
file.

     If you do not have a TeXpert at handy, here is a first
approximation to an efficient setup. By deletions reduce a copy
of your article to just a few lines before and after the figure.
Now label the figure, and finally, copy and paste the labelled
figure to the original article. Then copy the next figure to label
into this testbed and repeat. The TeXpert can improve the  speed
at which TeX starts up, by compiling a format specifically for
your article; just one caution: best NOT include in the format
ephemeral details of setup like \Set<mydriver>ArtSpecials (from
boxedeps.tex because this reads  figure dimensions which you may
change during your work session.

     An improved mechanism to repeatedly typeset troublesome
pieces is now available on the Macintosh; it is called LinoTeX;
see the same ftp sources.  It could be set up on many types
of computer.

     Before using labelfig.tex to attach labels to a graphics
object inserted using boxedeps.tex or BoxedArt.tex, make it a
firm rule to carefully adjust the bounding box using the trimming
commands of these packages, and also at least tentatively scale
and position the object. Beware of changing the grid inadvertently
after the labels have been positioned.  For example, correcting
the bounding box of a PostScript graphics object can foul up the
labels by changing the coordinate grid to which the labels are
attached. This is particularly true for the trimming  commands of
boxedeps.tex and BoxedArt.tex. However, as noted already, change
of scale is much less disruptive, and modest adjustments should be
well tolerated.

     Sometimes the labels protrude so far from the bounding box
of a figure that the figure has to be repositioned.  Best do this
by ad hoc spacing, say using \hglue and \vglue; altering the
bounding box would create a vicious circle.

     Remember that you are responsible for preventing labels
from overlapping. You are responsible for all label typography
including size and style. A label is really just about anything
that can be put in a TeX box. Note that spaces at the beginning
and end of labels will normally be suppressed; if you really want
them you must protect them with TeX braces.

     This package temporarily sets the \mathsurround parameter
of TeX to zero  while the labels are being affixed. This is done
because nonzero \mathsurround space would influence the position
of left and right aligned labels; then, when a texpert or printer
modifies mathsurround, diagram labeling might be disastrously
altered. There is a small price to pay involving labels that are
formatted as caption boxes including mathematics: you  may want or
need to specify an explicit mathsurround space within the caption
box; it will not influence anything outside.

     Those hostile to the use of * as separator between
the X and Y coordinates of label insertion points, are free to
impose another using \SetXYSeparator{<the new separator>}.  
Americans may prefer "," to "*" since they never use a 
comma as a decimal point; on the other hand, * may be more visible.

APPENDIX (I)  MERGING labelfig.tex LABELS INTO AN EPSF GRAPHICS OBJECT.

     As promised in the introduction, here is a recipe useful for
publishers. It works at least on Macintosh and at least for vectorized
graphics and Adobe type1 fonts.  (There is surely a similar recipe for
PCs under MSWindows.)

 (a)  Use boxedeps.tex utility to integrate the figure given by the eps
file, "x.eps" say, with a visible frame around it.  See
\ShowDisplacementBoxes command in boxedeps.tex.  To get precise results
automatically it is important to use the \Trim... commands of
boxedeps.tex making the "DisplacementBox" neatly fit the figure.

 (b)  Use the TeX printer driver and LaserWriter (versions >= 8.1.1) to
export to an EPSF the DVI page containing the integrated, labelled
figure. You now have an EPS file  "xx.eps"  that contains too much, and at
the wrong scale, and at wrong position.

 (c)  Convert the EPSF to an Adode Illustrator format EPSF using
the shareware utility called epsConvert by Sam Weiss
1993-- (currently $25).

 (d)  In Illustrator (or a compatible program), group the labels and the
"DisplacementBox"; copy them to the clipboard and paste them into "x.ps".
This step requires that all the label fonts be "visible to the Macintosh.

 (e)  Translate and scale the pasted group consisting of the labels plus
the "DisplacementBox" so as to make the "DisplacementBox" the bounding
box of (labelless) figure represented by "x.eps".  At this point the
labels will be correctly placed on the figure "x.eps".

 (f)  Ungroup and delete the "DisplacementBox".  The result is the
desired single EPS file, "x+.eps" say, It contains the original figure
plus its labels.  

     Using grouping and ungrouping appropriately in "x+.eps", a
publisher's staff can very efficiently improve label positions etc.

APPENDIX II)  SOME EXOTIC APPLICATIONS

     The grid of labelfig.tex is analogous to a light-table in
classical page makeup with wax or latex glue.  In principle, you
can use it to compose any page from its indivisible parts.  This
even has some of the artisanal charm of classical paste-up
provided you have a fast screen preview to make the process
"interactive".

     In practice labelfig.tex is a tool for nonstandard jobs.
Here are a few going beyond the labelling already discussed.

(I)  GRAPHICS INTEGRATION.

     This is accomplished by treating the imported graphics
objects as labels.  The underlying graphics object is then
typically an empty  \vbox to <dimension>{\vfill} in a TeX
\midinsert...\endinsert construction.  A label line
might be of the form

   (.1*.1) \\

The exact form of the special command varies from driver to
driver.  However, in the case of encapsulated PostScript graphics
(EPSF norm), by relying on boxedeps.tex, one can have the
following standard syntax (independant of driver  (see
boxedeps.doc for details.
  
  (.1*.1) \BoxedEPSF{MyFigure scaled <scale in mils>}\\

This may be slow since it requires TeX to read the PostScript
file to read bounding box using many complex macros.  So you
may want to try

  (.1*.1) \EPSFSpecial{MyFigure}{<scale in mils>}\\

which is fast and driver independant, but it squashes the
bounding box, normally to its lower left corner.

     Similarly for graphics of the Macintosh PICT norm ---
using BoxedArt.tex (same sources) in place of boxedeps.tex.

     This approach to integration is to be recommended when
one is assembling a composite graphics object.

 (II)  COMMUTATIVE DIAGRAM ENHANCEMENT

     Commutative diagrams or arrays of mathematical objects
connected by arrows of various sorts are common in mathematics.
The mathematical objects require the use of TeX.  Recently TeX
acquired a good collection of arrows of all slopes --- that of
LamSTeX --- plus pwerful macros to build the diagrams.

     However, even the LamSTeX collection is often
inadequate; it lacks for example double shafted arrows, dotted
arrows and curved arrows. Fortunately it is possible to produce
such arrows on an individual basis using sophisticated graphics
programs such as Illustrator and AldusFreehand (both serving
the EPSF norm) or using Metafont (with its public domain norm).
Since the creation of each new arrow is a work of love, you
probably want to limit the number of arrows by using LamSTeX
for most arrows. The 40K commutative diagram module of LamSTeX
has been adapted to work with AmSTeX and a copy may be posted
with LabelFig and related files. Unfortunately no one has yet
offered a version that works with Plain TeX or LaTeX.

       Suffice it here to say that when the exotic arrow has
been somehow imported into TeX, labelfig.tex treats it as a
label that one affixes to the commutative diagram.  Two other
steps will be treated in separate notes, namely the matter of
extracting the dimension specifications for the arrow and the
construction of the arrow --- for these steps are far from
unique and often depend intimately on your computer environment. 
Notes for the Macintosh-Textures-Illustrator combination are
found in the file ExoticArrows.doc.

 (III) NESTING 

Ingenuity pays off in exploiting labelfig.tex. One can
mix graphics and typography quite freely.  labelfig.tex is good
for freeform or overlapping arrangements, while boxedeps.tex (or
BoxedArt.tex) is best for regimented non-overlapping
arrangements --- and the two can be combined.

     The default behavior of labelfig.tex is not ideal 
for nesting objects, because to prevent trouble for beginners
the register for labels is globally cleared when \AffixLabels
concludes.  But there are switches available

      \LabelsGlobal      \LabelsLocal

which change this.  To understand this, extend the above test 
by something like:

 \LabelsLocal

 \SetLabels
    (.5*.5) AAA\\
 \endSetLabels

 {
 \SetLabels
    (.5*.5) ZZZ\\
 \endSetLabels
   \AffixLabels{\FirstQuadrant}
 }

   \AffixLabels{\FirstQuadrant}

     There are however potential pitfalls.  Neither
labelfig.tex nor boxedeps.tex has been tested under extreme
conditions. Problems may occur if their procedures are
indiscriminately nested. For boxedeps.tex (not labelfig.tex)
there is a precise cause for worry, namely many of its
variables are "global", which means that TeX braces will not
provide the protection one might expect.

COMMAND SUMMARY FOR labelfig.tex

  Here [...] means optional (one or zero)
       [...]* means any number of such constructs

  \SetLabels
    [[<P>](<X><Sep><Y>) <label> \\]*
  \endSetLabels
  \ShowGrid  
  \AffixLabels{<the figure>}

   --- <P> is tack position, one of eleven or empty
              order irrelevant

                   \L\T      \T      \R\T

                   \L\E      \E      \R\E

                     \L               \R

                   \L\B      \B      \R\B

   --- (<X><Sep><Y>) insertion point;
  <Sep> is separator, = * by default;
  \SetXYSeparator{<Sep>} changes it.
   <X> and <Y> are real numbers

  --- <label> a label to attach 

  --- <the figure> the figure to label 

  \GlobalLabels (default)     
  \LocalLabels  setting for nested constructs.

 \Grids makes ALL grids appear; \HideGrid then makes just next disappear.
 \noGrids returns to default.  The commands are always global.

 \GridLineWidth{<dimension>} adjusts width of grid lines. Default is very
small, to give "hairline" effect. If your grid lines are missing try
setting \GridLineWidth{1pt}.

 \Edges#1 globally changes the edge size of all grids to the numerical 
value #1, which must be 1, 10, 100, or 1000.  The default is 1.

VERSION HISTORY.
 --- Jan 1993: \Edges#1 and [??] option after \SetLabels
 --- July 1992: \Grids, \noGrids, \HideGrid;
       Gridlines become hairlines; \GridLineWidth{<dimension>}.
 --- Oct 1991, Jan 1992: \SetXYSeparator{<Sep>},  \LabelsGlobal,
       \LabelsLocal.
 --- July 1991: first release

Address for bugs and other feedback:

        Raymond S\'eroul
        IREM and Lab. de Typographie Informatise
        Univ. Rene Descartes
        Strasbourg

    Tel 33-88-41-63-45
    Email:  A18645@FRCCSC21.BITNET

        Laurent Siebenmann
        Mathematique, Bat. 425,
        Univ de Paris-Sud,
        91405-Orsay,
        France

    Tel 33-1-6941-7949; 
    Email: lcs@topo.math.u-psud.fr

\def\scalefig#1{\epsfxsize #1\textwidth}

\newcommand {\Ebb}{{\mathbb{E}}}
\newcommand {\Rbb}{{\mathbb{R}}}

\newcommand {\Nbb}{{\mathbb{N}}}

\newtheorem{theorem}{Theorem}

\newtheorem{definition}{Definition}

\newtheorem{algorithm}{Algorithm}

\title{\LARGE {A Joint Time-Invariant Filtering Approach to the Linear Gaussian  Relay Problem}}

\author{
Cheulsoon Kim, {\em Student~Member,~IEEE}, Youngchul
Sung$^\dagger$\thanks{$^\dagger$Corresponding author}, {\em
Senior~Member,~IEEE}, and\\ Yong H. Lee, {\em Senior~Member,~IEEE}
\thanks{The authors are with the Dept. of Electrical Engineering,  KAIST, Daejeon 305-701, South
Korea. E-mail:\{lighid@stein., ysung@ee. and yohlee@\}kaist.ac.kr.
This work was supported in part by the IT R \& D program of
MKE/KEIT [2008-F-004-02, ``5G mobile communication systems based
on beam division multiple access and relays with group
cooperation"].
 This research was also supported in part by the KCC (Korea
Communications Commission), Korea, under the R \& D program
supervised by the KCA (Korea Communications Agency)
(KCA-2011-11913-04001).}
}

\begin{document}

\maketitle

\begin{abstract}
In this paper, the linear Gaussian relay problem is considered.
Under the linear time-invariant (LTI) model the problem is
formulated in the frequency domain based on the Toeplitz
distribution theorem. Under the further assumption of realizable
input spectra, the LTI Gaussian relay problem is converted to a
joint design problem of source and relay filters under two power
constraints: one  at the source and the other at  the relay, and a
practical solution to this problem is proposed based on the
projected subgradient method. Numerical results show that the
proposed method yields a noticeable gain over the  instantaneous
amplify-and-forward (AF) scheme in inter-symbol interference (ISI)
channels. Also, the optimality of the AF scheme within the class
of one-tap relay filters is established in flat-fading channels.
\end{abstract}


\section{Introduction}

Relay networks have drawn extensive interest from research
communities because they play an important role in enlarging the
network coverage in wireless communications. Although the capacity
of relay networks is not exactly known yet, many ingenious
 coding strategies including decode-and-forward (DF),
compress-and-forward (CF), etc. beyond simple AF schemes have been
developed \cite{Cover&ElGamal:79IT,ElGamal&Aref:82IT}. Recently,
Zahedi et al. proposed an advanced linear scheme for relay
networks based on (strictly-)causal linear processing at the relay
to compromise the complexity and performance between the
complicated coding strategies and the simple AF\footnote{In this
paper, the AF scheme means the instantaneous AF scheme, which can
easily be implemented by simple analog processing.} scheme
\cite{Zahedi&Mohseni&ElGamal:04ISIT,ElGamal&Mohseni&Zahedi:06IT}.
While information theorists approached the problem from the
perspective of capacity and capacity-achieving schemes
\cite{Kramer&Gastpar&Gupta:05IT, ElGamal&Hassanpour&Mammen:07IT,
DelCoso&Ibars:09WC, Khormuji&Skoglund:10IT}, researchers in the
signal-processing community also tackled this problem based on
measures like the received signal-to-noise ratio (SNR) or minimum
mean square error (MMSE). Most of their results are based on the
setup in which
 linear processing is at the relay and destination but not at the
source, e.g.,
\cite{Chen&Gershman&Shahbazpanahi:10SP,Liang&Ikhlef&Gerstacker&Schober:11WC}.
Although these works provide  meaningful approaches to the relay
problem, it is not optimal not to have processing at the source
from the fundamental perspective of data-rate maximization.  To
this end the processing at the source such as the input covariance
function design should be incorporated together with the
processing at the relay. (Once the processing at the source and
relay is fixed, the optimal destination processing is
automatically given for several well-known criteria.) However, the
joint design of source and relay processing is a hard problem even
in the linear Gaussian case, as shown in
\cite{Zahedi&Mohseni&ElGamal:04ISIT,ElGamal&Mohseni&Zahedi:06IT}.
In
\cite{Zahedi&Mohseni&ElGamal:04ISIT,ElGamal&Mohseni&Zahedi:06IT},
the authors considered general time-varying linear processing at
the relay in Gaussian channels. Although they obtained the
capacity for frequency-division strictly-causal linear relaying,
the general linear relay case was not explored fully
\cite{Zahedi&Mohseni&ElGamal:04ISIT,ElGamal&Mohseni&Zahedi:06IT}.
In the general linear relay case, the problem is a sequence of
non-convex optimization problems, and it is seemingly intractable.
To circumvent such difficulty, in this paper we consider tractable
and practical LTI filtering at the source and relay. We find that
it is still a hard problem to obtain the capacity with a
single-letter characterization even in this case because the
search space still has countably infinite dimensions; optimal
source and relay filters may have  infinite impulse responses
(IIRs). However, we {\em provide a practical solution to design
the source and relay filters jointly to maximize the transmission
rate for general ISI Gaussian relay networks.}

Under the  LTI framework, the linear Gaussian relay problem can be
formulated in the frequency domain using the Toeplitz distribution
theorem \cite{Grenander&Szego:book,Brockwell&Davis:book}. When the
relay filter is given and there is no power constraint on the
relay, the problem reduces to the classical ISI channel problem
for which the optimal strategy is known as water-filling in the
frequency domain \cite[pp. 407 - 430]{Gallager:book}. However, the
freedom to design the relay filter and the power constraint at the
relay make the problem far more difficult than the classical ISI
channel problem, especially when  stability and causality
constraints are imposed on the source and relay filters. Our
approach to this problem is that we first convert the problem to a
{\em constrained optimization
 problem  in a finite dimensional space} by restricting the source and
relay filters to the class of finite impulse response (FIR)
filters as in most practical filtering applications, and then
apply the {\em projected subgradient method}, initially proposed
by Polyak \cite{Polyak:69USSR} and  fully developed by Yamada et
al. \cite{Yamada&Ogura:04NFAO,Slavakis&Yamada&Ogura:06NFAO}, to
this problem. Numerical results show that our  method performs
well and yields a noticeable gain over the AF scheme in ISI relay
channels.

\vspace{0.5em} \noindent {\textit{Notations and Organizations}}

We will make use of standard notational conventions.
    Vectors and matrices are written
    in boldface with matrices in capitals. All vectors are column
    vectors. For a scalar $a$, $a^*$ denotes its complex
    conjugate.
 For a matrix $\Abf$, $\Abf^T$, $\Abf^H$ and $\mbox{tr}(\Abf)$ indicate the transpose,  Hermitian
 transpose and trace
 of $\Abf$, respectively, and  $\Abf(m,n)$ denotes the $m$-th row and $n$-th column element of $\Abf$.
    $\mbox{diag}(d_1,\cdots,d_n)$ denotes a diagonal matrix with elements
    $d_1,\cdots,d_n$.  $\bf{I}_n$ stands for the identity matrix
of size $n$ (the subscript is omitted
    when unnecessary), and ${\mathbf {0}}$ denotes a vector of all zero elements.
    For a vector $\abf$, $||\abf||$ denotes its
    2-norm.      The notation $\xbf\sim
\Nc(\mubf,\Sigmabf)$ means that $\xbf$ is
   Gaussian-distributed with mean vector $\mubf$ and
    covariance matrix $\Sigmabf$. $\Ebb\{\cdot\}$ denotes the
    expectation. For two signal processes $x[n]$ and $y[n]$, $x[n]
    * y[n]$ denotes the convolution of the two processes.
    $\Rbb$,
      ${\mathbb{I}}$,  ${\mathbb{I}}_+$ and $\Nbb$ denote the
    sets of real numbers, integers, nonnegative integers and natural numbers,
    respectively. For two sets $A$ and $B$, $A\backslash B$
    denotes the set minus operation. $j=\sqrt{-1}$.

 This paper is organized as follows. The system
model and background are described in Section
\ref{sec:systemmodel}. In Section \ref{sec:RateFormula}, the rate
formula in the frequency domain is derived under the LTI model,
and the performance of LTI relaying in flat-fading channels is
investigated in Section \ref{sec:FlatFadingChannels}. In Section
\ref{sec:JointSourceRelayISIDesign}, a joint source and relay
filter design method is proposed based on the projected
subgradient method, and its performance in ISI channels is
examined in Section \ref{sec:numerical},  followed by conclusions
in Section \ref{sec:conclusions}.

\section{System Model and Background}
\label{sec:systemmodel}

\begin{figure}[htbp]
\centerline{
    \begin{psfrags}
    \psfrag{x}[r]{{ $x_s[n]$}}  %
    \psfrag{z1}[c]{{ $w_r[n]$}} %
    \psfrag{y1}[c]{{ $y_r[n]$}} %
    \psfrag{x1}[c]{{ $x_r[n]$}} %
    \psfrag{z}[c]{ $w_d[n]$} %
    \psfrag{y}[l]{ $y_d[n]$}
    \psfrag{hsr}[c]{$H_{sr}(z)$} %
    \psfrag{hrd}[c]{$H_{rd}(z)$} %
    \psfrag{hsd}[c]{$H_{sd}(z)$} %
     \scalefig{0.65}\epsfbox{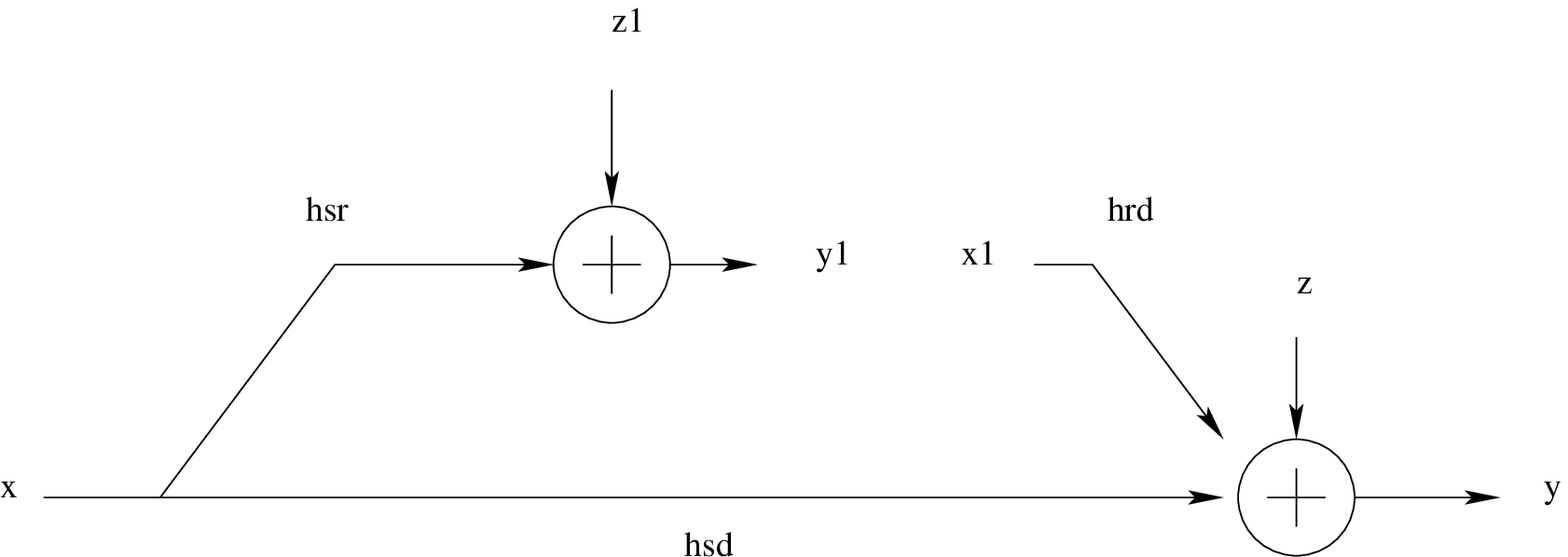}
    \end{psfrags}
} \caption{System model} \label{fig:systemModel}
\end{figure}

We consider the general discrete-time additive white Gaussian
noise (AWGN) relay network composed of source, relay and
destination nodes, as shown in Fig. \ref{fig:systemModel}, where
the source and relay nodes have maximum available average power
$P_s$ and $P_r$, respectively.  We assume that
 all propagation channels (i.e., the source-to-relay (S-R), relay-to-destination (R-D) and
 source-to-destination (S-D) channels) are linear, time-invariant and causal, and their impulse responses
 are  absolutely
 summable, i.e., $\sum_{l=0}^\infty |h_{sr}[l]| < \infty$, $\sum_{l=0}^\infty
|h_{rd}[l]| < \infty$ and $\sum_{l=0}^\infty |h_{sd}[l]| <
\infty$,  where $h_{sr}[l]$, $h_{rd}[l]$ and $h_{sd}[l]$ are the
S-R, R-D and S-D channel impulse responses, respectively. Due to
the absolute summability, the $z$-transforms of the propagation
channel impulse responses are well-defined and given by $H_{sr}(z)
= \sum_{l=0}^\infty h_{sr}[l]z^{-l}$, $H_{rd}(z) =
\sum_{l=0}^\infty h_{rd}[l] z^{-l}$ and  $H_{sd}(z) =
\sum_{l=0}^\infty h_{sd}[l] z^{-l}$.
  Then,
the received signals at the relay and destination at the $n$-th
symbol time  are given by
\begin{eqnarray}
y_r[n] &=&  h_{sr}[n] * x_s[n] + w_r[n],~~~\mbox{and}\\
y_d[n] &=&  h_{sd}[n] * x_s[n] + h_{rd}[n]*x_r[n] + w_d[n],
\end{eqnarray}
respectively, where $x_s[n]$ is the transmitted signal process at
the source; $x_r[n]$ and $y_r[n]$ are the transmitted and received
signal processes at the relay, respectively; $y_d[n]$ is the
received signal process at the destination; and  the  noise
processes $w_r[n]$ at the relay and $w_d[n]$ at the destination
are independent zero-mean white Gaussian processes with variance
$\sigma^2$.

\begin{figure}[htbp]
\centerline{
    \begin{psfrags}
    \psfrag{sc}[c]{{\footnotesize \textsf{Source} }}  %
    \psfrag{rel}[c]{{\footnotesize \textsf{Relay}}}  %
    \psfrag{sz}[c]{{ $T(z)$}}  %
    \psfrag{rz}[c]{{ $H(z)$}}  %
    \psfrag{xt}[c]{{ $\tilde{x}_s[n]$}}  %
    \psfrag{x}[c]{{ $x_s[n]$}}  %
    \psfrag{z1}[c]{{ $w_r[n]$}} %
    \psfrag{z}[c]{ $w_d[n]$} %
    \psfrag{y}[c]{ $y_d[n]$}
    \psfrag{a}[l]{{$H_{sr}(z)$}} %
    \psfrag{1}[c]{{$H_{sd}(z)$}} %
    \psfrag{b}[c]{{$H_{rd}(z)$}} %
     \scalefig{0.65}\epsfbox{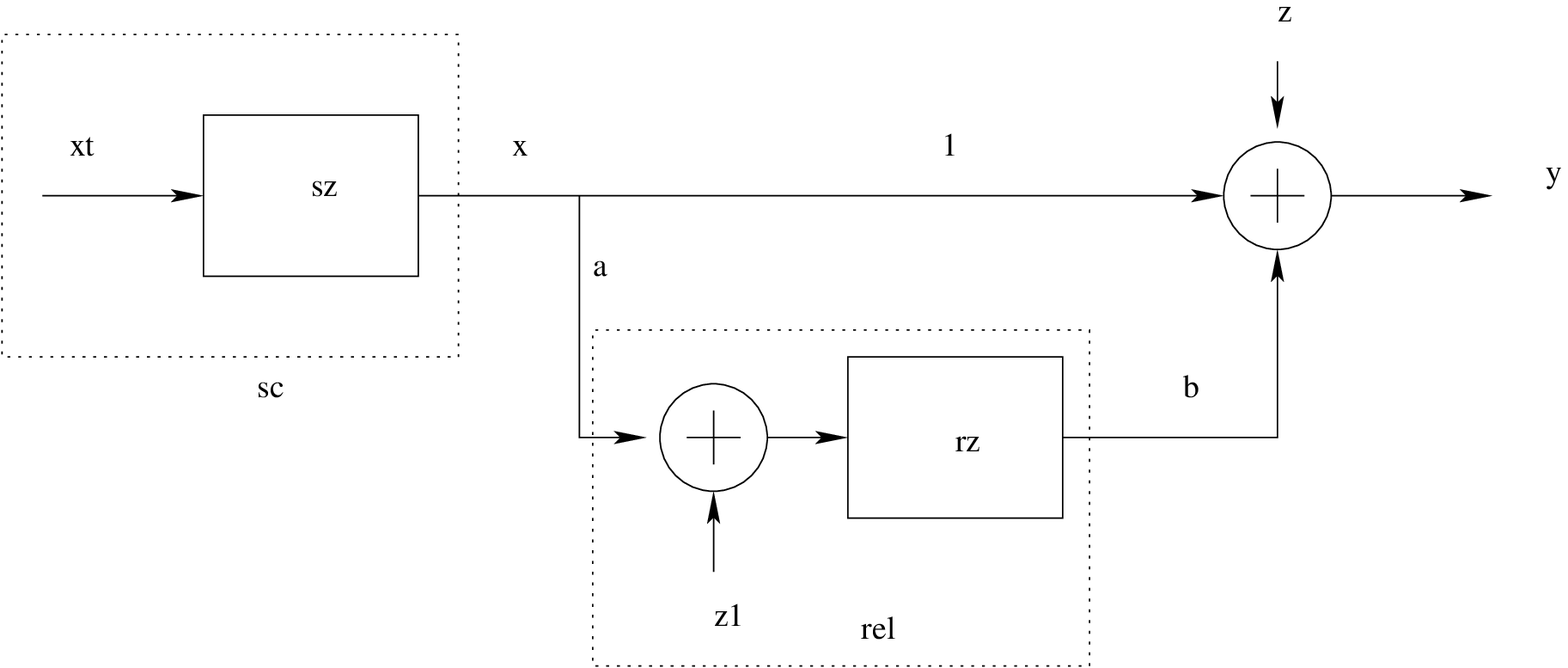}
    \end{psfrags}
} \caption{System model with linear time-invariant filters}
\label{fig:systemModelFilter}
\end{figure}

We consider  the linear and causal processing at the relay. The
general causal linear processing at the relay is given by
\begin{equation} \label{eq:LTVcase}
x_r[n]= \sum_{l \le n} d_{nl} y_{r}[l],
\end{equation}
for arbitrary linear combination coefficients $d_{nl}$, as
considered in
\cite{Zahedi&Mohseni&ElGamal:04ISIT,ElGamal&Mohseni&Zahedi:06IT}.
However, such linear processing requires time-varying filtering at
the relay, and is not readily realizable. Thus, in this paper, we
restrict ourselves to the case of LTI causal filtering at the
relay, as shown in Fig. \ref{fig:systemModelFilter}.
In this case, the relay output is given by
\begin{equation}  \label{eq:LTIfilteringRelay}
x_r[n] = \sum_{l=0}^\infty  h_l y_r[n-l],
\end{equation}
where $[h_0,h_1,h_2,\cdots]$ is the time-invariant  impulse
response of the relay filter and its $z$-transform is given by
$H(z) = \sum_{l=0}^\infty h_l z^{-l}$.  (In the case of  strict
causality, we have $h_0=0$.) The received signal
\eqref{eq:LTIfilteringRelay} at the relay can be written in
matrix form as (\ref{eq:LTIfiltering}), and the filtering matrix
$\Hbf_n$ in (\ref{eq:LTIfiltering})  has a Toeplitz structure.
\begin{equation} \label{eq:LTIfiltering}
 \left[
\begin{array}{c}
  x_r[0]   \\
  x_r[1]   \\
  \vdots \\
  x_r[n-1]   \\
\end{array}
\right]  = \underbrace{\left[
\begin{array}{ccccc}
  h_0      & 0     &  \cdots   & \cdots & 0     \\
  h_1      & h_0   &  0     &  \cdots   & 0 \\
  h_2      & h_1   & h_0     & 0      &  \vdots \\
  \vdots   & \ddots     &  \ddots      & \ddots &  0      \\
  h_{n-1}         & \cdots &  h_2      &   h_1    &  h_0      \\
\end{array}
\right]}_{=:\Hbf_n} \left[
\begin{array}{c}
  y_r[0]   \\
  y_r[1]   \\
  \vdots \\
  y_r[n-1]   \\
\end{array}
\right] + \left[
\begin{array}{c}
  w_r[0]   \\
  w_r[1]   \\
  \vdots \\
  w_r[n-1]   \\
\end{array}
\right].
\end{equation}
We assume the stability (i.e., $\sum_{l=0}^\infty |h_l| < \infty$)
and {\em realizability}\footnote{It means that the LTI response
$H(z)$ has a rational transfer function and it can readily be
implemented by an autoregressive  moving-average (ARMA) filter
\cite{Chen:book}.} for the relay filter. Since all processing from
the source and to the destination is linear and time-invariant,
the received signal at the destination in the $z$-domain is given
by
\begin{eqnarray}  \label{eq:SystemReceptionModel}
Y_d(z) &=& (H_{sd}(z)+H_{rd}(z)H(z)H_{sr}(z))X_s(z)  +
H_{rd}(z)H(z)W_r(z) + W_d(z),
\end{eqnarray}
where $W_r(z)$ and $W_d(z)$ are the $z$-transforms of noise
processes $w_r[n]$ and $w_d[n]$, respectively.

\subsection{Background} \label{subsec:background}

In this subsection, we briefly summarize some relevant results
including the eigen-structure of Toeplitz matrices and the
spectral factorization for the development in later sections. For
a zero-mean\footnote{ In the case of a known non-zero mean, the
mean of the process can be subtracted and the result can still be
applied.} stationary random process $y[n]$, the covariance
sequence and its $z$-spectrum are given by
\begin{equation}
r_y[k] = \Ebb\{ y[n] y^*[n-k] \} = r_y^*[-k] ~~~\mbox{and}~~~
S_y(z) := \sum_{k=-\infty}^{\infty} r_y[k]z^{-k} = \Ebb\{
Y(z)y[0]\},
\end{equation}
respectively, where $Y(z)= \sum_n y[n]z^{-n}$.  The covariance
matrix of a finite collection $\ybf_n :=[y[0],y[1],\cdots,$
$y[n-1]]^T$ is given by
\begin{equation} \label{eq:StationaryCovMtx}
\Sigmabf_{n}^y := \Ebb\{\ybf_n \ybf_n^H\}= \left[
\begin{array}{cccc}
  r_y[0]   &  r_y[-1]       & \cdots &  r_y[-n+1]  \\
  r_y[1]   &  r_y[0]        &        &  \vdots   \\
  \vdots   &  \vdots        & \ddots &  r_y[-1]  \\
  r_y[n-1] &  r_y[n-2]      & \cdots &  r_y[0]   \\
\end{array}
\right].
\end{equation}

\vspace{0.5em}
\begin{theorem} [Asymptotic eigen-structure of
Toeplitz covariance matrices \cite{Brockwell&Davis:book}, p. 135]
\label{theo:asymEigenToep}  Let $r_y[k]$ be an absolutely summable
autocovariance sequence of a stationary process $y[n]$, let
$S_y(e^{j\omega})$ be its power spectral density (PSD), i.e.,
$S_y(e^{j\omega})=S_y(z)|_{z=e^{j\omega}}$,  and let $\Dbf_n$ be
the $n\times n$ matrix,
\[
\Dbf_n =
\mbox{diag}(S_y(e^{j0}),S_y(e^{j\omega_1}),S_y(e^{-j\omega_1}),\cdots,
S_y(e^{\omega_{(n-1)/2 }}),S_y(e^{-\omega_{ (n-1)/2 }})),
\]
where $\omega_l= \frac{2\pi l}{n}$. Then, for the covariance
matrix $\Sigmabf_{n}^y = [r_y[k-l]]_{k,l=1}^n$,  the components of
$\Xbf^{(n)}:= \Wbf\Sigmabf_{n}^y\Wbf^{-1} -  \Dbf_n$ converge to
zero uniformly  as $n\rightarrow \infty$ (i.e. $\sup_{1\le k,l \le
n} |\Xbf^{(n)}(k,l)| \rightarrow 0$), where $\Wbf$ is the discrete
Fourier transform (DFT) matrix.
\end{theorem}
\vspace{1em}

\noindent For even $n$, we have a similar result with a slight
modification. Theorem \ref{theo:asymEigenToep} simply states that
the eigenvalues of the Toeplitz covariance matrix of a stationary
process are  the uniform samples of its spectrum. Using Theorem
\ref{theo:asymEigenToep}, the following can easily be shown.

\vspace{0.5em}
\begin{theorem}[Toeplitz distribution theorem
\cite{Grenander&Szego:book}, p. 65] \label{theo:ToeplitzDist} Let
$\{\lambda_i^{(n)}\}$ be the eigenvalues of the Toeplitz
covariance matrix $\Sigmabf_n^y$ of a stationary process $y[n]$.
Then,
\begin{equation}
\lim_{n\rightarrow\infty} \frac{1}{n}\sum_{i=1}^n
f(\lambda_i^{(n)})= \frac{1}{2\pi} \int_{-\pi}^{\pi}
f(S_y(e^{j\omega}))d\omega
\end{equation}
for any continuous function $f(\cdot)$.
\end{theorem}

\vspace{0.5em} \noindent In addition to the asymptotic
eigen-structure of Toeplitz covariance matrices, we need some
background in the spectral theory for stationary random processes,
especially canonical spectral factorization.

\vspace{0.5em}
\begin{definition}[Canonical Spectral Factorization \cite{Kailath:book}, p. 197]
Let $S_y(z)$ be a rational z-spectrum of a finite power process
and assume that $S_y(z)$ is strictly positive. Then, the canonical
spectral factorization of $S_y(z)$ is given by
\begin{equation} \label{eq:spectralfactor}
S_y(z) = L(z) \gamma_e L^\sharp(z),
\end{equation}
where $L(z)=\sum_{i=0}^\infty l_i z^{-i}$ is a unique stable,
causal, monic and minimum-phase
 (SCAMP) filter (i.e., the zeros and poles of $L(z)$ are strictly inside the unit
circle and $L(\infty)=1$ (or equivalently $l_0=1$)), and $\gamma_e
> 0$. Here, $L^\sharp(z):= L^*(z^{-*})$ denotes the para-Hermitian
conjugate.
\end{definition}

\section{The Rate Formula in Frequency-Domain for LTI Relays}
\label{sec:RateFormula}

First, note that the overall channel model
\eqref{eq:SystemReceptionModel} with LTI relay filtering is still
a linear additive {\em stationary} Gaussian noise channel. Thus,
for a given relay filter, the overall channel with the LTI relay
filter reduces back to the classical ISI channel with stationary
Gaussian noise\footnote{However, the major difference between the
two problems is that in the relay problem we even have to design
the overall channel by properly choosing the relay filter and the
power constraints at the source and relay are interwined.}. In
this case, stationary Gaussian signal processes with well-defined
spectra are sufficient to achieve the capacity \cite[pp. 407 -
430]{Gallager:book}. Hence, we assume that the source (or input)
process $x_s[n]$ is a stationary Gaussian process. By
concatenating symbols at the source up to time $n-1$, we have
\begin{equation}
\xbf_n^{s} := [x_s[0],x_s[1],\cdots,x_s[n-1]]^T \sim \Nc({\mathbf
0}, \Sigmabf_n^{x_s}),
\end{equation}
and vectors  $\ybf_n^r$, $\xbf_n^r$ and $\ybf_n^d$ are constructed
similarly for the relay and destination nodes. Then, the power
constraints for the source and relay are respectively given by
\begin{eqnarray}
(1/n) \mbox{tr}( \Sigmabf_n^{x_s}) &\le& P_s, ~~~\mbox{and}\label{eq:MatPwrConst1}\\
(1/n) \Ebb\{ \mbox{tr} (\Hbf_n\ybf_r^n (\Hbf_n \ybf_r^n)^H) \} &=&
(1/n) \mbox{tr}(\Hbf_n(\Hbf_{n}^{sr}\Sigmabf_n^{x_s} \Hbf_{n}^{sr}
+ \sigma^2 \Ibf) \Hbf_n^H ) \le P_r,  \label{eq:MatPwrConst2}
\end{eqnarray}
 where $\Hbf_{n}^{sr}$ is the filtering matrix for the S-R
channel
 constructed based on $\{h_{sr}[l]\}$ similar to $\Hbf_n$ in
\eqref{eq:LTIfiltering}.  Thus, the maximum rate with LTI relaying
for block size $n$ is given by maximizing the mutual information
between $\xbf_n^s$ and $\ybf_n^d$ over $\Sigmabf_n^{x_s}$ and
$\Hbf_n$ under power constraints \eqref{eq:MatPwrConst1} and
\eqref{eq:MatPwrConst2}, and the capacity with LTI relaying is
given by its limit
\begin{equation}  \label{eq:CLTIformula}
C_{LTI} = \lim_{n\rightarrow\infty} \sup_{\Sigmabf_n^{x_s},\Hbf_n}
\frac{1}{n}I(\xbf_n^s;\ybf_n^d),
\end{equation}
as $n\rightarrow \infty$ \cite{ElGamal&Mohseni&Zahedi:06IT}, where
\begin{eqnarray}
I(\xbf_n^s;\ybf_n^d)&=& H(\ybf_n^d) -
H(\ybf_n^d|\xbf_n^s),\nonumber\\
&=& \log |\sigma^2\Hbf_n^{rd}\Hbf_n \Hbf_n^H (\Hbf_n^{rd})^H +
\sigma^2\Ibf +
(\Hbf_n^{sd}+\Hbf_n^{rd}\Hbf_n\Hbf_n^{sr})\Sigmabf_n^{x_s}(\Hbf_n^{sd}+\Hbf_n^{rd}\Hbf_n\Hbf_n^{sr})^H|
\nonumber\\
&& ~~~~~~~~~~~~~~ - \log |\sigma^2\Hbf_n^{rd}\Hbf_n \Hbf_n^H
(\Hbf_n^{rd})^H + \sigma^2\Ibf|.
\end{eqnarray}
 Here, $\Hbf_n^{sd}$ and $\Hbf_n^{rd}$ are the filtering matrices for the S-D and R-D channels, respectively.
  Note
 that \eqref{eq:CLTIformula} is still valid for general linear
time-varying relay filtering with $\Hbf_n$ given by  an arbitrary
lower triangular matrix.  As mentioned in
\cite{ElGamal&Mohseni&Zahedi:06IT}, the computation of capacity
and the design of capacity-achieving (or at least reasonable)
$\Sigmabf_n^{x_s}$ and $\Hbf_n$ are difficult problems in the case
of general linear causal relay filtering.  In the time-varying
case, if we increase $n$ by one, at least $2n$ new variables
$\{\Sigmabf^{x_s}_n(n,1),\Sigmabf^{x_s}_n(n,2), \cdots,
\Sigmabf^{x_s}_n(n,n),$ $d_{n1},d_{n2},\cdots, d_{nn} \}$ appear
(see \eqref{eq:LTVcase}), and thus the complexity of the problem
increases with the order of $n!$ to make the problem difficult
\cite{Zahedi&Mohseni&ElGamal:04ISIT,ElGamal&Mohseni&Zahedi:06IT}.
 In the LTI case with a stationary source process, however,  we have only two new variables
$r_{x_s}[n-1]$ and $h_{n-1}$ for the increase of the problem size
from $n-1$ to $n$ because of the Toeplitz structure of the
covariance matrix in \eqref{eq:StationaryCovMtx} and the filtering
matrix in \eqref{eq:LTIfiltering}.  Following the best input
covariance matrix and relay filter for the problem size $n$ is
equivalent to designing the best infinitely long autocovariance
sequence $\{r_{x_s}[k], ~k=0,1,\cdots\}$ and infinitely long relay
filter $\{h_l, ~l=0,1,\cdots\}$ first and then increasing the
problem size. Thus, in the LTI case, we have
\begin{equation}
C_{LTI} = \sup_{\{r_{x_s}[k]\}, H(z)} \lim_{n\rightarrow\infty}
\frac{1}{n}\left[ I(\xbf_n^s;\ybf_n^d) \vert
\Sigmabf_n^{x_s}(\{\gamma_{x_s}[k]\}), \Hbf_n(H(z))\right],
\end{equation}
where the respective dependence of $\Sigmabf_{x_s}^n$ and $\Hbf_n$
on $\{r_{x_s}[k]\}$ and $H(z)$ is explicitly shown.
  Here,
taking the limit of $n$ simplifies the problem significantly  due
to Theorem \ref{theo:ToeplitzDist} since the eigenvalues  are
strictly positive due to the additive noise term and since
$f(t)=\log(t)$ is a continuous function of $t$ for $t
> 0$. By Theorems \ref{theo:asymEigenToep} and
\ref{theo:ToeplitzDist} we have
\begin{equation}  \label{eq:CLTIfd}
C_{LTI} = \sup_{S_{x_s}(e^{j\omega}), H(z)} \frac{1}{2\pi}
\int_{-\pi}^\pi \frac{1}{2}\log_2\left( 1 +
\frac{|H_{sd}(e^{j\omega})+H_{sr}(e^{j\omega}) H(e^{j\omega})
H_{rd}(e^{j\omega})|^2}{\sigma^2(|H_{rd}(e^{j\omega})
H(e^{j\omega})|^2 + 1)}  S_{x_s}(e^{j\omega}) \right) d\omega,
\end{equation}
where the input spectrum $S_{x_s}(e^{j\omega}) =
\sum_{k=-\infty}^\infty r_{x_s}[k]e^{-j\omega k}$, since the
eigenvalues of a covariance matrix are the samples of its spectrum
and the determinant of a covariance matrix is the product of its
eigenvalues.  Here, we define the overall channel-to-noise power
ratio  (CNR) density as
\begin{equation} \label{eq:CNRformula}
\mbox{CNR} (e^{j\omega})
:=\frac{|H_{sd}(e^{j\omega})+H_{sr}(e^{j\omega}) H(e^{j\omega})
H_{rd}(e^{j\omega})|^2}{\sigma^2(|H_{rd}(e^{j\omega})
H(e^{j\omega})|^2 + 1)}=\frac{N(e^{j\omega})}{D(e^{j\omega})},
\end{equation}
where $N(e^{j\omega})$ and $D(e^{j\omega})$ are the numerator and
denominator of the CNR density, respectively. Note that the CNR
density captures the overall channel response from source to
destination. When the CNR density is multiplied by the input
signal PSD, the product becomes the overall SNR density at the
destination. (This quantity will be used in later sections.) In
addition to the rate formula \eqref{eq:CLTIfd} in the frequency
domain, the power constraints can also be expressed in the
frequency domain as $n\rightarrow \infty$. As $n\rightarrow
\infty$, again by Theorems \ref{theo:asymEigenToep} and
\ref{theo:ToeplitzDist}, the power constraints
\eqref{eq:MatPwrConst1} and \eqref{eq:MatPwrConst2} are
respectively given by
\begin{eqnarray}
\frac{1}{2\pi} \int_{-\pi}^\pi S_{x_s}(e^{j\omega}) d\omega &\le&
P_s, ~~~\mbox{and} \label{eq:powerConstSpec1}\\
\frac{1}{2\pi} \int_{-\pi}^\pi
|H(e^{j\omega})|^2(|H_{sr}(e^{j\omega})|^2S_{x_s}(e^{j\omega})+\sigma^2)
d\omega &\le& P_r, \label{eq:powerConstSpec2}
\end{eqnarray}
 since the trace of a matrix is the sum of its eigenvalues. Thus,
the LTI relay problem is summarized by \eqref{eq:CLTIfd},
\eqref{eq:powerConstSpec1} and \eqref{eq:powerConstSpec2}. Note
that for a given relay filter $H(z)$ the problem without the power
constraint \eqref{eq:powerConstSpec2} reduces to the well-known
ISI channel problem and the solution of $S_{x_s}(e^{j\omega})$ is
given by water-filling in the frequency domain
\cite{Cover&Thomas:book2}. However, the freedom to design $H(z)$
and the relay power constraint \eqref{eq:powerConstSpec2} make the
problem far more difficult than the simple ISI channel problem. To
construct a practical method to solve this problem, we further
assume that the input spectrum $S_{x_s}(e^{j\omega})$ is also {\em
realizable}. That is, its canonical spectral factorization is
given by
\begin{equation}  \label{eq:sourceSpecFactor}
S_{x_s}(z) = \alpha \tilde{T}(z)\tilde{T}^\sharp (z)=T(z)T^\sharp
(z), ~~~~~(T(z)=\sqrt{\alpha}\tilde{T}(z)),
\end{equation}
where the SCAMP filter $\tilde{T}(z)$ has a rational transfer
function and, thus, $S_{x_s}(z)$ is a rational spectrum. In this
case, the source process $x_s[n]$ can be modelled as the output of
the  stable and causal ARMA filter $T(z)$ driven by a white
Gaussian process $\tilde{x}_s[n]$ with unit variance, as seen in
Fig. \ref{fig:systemModelFilter}.  Thus, the rate maximization
problem under {\em LTI relaying with realizable input spectra} now
reduces to a {\em joint design problem of LTI source and relay
filters}. Obtaining the capacity in a closed form still seems to
be a difficult problem even in the LTI relay case. However, we
propose a very effective and practical solution to this joint
filter design problem in Section
\ref{sec:JointSourceRelayISIDesign}. Before we tackle this
problem, we investigate the problem in the case that all S-D, S-R
and R-D channels have flat frequency responses in the next
section.

\section{Examination of LTI Relaying in Flat-Fading Channels}
\label{sec:FlatFadingChannels}

In the case of flat fading, we have the system model
\eqref{eq:SystemReceptionModel} in which  each of S-R, R-D and S-D
channels has only one tap, i.e., $H_{sd}(z) = 1$,  $H_{sr}(z) = a$
and $H_{rd}(z) = b$,  as considered in
\cite{Zahedi&Mohseni&ElGamal:04ISIT,ElGamal&Mohseni&Zahedi:06IT}.
Then, the received signal model in the $z$-domain is given by
\begin{equation}  \label{eq:scalarOneTapDataModel}
Y_d(z) = (1+abH(z))X_s(z) + bH(z)W_r(z) + W_d(z).
\end{equation}

\subsection{The One-Tap Relay Filter Case}
\label{subsec:onetaprelayfilter}

 First,
consider the well-known AF relaying. In this case, we have
\[
x_r[n] = d y_r[n],
\]
where $\Ebb\{x_s^2\}=P_s$ and $0 \le d \le
\sqrt{\frac{P_r}{a^2P_s+\sigma^2}}=:d_{max}$ to satisfy the power
constraints, and the received signal model is given by
\begin{equation} \label{eq:AFrelayModel}
y_d[n]= (1+abd) x_s[n] + bdw_r[n]+w_d[n].
\end{equation}
Due to the simple data model \eqref{eq:AFrelayModel}, the
achievable rate in this case is known and given by
\begin{equation}   \label{eq:InstAFachievRate}
R_{AF} = \max_{0\le d\le d_{max}} \frac{1}{2}\log \left( 1+
\frac{(1+abd)^2}{b^2d^2+1}\cdot\frac{P_s}{\sigma^2} \right),
\end{equation}
and the optimal value of $d$  is explicitly given by
\begin{equation} \label{eq:optdAF}
d^* =
\min\left\{\frac{a}{b},\sqrt{\frac{P_r}{a^2P_s+\sigma^2}}\right\}.
\end{equation}

Now consider the one-tap LTI relay filter with an arbitrary delay:
\begin{equation}
H(z) = d z^{-\Delta}
\end{equation}
for some integer $\Delta \ge 0$\footnote{Note that strict
causality implies $\Delta
>0$.}, where the relay gain $d$ can be
optimized under the power constraints. Note in the system model
\eqref{eq:scalarOneTapDataModel} that the relay filter affects
both the channel gain and noise spectrum. However, in the one-tap
relay filter case, the problem is simplified because the overall
noise spectrum is white.  In this case, the overall channel gain
is given by $1+abH(z) = 1 + abd z^{-\Delta}$ and the overall noise
spectrum is given by
\begin{eqnarray*}
b^2 H(z) H^\sharp(z) \sigma^2+ \sigma^2
&=&b^2\sigma^2(dz^{-\Delta})(dz^{-\Delta})^\sharp + \sigma^2,\\
&=& (b^2d^2+1)\sigma^2,
\end{eqnarray*}
 since $(z^{-\Delta})^\sharp = z^\Delta$. Note that
the overall noise process in this case is white and equivalent to
that in the AF data model \eqref{eq:AFrelayModel}; both have the
same variance $(b^2d^2+1)\sigma^2$. Thus, the spectrum of $Y_d(z)$
is given by
\begin{equation}
S_{y_d}(e^{j\omega}) = |1+abH(e^{j\omega})|^2 S_{x_s}(e^{j\omega})
+ (b^2d^2+1)\sigma^2,
\end{equation}
and the channel frequency response is explicitly given by a
raised-cosine function:
\begin{eqnarray}
|1+abH(e^{j\omega})|^2 &=& (1 +ab d e^{-j\omega\Delta})(1 + abd
e^{j\omega\Delta}),  \nonumber \\
&=& 1+ 2 abd \cos (\omega \Delta) + a^2b^2d^2 \ge 0.
\end{eqnarray}
Since  $|H(e^{j\omega})|^2=d^2$ for $H(z)=dz^{-\Delta}$, from
\eqref{eq:powerConstSpec1} and \eqref{eq:powerConstSpec2}
 the
power constraints are given by
\begin{equation}
\frac{1}{2\pi}\int_{-\pi}^{\pi} S_{x_s}(e^{j\omega})d\omega \le
 P_s, ~~~\mbox{and} \label{eq:scalarOneTapPC1}
\end{equation}
\begin{equation} \label{eq:scalarOneTapPC2}
   \frac{d^2}{2\pi}\int_{-\pi}^{\pi}  ( a^2S_{x_s}(e^{j\omega}) + \sigma^2) d\omega    =d^2 \left( a^2 \frac{1}{2\pi}\int_{-\pi}^{\pi}  S_{x_s}(e^{j\omega})  d\omega + \sigma^2 \right) \le
 P_r,
\end{equation}
which are the same as those of the  AF scheme with $\Delta =0$.

The problem with the given relay filter $H(z)=dz^{-\Delta}$
reduces to the simple ISI channel problem, and the optimal input
spectrum $S_{x_s}^*(e^{j\omega})$ is obtained by water-filling
under the two simple power constraints (\ref{eq:scalarOneTapPC1})
and (\ref{eq:scalarOneTapPC2}). In the following theorem, we
establish the optimality of the AF scheme within the class of all
one-tap relay filters.

\vspace{0.5em} \begin{theorem} \label{theo:AFopt} Among all
one-tap linear relay filters, i.e., $H(z) = dz^{-\Delta}$ with
$\Delta \in \mathbb{I}$, the AF scheme with $\Delta=0$ maximizes
the achievable rate.
\end{theorem}
\vspace{0.5em}

\noindent {\em Proof:} For a given $\Delta  \ne 0$, let
\begin{equation}
 (S_{x_s}^*(e^{j\omega}), d^*) = \mathop{\arg \max}_{S_{x_s}(e^{j\omega}),d}  \frac{1}{2\pi}\int_{-\pi}^{\pi} \frac{1}{2}
 \log \left(  1+\frac{|1+abd e^{-j\omega \Delta}|^2}{(b^2d^2+1)\sigma^2} S_{x_s}(e^{j\omega}) \right )
 d\omega.
\end{equation}
Then, {\small
\begin{eqnarray}
&&\frac{1}{2\pi}\int_{-\pi}^{\pi} \frac{1}{2}
 \log \left(  1+\frac{|1+abd^* e^{-j\omega \Delta}|^2}{(b^2d^{*2}+1)\sigma^2} S_{x_s}^*(e^{j\omega}) \right) d\omega \nonumber\\
&\le& \frac{1}{2\pi}\int_{-\pi}^{\pi} \frac{1}{2}
 \log \left(  1+\frac{(1+abd^*)^2}{(b^2d^{*2}+1)\sigma^2} S_{x_s}^*(e^{j\omega}) \right) d\omega, \label{eq:Delta0opt1}\\
&\le& \sup_{S_{x_s}(e^{j\omega}),d}
\frac{1}{2\pi}\int_{-\pi}^{\pi} \frac{1}{2}
 \log \left(  1+\frac{(1+abd)^2}{(b^2d^{2}+1)\sigma^2} S_{x_s}(e^{j\omega}) \right) d\omega,\label{eq:Delta0opt2}\\
&\le& \sup_{S_{x_s}(e^{j\omega}),d}
 \frac{1}{2}
 \log \left(  1+\frac{(1+abd)^2}{(b^2d^{2}+1)\sigma^2} \frac{1}{2\pi}\int_{-\pi}^{\pi} S_{x_s}(e^{j\omega})d\omega \right),\label{eq:Delta0opt3}\\
&=& R_{AF}. \label{eq:Delta0opt4}
\end{eqnarray}
} Here,  \eqref{eq:Delta0opt1} is obtained because $|1+abd^*
e^{-j\omega\Delta}|^2 \le (1+abd^*)^2$.  \eqref{eq:Delta0opt2} is
obtained because the feasible set $(S_{x_s}(e^{j\omega}),d)$
satisfying the power constraint for $\Delta \ne 0$ is the same as
that for $\Delta=0$ when $H(z) =d z^{-\Delta}$. (See
 (\ref{eq:scalarOneTapPC1})
and (\ref{eq:scalarOneTapPC2}).) \eqref{eq:Delta0opt3} is obtained
by Jensen's inequality. Finally, \eqref{eq:Delta0opt4} is obtained
by the definition of $R_{AF}$ in \eqref{eq:InstAFachievRate}.
\hfill{$\blacksquare$} \vspace{0.5em}

\noindent Theorem \ref{theo:AFopt} states that the AF scheme with
$\Delta =0$ performs best within the class of one-tap relay
filters with arbitrary delays.  This is because the AF scheme
achieves coherent signal combining between the two signal paths
S-D and S-R-D. Instead of using the optimal water-filling source
filter, we can also consider a simple
 channel-equalizing source filter. However, the performance in this case is bad, as shown in the following
 theorem.

\vspace{0.5em}
\begin{theorem}  \label{theo:EqualizingSourceFilter}
The achievable rate by an equalizing source filter for the one-tap
relay filter $H(z) = d z^{-\Delta}$ is given by
\begin{equation}  \label{eq:RateEqualizingFilter}
 R_{1-tap,EQ} =
\sup_{0\le d < \min\left\{d_{max},\frac{1}{ab}\right\}}
\frac{1}{2}\log \left( 1 + \frac{1-(abd)^2}{b^2d^2
+1}\cdot\frac{P_s}{\sigma^2} \right)
\end{equation}
regardless of the value of $\Delta > 0$. Further,  the supremum is
given by $R_{1-tap,EQ} = \frac{1}{2}\log \left( 1 +
\frac{P_s}{\sigma^2} \right)$ achieved  when $d=0$.
\end{theorem}

\noindent {\em Proof:} We have $X_s(z) = \sqrt{\alpha}
\tilde{T}(z) \tilde{X}_s(z)$, where $\tilde{X}_s(z)$ is the
$z$-transform of the white Gaussian process $\tilde{x}_s[n]$ with
unit variance,  and the equalizing source filter is given by
\[
\tilde{T}(z) =
\frac{1}{1+abH(z)}=\frac{1}{1+abdz^{-\Delta}}=1-abdz^{-\Delta}+(abd)^2z^{-2\Delta}-(abd)^3z^{-3\Delta}+
\cdots.
\]
  When $0 \le d <
\frac{1}{ab}$, the overall channel response $1+abH(z)$ is SCAMP
and, thus,  the channel-equalizing source filter $\tilde{T}(z)$ is
also SCAMP. By the power constraint at the source,  we have
\begin{equation}
 \frac{1}{2\pi}\int_{-\pi}^{\pi} S_{x_s}(e^{j\omega})d\omega
 = \frac{\alpha}{2\pi}\int_{-\pi}^{\pi} \frac{1}{1+ 2 abd \cos (\omega \Delta) + (abd)^2}d\omega =
 P_s
\end{equation}
because $S_{x_s}(z) = \alpha \tilde{T}(z)\tilde{T}^\sharp(z)$.
Since $\int_{-\pi}^{\pi} \frac{1}{1+ 2 abd \cos (\omega \Delta) +
(abd)^2}d\omega=\frac{2\pi}{1-(abd)^2}$ for every integer $\Delta
> 0$, $\alpha = (1-(abd)^2)P_s$ regardless of the value of $\Delta
>0$. With the channel-equalizing source filter
$T(z)=\sqrt{\alpha}\tilde{T}(z)$, the data model is given by
$y_d[n] = \sqrt{\alpha}\tilde{x}_s[n] + w_{eff}[n]$,
 where
$\tilde{x}_s[n] \sim \Nc(0, 1)$ and $w_{eff}[n] \sim \Nc(0,
(b^2d^2+1) \sigma^2 )$, and the corresponding achievable rate is
given by \eqref{eq:RateEqualizingFilter}. Now consider
$\mbox{CNR}(d)=\frac{1-(abd)^2}{b^2d^2 +1}\frac{1}{\sigma^2}$ in
\eqref{eq:RateEqualizingFilter}. Its derivative with respect to
(w.r.t.) $d$ is given by $ \mbox{CNR}^\prime(d)= [ -
2a^2b^2d(b^2d^2+1)-(1-a^2b^2d^2)2b^2d]/[(b^2d^2+1)^2\sigma^2] \le
0$ for all $d\ge 0$. Thus, the rate is maximized when $d=0$.
\hfill{$\blacksquare$}

\begin{figure}[htbp]
\centerline{ \SetLabels
\L(0.25*-0.1) (a) \\
\L(0.76*-0.1) (b) \\
\endSetLabels
\leavevmode
\strut\AffixLabels{ \scalefig{0.5}\epsfbox{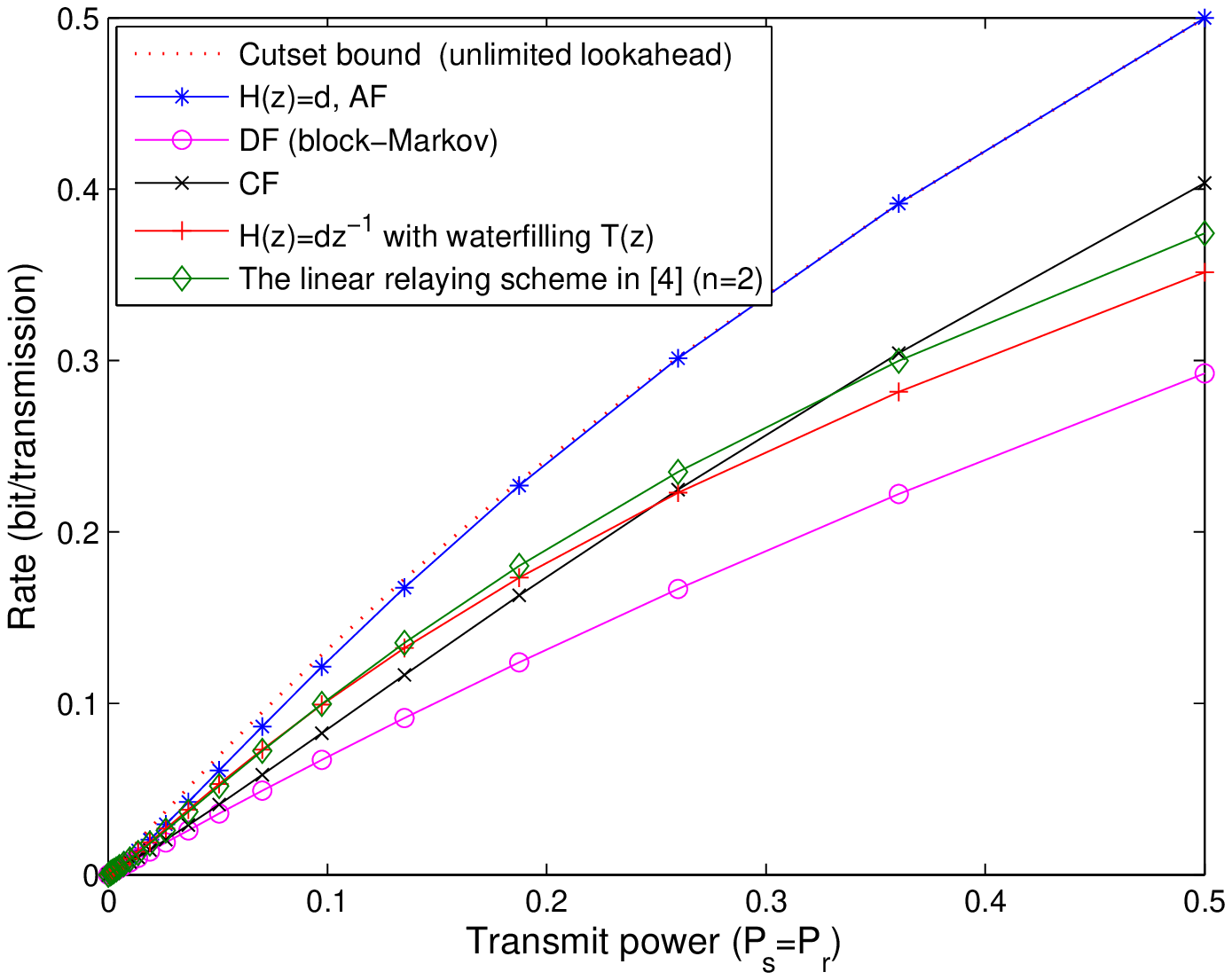}
\hspace{0.3cm} \scalefig{0.5}\epsfbox{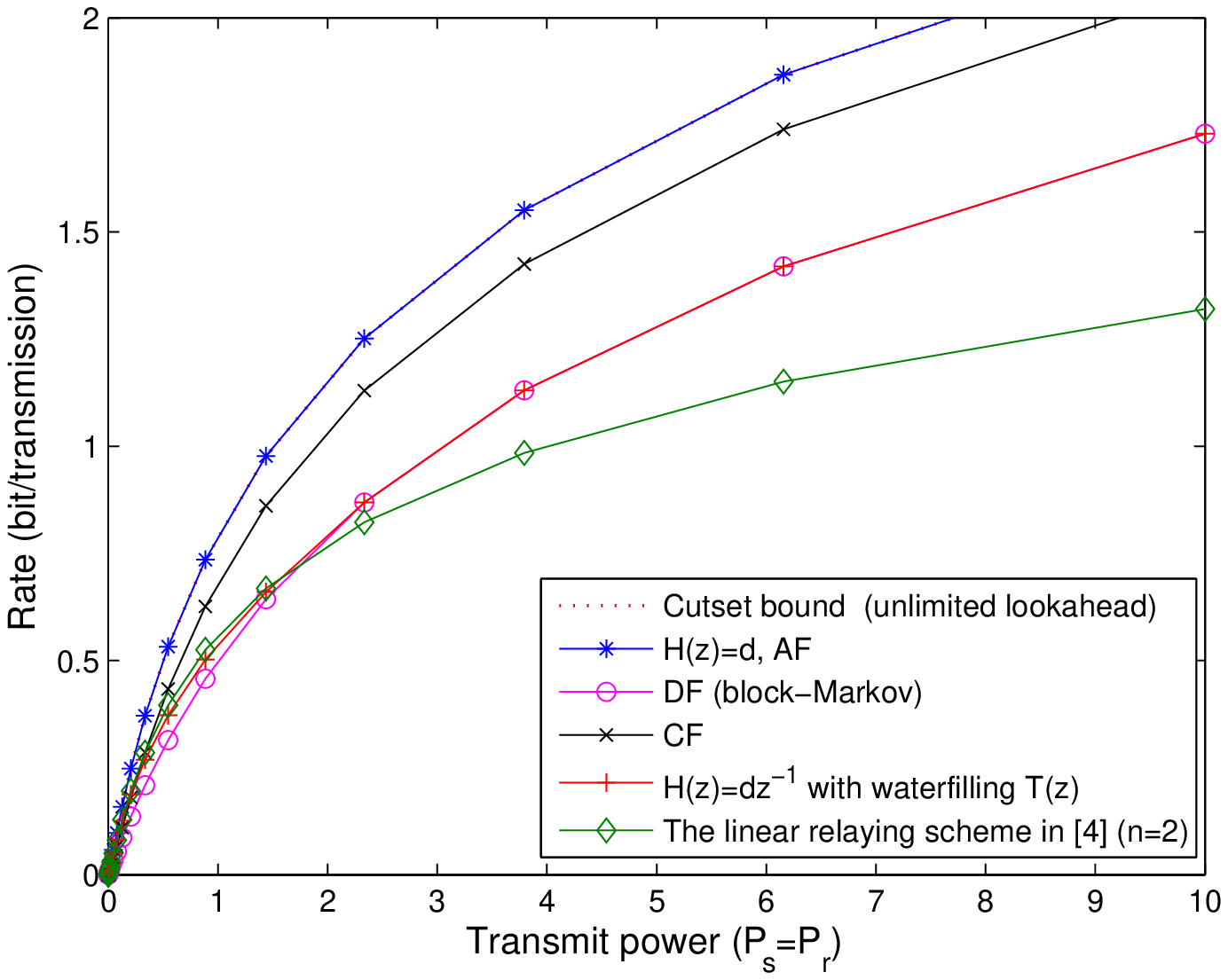} } }
\vspace{0.5cm} \centerline{ \SetLabels
\L(0.25*-0.1) (c) \\
\L(0.76*-0.1) (d) \\
\endSetLabels
\leavevmode
\strut\AffixLabels{ \scalefig{0.5}\epsfbox{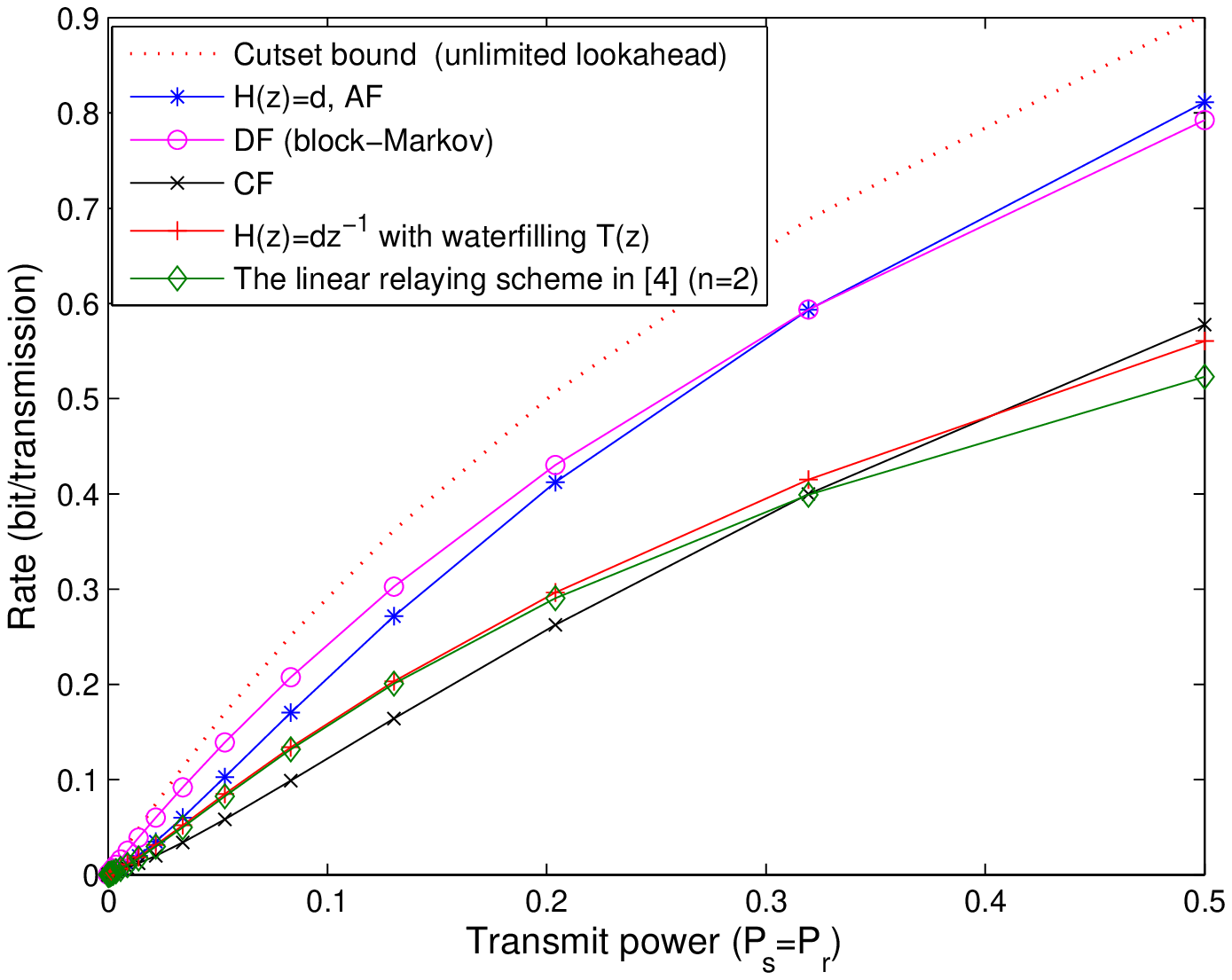}
\hspace{0.3cm} \scalefig{0.5}\epsfbox{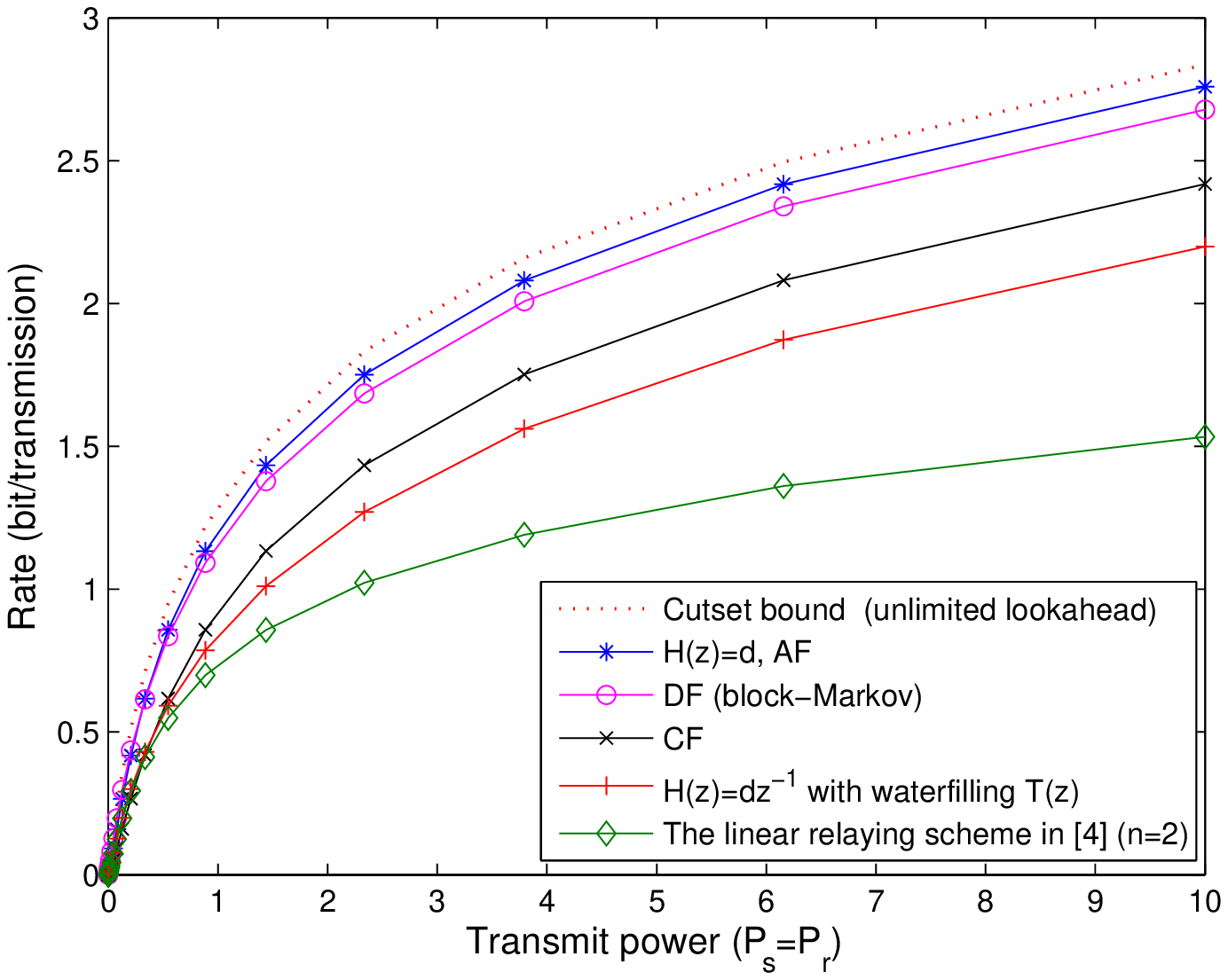} } }
\vspace{0.5cm}  \caption{Achievable rates of several schemes in
flat-fading channels ($P_s=P_r$, $\sigma^2=1$) - (a) and (b):
$a=1$, $b=2$, and (c) and (d): $a=2$, $b=2$}
\label{fig:flatFadingPlot}
\end{figure}

\noindent Theorem \ref{theo:EqualizingSourceFilter} states that it
is optimal to turn off the (one-tap) relay filter when the
channel-equalizing filter is to be used at the source. Thus, using
the channel-equalizing source filter is not a proper choice for
relay networks.

Fig. \ref{fig:flatFadingPlot} shows the achievable rates of
several relaying schemes.  For Fig.  \ref{fig:flatFadingPlot} (a)
and (b), which show the same curves with two different x-axis
ranges, we set $a=1$ and $b=2$, as in
\cite{ElGamal&Mohseni&Zahedi:06IT}. It is seen that
 simple
linear strictly causal schemes (one based on the filtering matrix $\left[\begin{array}{cc} 0 & 0 \\
d & 0\end{array}\right]$ in \cite{ElGamal&Mohseni&Zahedi:06IT} and
the other based on one-tap filtering $H(z) = dz^{-1}$) can
outperform the CF scheme in the low SNR region, as already known
from \cite{ElGamal&Mohseni&Zahedi:06IT}. In this case of $a=1$ and
$b=2$, the AF scheme  achieves the cut-set upperbound for $P_s=P_r
\ge 1/3$ \cite[Proposition 9]{ElGamal&Hassanpour&Mammen:07IT}. It
is interesting to observe that the simple linear scheme in
\cite{ElGamal&Mohseni&Zahedi:06IT} with $n=2$ performs better than
$H(z)=dz^{-1}$ filtering in some low SNR values, although the
latter outperforms the former  eventually at high SNR. Fig.
\ref{fig:flatFadingPlot} (c) and (d), again showing the same
curves in two different x-axis ranges, show the achievable rates
when $a=b=2$. In this case, it is seen that there is a gap between
the cut-set bound and the AF scheme. In all the cases, it is seen
that the two strictly causal linear schemes (one based on
two-symbol concatenation in \cite{ElGamal&Mohseni&Zahedi:06IT} and
the other based on one-tap LTI filtering $H(z)=dz^{-1}$) do not
outperform the AF scheme, as expected by Theorem \ref{theo:AFopt}.

\subsection{The Multiple-Tap Relay Filter Case: Insights from Ideal Low-Pass Filtering Relays}

\begin{figure}[htbp] \centerline{
\begin{psfrags}
    \psfrag{S_R}[l]{{\hspace{0cm}\small{$P_{stop}=0$}}}
    \psfrag{State1_1}[l]{{\hspace{-0.35cm}\small{Type 1-1}}}
    \psfrag{State1_2}[l]{{\hspace{-0.37cm}\small{Type 1-2}}}
    \psfrag{State2}[l]{{\hspace{-0.2cm}\small{Type 2}}}
    \psfrag{State3}[l]{{\hspace{-0.2cm}\small{Type 3}}}
    \psfrag{delta_condition}[l]{{\hspace{0.2cm}\small{$\delta^*=\frac{a}{b}$}}}  %
    \psfrag{obj}[l]{{\small{\hspace{-0.4cm} max $\frac{\omega_c}{\pi}\log_2\left( 1+  \frac{(1+ab\delta)^2}{ (b^2\delta^2+1)\sigma^2} \cdot \frac{\pi}{\omega_c}P_{pass} \right) + \frac{\pi-\omega_c}{\pi}\log_2\left( 1+  \frac{1}{\sigma^2} \cdot \frac{\pi}{\pi-\omega_c}P_{stop}  \right)$}}}  %
    \psfrag{constr0}[l]{{\small{s.t. $P_{pass} +  P_{stop} = P_s$, ~$P_{pass}\geq0$ and $\delta \le  \sqrt{\frac{P_r}{a^2P_{pass}+\frac{\omega_c}{\pi}\sigma^2}}$}}}
    \psfrag{omega_c}[l]{{\footnotesize{$\omega_c$}}}  %
    \psfrag{pi}[l]{{\footnotesize{$\pi$}}}  %
    \psfrag{l1}[l]{{\footnotesize{$l_1$}}}  %
    \psfrag{l2}[l]{{\footnotesize{$l_2$}}}  %
    \psfrag{sig}[l]{{\footnotesize{$\sigma^2$}}}  %
    \psfrag{nu}[l]{{\footnotesize{$\eta_{pass}$}}}  %
    \psfrag{YES}[l]{{\small{YES}}}  %
    \psfrag{NO}[l]{{\small{NO}}}  %
    \psfrag{lambda_mu_+_0}[l]{{\scriptsize{$(\lambda,\nu)=(+,0)$}}}  %
    \psfrag{lambda_mu_+_+}[l]{{\scriptsize{$(\lambda,\nu)=(+,+)$}}}  %
    \psfrag{P_pass_stop_+_0}[l]{{\scriptsize{$(P_{pass},P_{stop})=(+,0)$}}} %
    \psfrag{P_pass_stop_+_+}[l]{{\scriptsize{$(P_{pass},P_{stop})=(+,+)$}}}
    \centerline{  \scalefig{0.95}\epsfbox{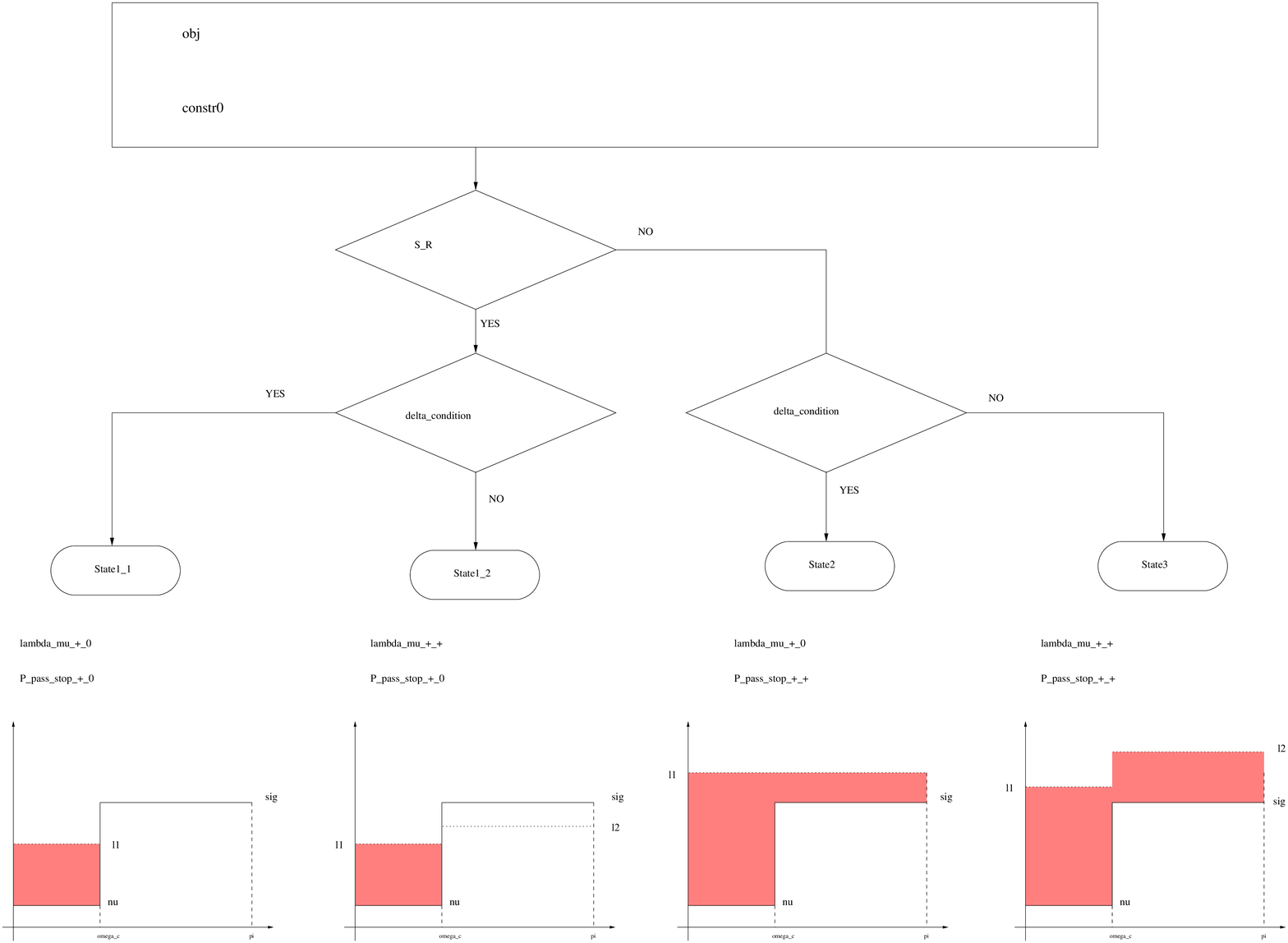} }
\end{psfrags}
} \caption{Different types of the solution to the KKT conditions
($l_1 = \frac{1}{(2\ln2)(\lambda+\nu a^2 \delta^2)}$ and $l_2 =
\frac{1}{(2\ln2)\lambda}$)  } \label{fig:flowchart}
\end{figure}

In Section \ref{subsec:onetaprelayfilter}, it is shown that
one-tap relay filters do not outperform the AF scheme in
flat-fading channels. This is because any one-tap relay filter
with a causal or non-causal non-zero delay cannot change  the
noise spectrum, but destroys the coherent signal combining that is
available in the AF scheme. However, this is not the case when the
relay filter has multiple taps. In this case, the overall noise
spectrum as well as the channel gain spectrum in
\eqref{eq:scalarOneTapDataModel} can be shaped by the relay
filter, and the LTI relaying scheme with multiple taps can
outperform the AF scheme in flat-fading channels. However, the
performance analysis in this case is far more difficult than that
in the one-tap relay case, especially when the causality
constraint is imposed on the relay filter. To circumvent this
difficulty, in this subsection we relax the causality constraint
on $H(z)$ and consider the tractable ideal
low-pass\footnote{Different types of ideal filters, i.e.,
high-pass, band-pass, band-stop filters, yield essentially the
same result as the ideal low-pass filters when the bandwidth of
passband is the same.} relay filtering to gain insights into the
interaction between the source and relay filters and to assess the
rate gain that can be obtained by multiple-tap relay filtering.
The frequency response of the ideal low-pass relay filter is given
by
\begin{equation}
        H(e^{j\omega}) =\left\{
                          \begin{array}{cc}
                                \delta, & {|\omega|<\omega_c}, \\
                                0, & {\omega_c<|\omega|<\pi,}
                          \end{array}
                        \right.
\end{equation}
where $\delta$ is the passband gain and $\omega_c$ is the cutoff
frequency.  For a given $\omega_c$, the optimization problem
(\ref{eq:CLTIfd},\ref{eq:powerConstSpec1},\ref{eq:powerConstSpec2})
with the ideal low-pass relay filter is expressed as {\small
\begin{equation} \label{eq:finalOptProb_lowpass}
\max_{\delta,S_{x_s}(e^{j\omega})}  \left[
\frac{1}{\pi}\int_{0}^{\omega_c} \frac{1}{2} \log_2\left(
1+\frac{(1+ab\delta)^2}{(b^2 \delta^2 +1)\sigma^2}
S_{x_s}(e^{j\omega}) \right) d\omega +
\frac{1}{\pi}\int_{\omega_c}^{\pi} \frac{1}{2} \log_2\left(
1+\frac{1}{\sigma^2}  S_{x_s}(e^{j\omega}) \right) d\omega \right]
\end{equation}}
subject to
\begin{eqnarray}
\frac{1}{\pi}\int_{0}^{\pi} S_{x_s}(e^{j\omega})d\omega - P_s &\le& 0 \label{eq:powerConstSpec1_lowpass}\\
  \frac{1}{\pi}\int_{0}^{\omega_c} \delta^2 ( a^2 S_{x_s}(e^{j\omega}) + \sigma^2) d\omega  - P_r &\le& 0
  \label{eq:powerConstSpec2_lowpass}\\
 S_{x_s}(e^{j\omega}) &\ge& 0 ,~~\forall \omega\in[0,\pi],  \label{eq:powerConstSpec3_lowpass}
\end{eqnarray}
where the even symmetry of spectra is used. Note that the problem
is not jointly convex w.r.t. $\delta$ and $S_{x_s}(e^{j\omega})$
for a given $\omega_c$. However, we can still apply the
Karush-Kuhn-Tucker (KKT) conditions to this problem to obtain the
necessary conditions for optimality \cite{Boyd:04cvx:book}. The
Lagrangian of this problem is given by {\small
\begin{eqnarray}
 {\mathcal{L}}&=& -\frac{1}{\pi}\int_{0}^{\omega_c} \frac{1}{2} \log_2\left(  1+\frac{(1+ab\delta)^2}{(b^2 \delta^2 +1)\sigma^2} S_{x_s}(e^{j\omega}) \right) d\omega
    - \frac{1}{\pi}\int_{\omega_c}^{\pi} \frac{1}{2} \log_2\left(1+\frac{1}{\sigma^2}  S_{x_s}(e^{j\omega}) \right) d\omega \\
   && ~~ + \lambda \left( \frac{1}{\pi}\int_{0}^{\pi} S_{x_s}(e^{j\omega})d\omega - P_s \right) \label{eq:CompSlack1}\\
   && ~~ + \nu \left( \frac{1}{\pi}\int_{0}^{\omega_c} \delta^2 ( a^2S_{x_s}(e^{j\omega}) + \sigma^2) d\omega  - P_r
   \right),\label{eq:CompSlack2}
\end{eqnarray}}
where $\lambda$ and $\nu$ are non-negative dual variables. Due to
 the complementary slackness,  either
$\lambda=0$ or the source uses full power, i.e.,
$\frac{1}{\pi}\int_{0}^{\pi} S_{x_s}(e^{j\omega})d\omega = P_s$;
and either $\nu=0$ or the relay uses full power. Suppose that the
source does not use full power. Then, the source can increase the
PSD over the frequency band $[\omega_c, \pi]$ without changing the
PSD over $[0, \omega_c)$. Then, the power constraint
\eqref{eq:powerConstSpec2_lowpass}  at the relay  is not affected
and the rate in \eqref{eq:finalOptProb_lowpass} increases. Thus,
the source should use full power to yield maximum rate, and we
have $\lambda>0$ due to the complementary slackness. However, the
relay may or may not use full power depending on the channel
condition. The source PSD solution to the KKT conditions is given
by a modified water-filling method:
\begin{equation}
S_{x_s}(e^{j\omega}) = \left\{
\begin{array}{lc}
 \left(\frac{1}{(2\ln2)(\lambda+\nu a^2\delta^2)}- \frac{(b^2 \delta^2+1)\sigma^2}{(1+ab\delta)^2} \right)^+, & {|\omega|<\omega_c}, \\
 \left(\frac{1}{(2\ln2)\lambda}- \sigma^2\right)^+, & {\omega_c \le |\omega| \le
 \pi},
\end{array}
\right. \label{eq:ILPFpsdSolution}
\end{equation}
where $\gamma^+:=\max\{0, \gamma\}$. The optimal source PSD is
given by the difference between the `water level' and the overall
effective `noise level'. The water level is a function of two dual
variables $\lambda$ and $\nu$, and we may have two water levels if
$\nu>0$, i.e., the relay also uses full power.  The effective
noise level $\eta$ is defined as the inverse of the overall CNR
density, and it
 is given by $\eta_{pass}
=\frac{b^2\delta^2+1}{(1+ab\delta)^2}\sigma^2$ and
$\eta_{stop}=\sigma^2$ for $[0,\omega_c)$ and $[\omega_c,\pi]$,
respectively. From here on, we will consider only the case
$\eta_{pass}   < \sigma^2$. (Otherwise, it is better not to use
the relay ($\delta =0$) since $\sigma^2$ over $[0,\pi]$ is the
noise level without the relay.)
 To obtain optimal $\delta$, we need to consider both
\eqref{eq:finalOptProb_lowpass} and
\eqref{eq:powerConstSpec2_lowpass}. By differentiating
$\eta_{pass}(\delta)=\frac{(b^2 \delta^2
+1)\sigma^2}{(1+ab\delta)^2}$ w.r.t. $\delta$ and setting the
derivative to zero, we obtain $\delta=\frac{a}{b}$ for the minimum
noise level $\eta^*_{pass} = (1+a^2)\sigma^2/(1+a^2)^2$. However,
we have the relay power constraint
\eqref{eq:powerConstSpec2_lowpass}, yielding $\delta \le
\sqrt{P_r/(\frac{1}{\pi}\int_{0}^{\omega_c}a^2S_{x_s}(e^{j\omega})
d\omega+\frac{\omega_c}{\pi}\sigma^2})= \sqrt{P_r/(a^2P_{pass}+
(\omega_c/\pi)\sigma^2)}$,  where $P_{pass}$ is the portion of the
source power allocated to the relay's passband $[0, \omega_c)$ and
$P_{stop}:=P_s - P_{pass}$.  Thus, the optimal $\delta^*$ is given
by
\begin{equation}  \label{eq:optdeltaLPF}
\delta^* = \min\left\{~\frac{a}{b},
~\sqrt{\frac{P_r}{a^2P_{pass}+\frac{\omega_c}{\pi}\sigma^2}}
~\right\},
\end{equation}
since $\eta_{pass}(\delta)$ is monotonically increasing as
$\delta$ decreases  from $a/b$ to zero.   With the optimal
$\delta^*$, the multiple-tap relay filter generates a well in the
noise level, as shown in Fig. \ref{fig:flowchart}. The effective
noise level $\eta_{pass}(\delta)$ of this well is equal to or
lower than that of the AF scheme because the effective noise level
for the AF scheme is $\eta_{AF}(d) = \frac{(b^2 d^2
+1)\sigma^2}{(1+ab d)^2}$ for $-\pi \le \omega \le \pi$ (see
\eqref{eq:InstAFachievRate}) and because for the same source power
$P_s$ and relay power $P_r$ the upperbound for $\delta$ in
\eqref{eq:optdeltaLPF} is larger than that for $d$ in
 \eqref{eq:optdAF}. (Note that $P_{pass} \le P_s$ and $\omega_c \le
 \pi$.)  Thus, when $P_r$ is small or $b << a$ such that (s.t.)  $d^* < a/b$
 (consequently, $\eta_{pass}(\delta) < \eta_{AF}(d)$),  and $P_s$ is also small enough
 to be confined within the passband well, the ideal low-pass
filtering  outperforms the AF scheme. (The amount of gain will be
evaluated numerically shortly.)

The structure of solution $S_{x_s}(e^{j\omega})$ to the KKT
conditions can be classified into four types depending on whether
the source power is allocated to the relay's stopband $[\omega_c,
\pi]$ or not, and whether the relay uses full power or not
(equivalently, $\delta^* \ne a/b$ or not). Fig.
\ref{fig:flowchart} shows the solution types.
A solution of Type 1 occurs when all source power is allocated to
the passband of the relay filter. We can further distinguish Type
1 depending on the power use of the relay. When the relay uses
full power, both Lagrange dual variables $\lambda$ and $\nu$ are
positive, and there exist two water levels, although no water or
power is allocated in the stop band, i.e., $l_2 \le \sigma^2$
(Type 1-2). Otherwise, we have Type 1-1 in which only $\lambda$ is
non-zero. A solution of Type 2 occurs when the relay does not use
full power and the source power is allocated over the entire band;
there is one water level common to both the passband and stopband.
A Type 3 solution
 occurs when the relay uses full power and the source
power is distributed over the entire band; both Lagrange dual
variables are positive and the water levels at the passband and
stopband are different. Different types of solutions occur for
different combinations of parameters, $P_s, P_r, \sigma^2, a, b$,
and $\omega_c$.

\begin{figure}[htbp]
\centerline{ \SetLabels
\L(0.17*-0.15) (a) \\
\L(0.50*-0.15) (b) \\
\L(0.83*-0.15) (c) \\
\endSetLabels
\leavevmode
\strut\AffixLabels{ \scalefig{0.35}\epsfbox{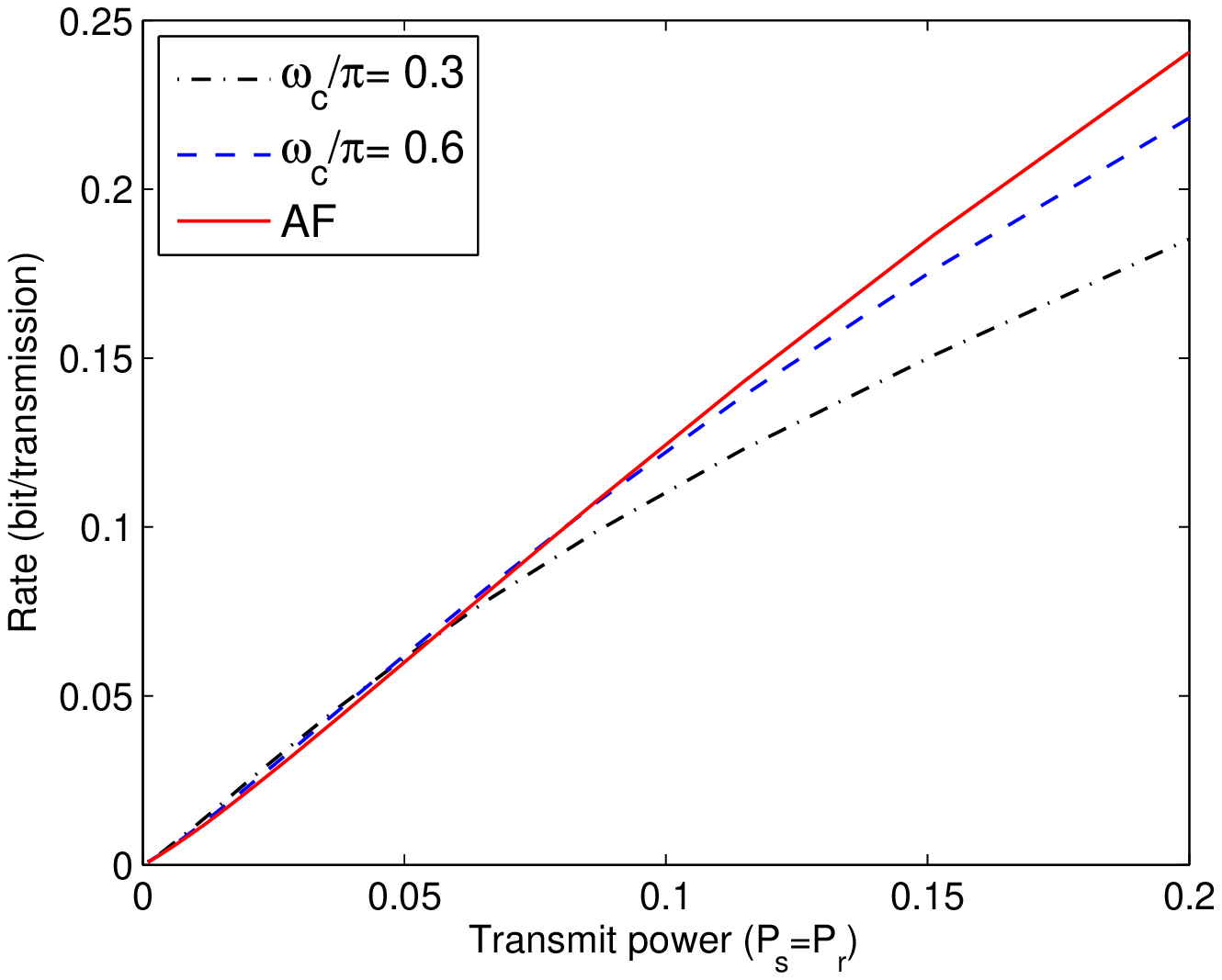}
\scalefig{0.35}\epsfbox{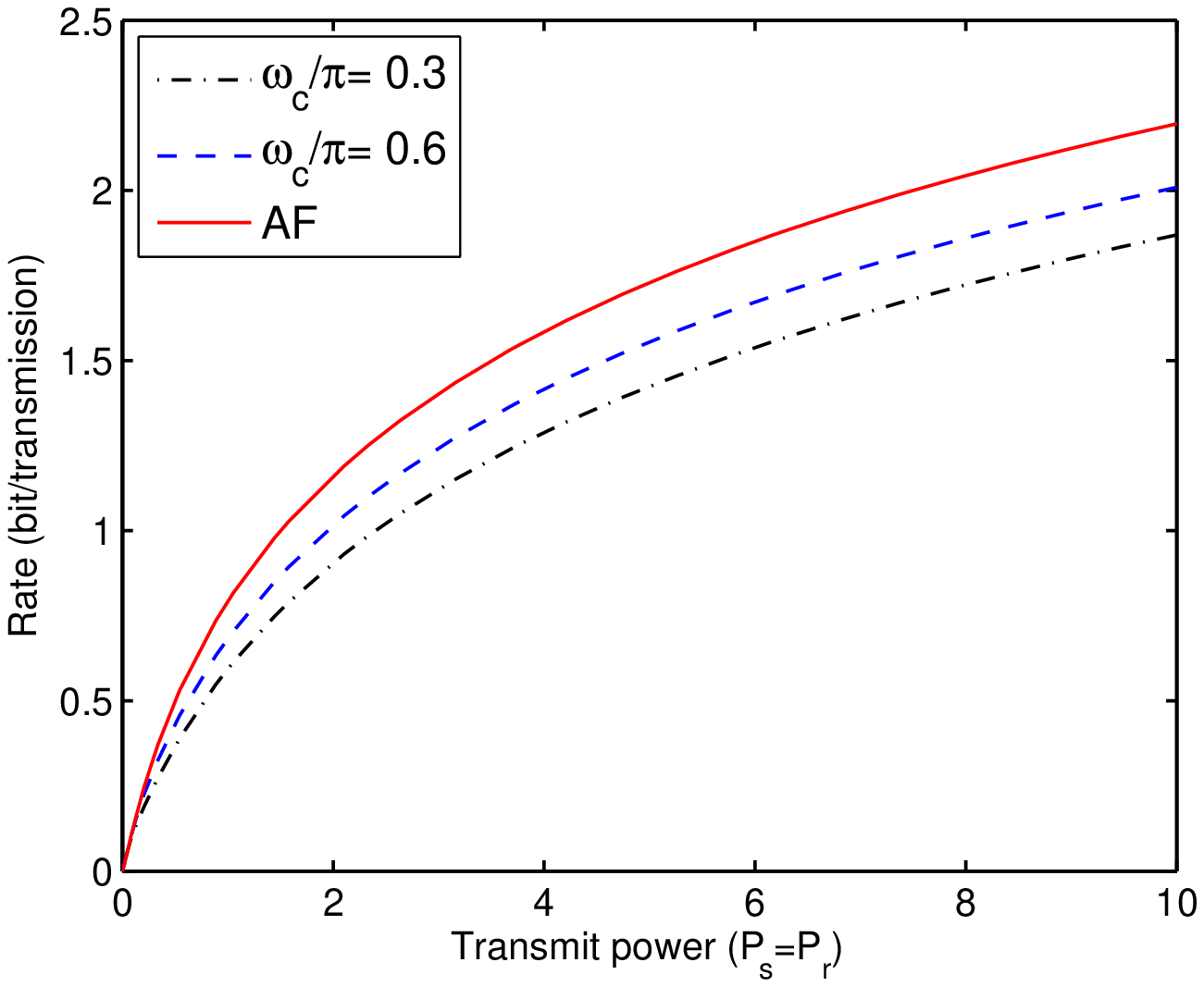}
\scalefig{0.35}\epsfbox{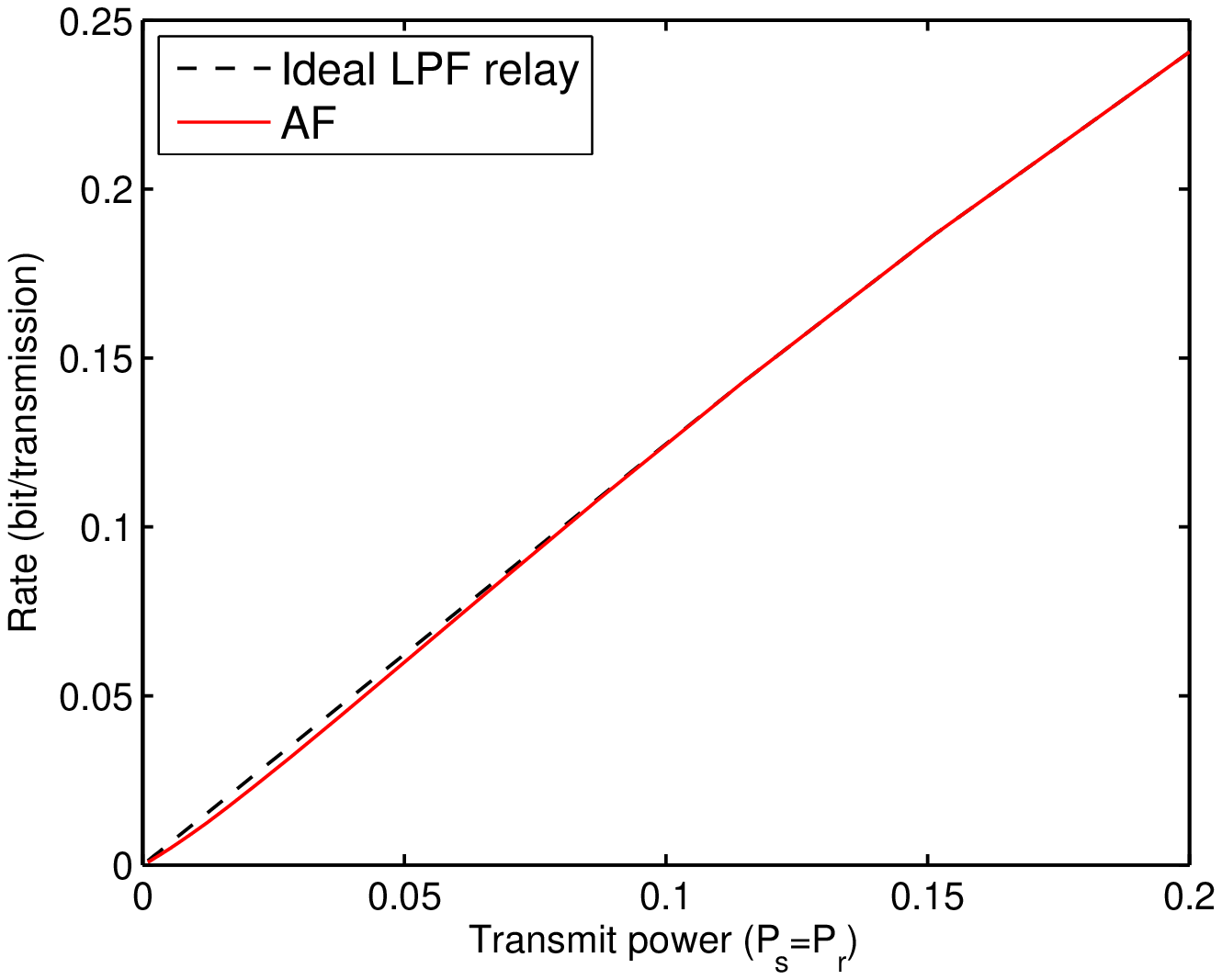} } }
 \vspace{0.3cm} \caption{Rate performance: Ideal low-pass relay filtering versus AF ($P_s=P_r$, $\sigma^2=1$, $a=1$ and $b=2$):
 (a) fixed $\omega_c$ (at low $P_s$), (b) fixed $\omega_c$, and (c) optimized $\omega_c$}
\label{fig:LTIF_IAF}
\end{figure}

\begin{figure}[htbp]
\centerline{ \SetLabels
\L(0.17*-0.15) (a) \\
\L(0.50*-0.15) (b) \\
\L(0.83*-0.15) (c) \\
\endSetLabels
\leavevmode
\strut\AffixLabels{ \scalefig{0.35}\epsfbox{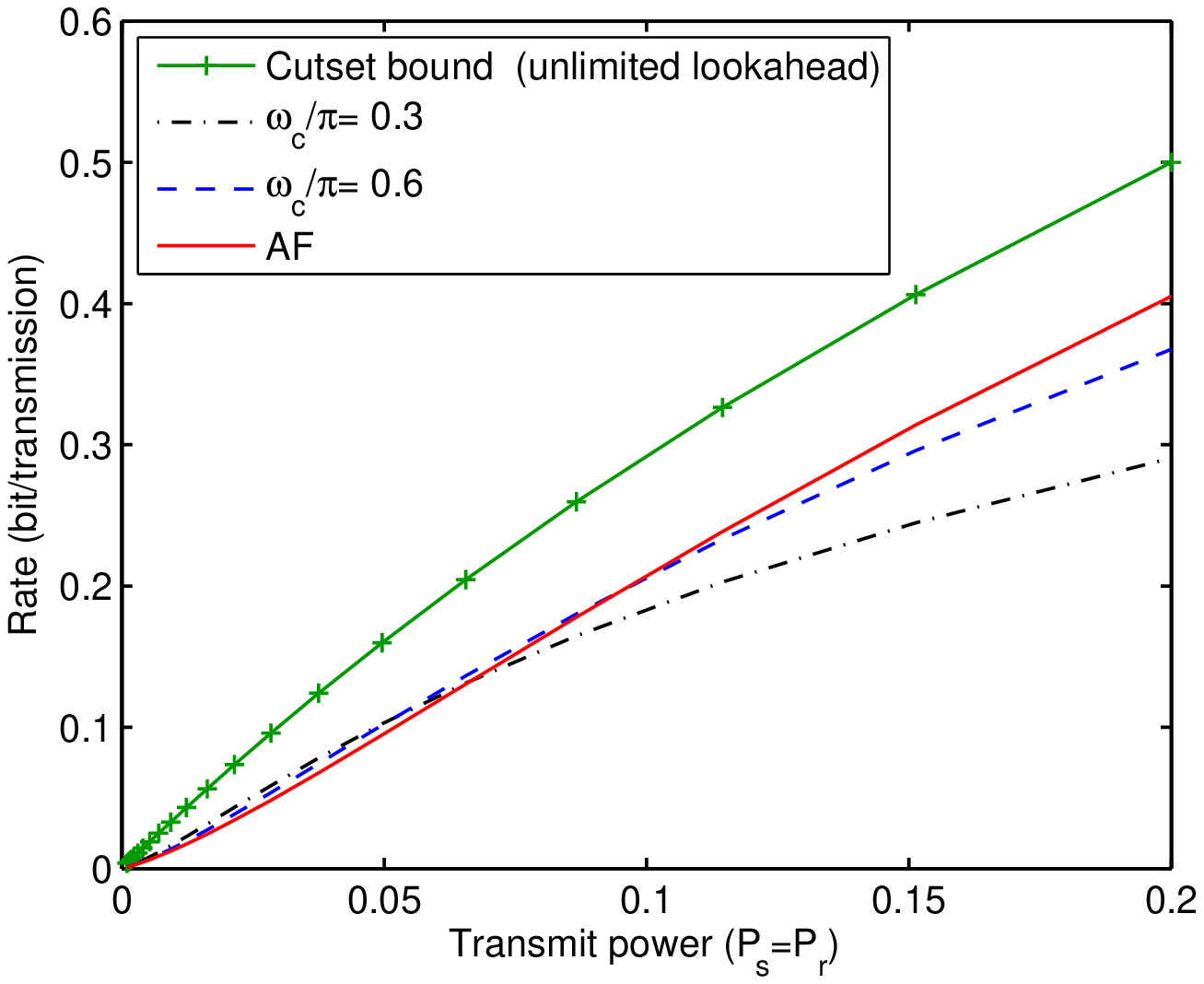}
\scalefig{0.35}\epsfbox{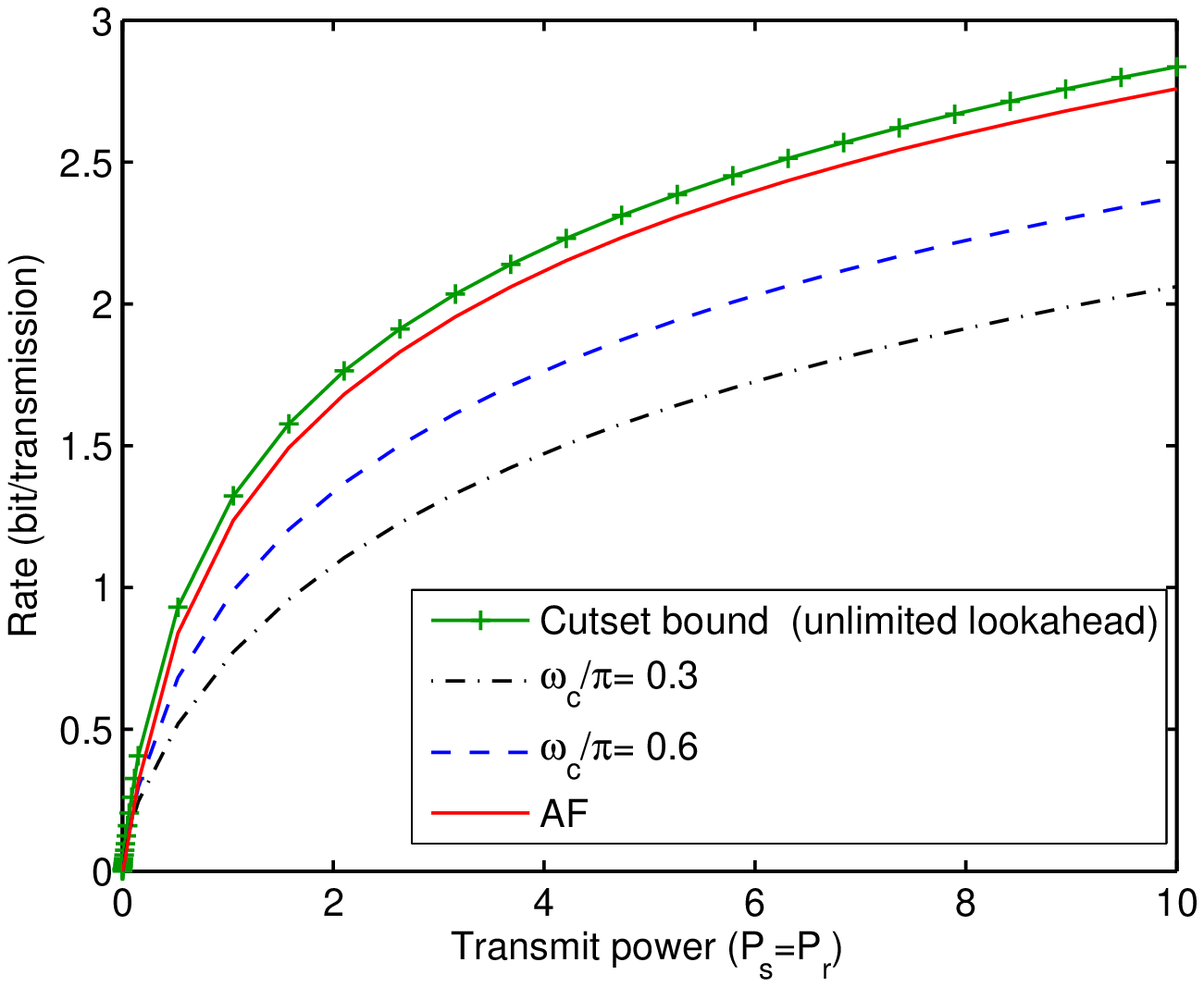}
\scalefig{0.35}\epsfbox{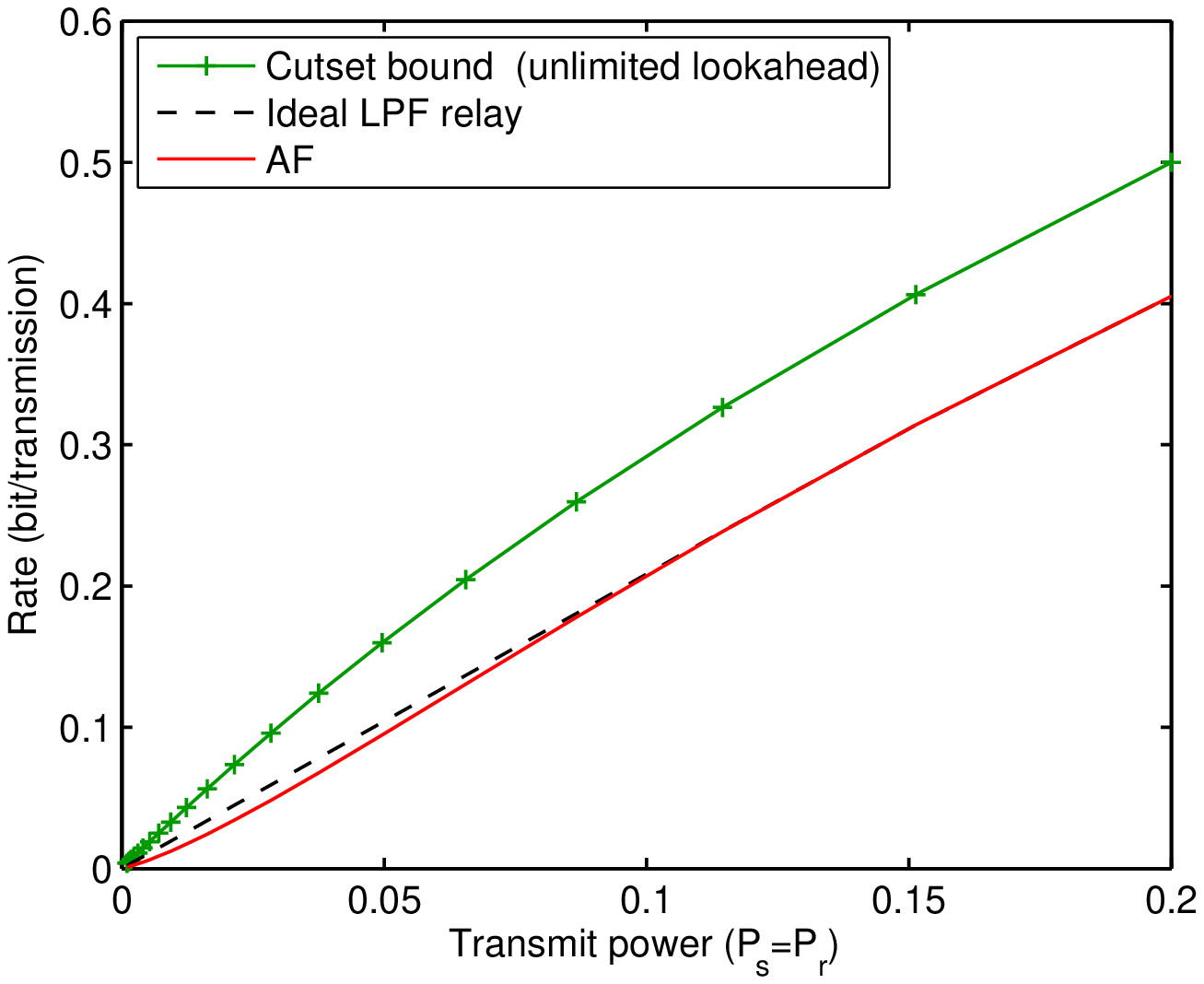} } }
 \vspace{0.3cm} \caption{Rate performance: Ideal low-pass relay filtering versus AF ($P_s=P_r$, $\sigma^2=1$, $a=2$ and $b=2$):
 (a) fixed $\omega_c$ (at low $P_s$), (b) fixed $\omega_c$, and (c) optimized $\omega_c$}
\label{fig:LTIF_IAF2}
\end{figure}

To evaluate the performance of the ideal low-pass filtering at the
relay, the optimization problem (\ref{eq:finalOptProb_lowpass} -
\ref{eq:powerConstSpec3_lowpass}) was  solved numerically using a
commercial tool with $\delta$ and $P_{pass}$ as variables for
given  $P_s$, $P_r$, $\sigma^2$, $a$, $b$ and $\omega_c$. First,
we fixed $a=1$, $b=2$ and $\sigma^2=1$ (the same as for Fig.
\ref{fig:flatFadingPlot} (a) and (b)),
 and swept $P_s=P_r$ for each $\omega_c\in\{0.3\pi,0.6\pi\}$, and
 compared the performance of the ideal low-pass filtering with that of the AF scheme.
 The performance is shown in Fig. \ref{fig:LTIF_IAF} (a) and
(b).  (The same rate curves are plotted with two different x-axis
ranges.) It is seen in Fig. \ref{fig:LTIF_IAF} (a) that indeed
there is a gain over the AF scheme at the very low source power
values. This is because the low-pass relay filter intentionally
generates a well in the noise level in the passband and because
all source power is allocated into this well at the very low
source power values, as explained already. In this way, the signal
is transmitted through a narrowband channel that has a lower noise
level than that of the AF scheme. As $P_s$ increases, however, the
source power spills over the relay's stopband. At high SNR, the
effect of the intentional noise level shaping is negligible (the
effect of noise itself becomes negligible), and the uniform source
power distribution over the entire frequency band is optimal at
high SNR. Consequently, filtering out the signal existing at the
stopband $[\omega_c,\pi]$ at the relay is detrimental to the
performance at high SNR, as shown in  Fig. \ref{fig:LTIF_IAF} (a)
and (b). So, we optimized the cut-off frequency $\omega_c$
together with $\delta$ and $P_{pass}$, and the result is shown in
Fig. \ref{fig:LTIF_IAF} (c). It is seen that the optimized
solution shows the performance gain over the AF scheme at low SNR
and eventually converges to the AF scheme as $P_s$ increases. In
this case of $a=1$ and $b=2$, the performance gain of the LPF
relay over the AF scheme is not significant since the AF scheme
performs well already, as shown in Fig. \ref{fig:flatFadingPlot}
(a) and (b) \cite[Proposition 9]{ElGamal&Hassanpour&Mammen:07IT}.
Thus, we tried another case of $a=b=2$ and $\sigma^2=1$ in which
the AF scheme has a noticeable loss from the cut-set bound, as
shown in Fig. \ref{fig:flatFadingPlot} (c) and (d). Fig.
\ref{fig:LTIF_IAF2} shows the achievable rate in this case. The
figure shows similar behavior to the case of $a=1$ and $b=2$.
 One important fact  here is that the performance gain
of the ideal low-pass (multiple-tap) relay filtering (even
breaking the causality constraint) over the AF scheme is not so
significant in both of the cases. Thus, it seems that,
practically, there is no need for complicated linear
(time-invariant) filtering at the relay beyond the AF scheme in
flat-fading channels. However, the insignificant gain is only for
flat-fading channels, and this is not the case in realistic ISI
channels. In the next section, we will tackle the LTI relay
problem (\ref{eq:CLTIfd}, \ref{eq:powerConstSpec1},
\ref{eq:powerConstSpec2}, \ref{eq:sourceSpecFactor}) in ISI
channels, propose a practical method to solve this problem, and
show that the LTI relay filtering yields a noticeable gain over
the AF scheme in ISI channels.

\section{Joint Source and Relay Filter Design in ISI Channels}
\label{sec:JointSourceRelayISIDesign}

The LTI relay problem (\ref{eq:CLTIfd}, \ref{eq:powerConstSpec1},
\ref{eq:powerConstSpec2}, \ref{eq:sourceSpecFactor}) in ISI
channels is basically a constrained optimization problem.  Our
approach to this problem is based on a powerful tool of the {\em
projected subgradient method} developed by Polyak
\cite{Polyak:69USSR} and Yamada et al.
\cite{Yamada&Ogura:04NFAO,Slavakis&Yamada&Ogura:06NFAO}.  In the
next subsection, we will briefly introduce their results that are
relevant to our problem, and then move on to the original LTI
relay problem.

\subsection{Background: The Adaptive Projected Subgradient Method
(APSM) \cite{Yamada&Ogura:04NFAO,Slavakis&Yamada&Ogura:06NFAO}}

The projected subgradient method was initially proposed by Polyak
\cite{Polyak:69USSR} to solve  cost minimization problems over
convex constraint sets. Recently, Yamada et al. fully generalized
the method even
 to adaptive situations with time-varying cost functions, and
proved strong convergence of the method
\cite{Yamada&Ogura:04NFAO,Slavakis&Yamada&Ogura:06NFAO}.

\begin{definition}[Subdifferential]
Let $\phi:{\mathcal{H}}\rightarrow {\mathbb{R}}$ be a continuous
convex function from a Hilbert space $\Hc$ to the set
${\mathbb{R}}$ of real numbers.  Then, the
\textit{subdifferential} of $\phi$ at $\ubf$ is defined as  the
set of all the subgradients of $\phi$ at $\ubf$:
\begin{equation} \label{eq:subgradientDef}
\partial \phi(\ubf):= \{\gbf \in \Hc | < \vbf-\ubf,\gbf > ~+~ \phi(\ubf) \le \phi(\vbf), ~\forall \vbf \in \Hc
\},
\end{equation}
where $<\cdot, \cdot>$ is the inner product between two vectors.
\end{definition}

\noindent Here, $< \vbf-\ubf,\gbf > ~+~ \phi(\ubf)=r$ is a
hyperplane with coordinates $(\vbf,r)$ passing $(\ubf,
\phi(\ubf))$ in the space $\Hc \times \Rbb$.  Thus,
\eqref{eq:subgradientDef} implies that the hyperplane $<
\vbf-\ubf,\gbf > ~+~ \phi(\ubf)=r$ is below the cost surface
$\phi(\vbf)=r$.  It is known that $\partial \phi(\ubf)$ is
nonempty, and by  definition we have ${\mathbf {0}} \in
\partial\phi(\ubf)\Leftrightarrow
\phi(\ubf)=\min_{\vbf \in \Hc}\phi(\vbf)$. In the differentiable
case, the gradient is a unique subgradient.

\begin{definition}[Metric projection] The metric projection $P_K(\ubf)$
of a point $\ubf$ onto a closed convex set $K$ is the closest
point of $\ubf$ in $K$, i.e., $|| P_K(\ubf) - \ubf|| \le || \vbf -
\ubf||, ~~\forall~ \vbf \in K$.
\end{definition}

\begin{definition}[Subgradient projection]
Suppose that a continuous convex function $\phi:\Hc\rightarrow
\Rbb$ satisfies $\mbox{lev}_{\leq 0}\phi := \{\vbf \in\Hc |
\phi(\vbf)\leq 0\} \neq \emptyset$, i.e., the zero-level set is
not empty. Let $\phi':\Hc\rightarrow\Hc$ be a selection of the
subdifferential $\partial\phi$, i.e.,
$\phi'(\ubf)\in\partial\phi(\ubf), ~~\forall \ubf \in \Hc$. Then,
a mapping $T_{sp(\phi)}: \Hc \rightarrow \Hc$ defined by
\begin{equation}\label{eq:TspFormula}
T_{sp(\phi)}:\ubf\longmapsto \left\{
\begin{array}{ll}
\ubf-\frac{\phi(\ubf)}{||\phi'(\ubf)||^2}\phi'(\ubf), \hspace{1em} & \mbox{if} ~~~\phi(\ubf)>0 \\
\ubf, &  \mbox{if} ~~~\phi(\ubf)\leq 0,
\end{array}
\right.
\end{equation}
is called a \textit{subgradient projection relative to $\phi$}.
\end{definition}

\noindent Here, the normalization of the subgradient by factor
$\frac{\phi(\ubf)}{||\phi'(\ubf)||^2}$ is crucial to determining
the step size parameter.  Basically, the subgradient projection
moves the current point $\ubf$ to its metric projection onto the
intersection of two hyperplanes $r=0$ and $< \vbf-\ubf,\gbf
> ~+~ \phi(\ubf)=0$ in the space of $\Hc \times \Rbb$ with coordinates $(\vbf,r)$. (See \cite[Fig.16]{Theodoridis&Slavakis&Yamada:01SPM}.)
It is known that the subgradient projection always moves the
original point closer to every point in the zero-level set, which
is known as the
 Fej\'er monotone property with respect to the zero-level set \cite{Slavakis&Yamada&Ogura:06NFAO,Censor&Zenios:book}. It is also known that the subgradient projection
belongs to the class of firmly quasi-nonexpansive mappings
\cite{Slavakis&Yamada&Ogura:06NFAO,Censor&Zenios:book}. Now, we
state the adaptive projected subgradient theorem by Slavakis,
Yamada and Ogura.

\begin{theorem}[APSM,\cite{Yamada&Ogura:04NFAO,Slavakis&Yamada&Ogura:06NFAO}]
\label{theo:APSM}
 Let $\phi_n:\Hc\rightarrow[0,\infty)\,~(\forall n
\in \Nbb)$ be a sequence of continuous convex functions and $K
\subset \Hc$ a nonempty closed convex set. For an arbitrarily
given $\ubf_0\in K$, the sequence $\{\ubf_n\}_{n\in\Nbb} \subset
K$ generated by the adaptive projected subgradient method:
\begin{equation} \label{eq:APSMupdate}
\ubf_{n+1}=\left\{
\begin{array}{ll}
P_K\left(\ubf_n-\mu_n\frac{\phi_n(\ubf_n)}{||\phi_n'(\ubf_n)||^2}\phi_n'(\ubf_n)\right), & \mbox{if}~\phi_n'(\ubf_n)\neq 0, \\
\ubf_n, & \mbox{otherwise}, \\
\end{array}
\right.
\end{equation}
where $\phi_n'(\ubf_n)\in\partial\phi_n(\ubf_n)$ and $0\leq
\mu_n\leq 2$, satisfies the following.

\begin{itemize}
\item[(Monotome approximation)] Suppose
\begin{equation}\label{eq:APSM_monotone}
\ubf_n \notin \Omega_n:=\{\ubf \in K
|\phi_n(\ubf)=\phi_n^*:=\inf_{\vbf \in
K}\phi_n(\vbf)\}\neq\emptyset.
\end{equation}
Then, by using $\forall \mu_n\in
\left(0,2(1-\frac{\phi_n^*}{\phi_n(\ubf_n)})\right)$, we have  $||
\ubf_{n+1} - \ubf_{(n)}^*|| < || \ubf_n - \ubf_{(n)}^*||$ for all
$\ubf_{(n)} \in \Omega_n$.

\item[(Asymptotic optimality)]  Suppose
\begin{equation}\label{eq:AMSM_asymOpt}
\exists N_0\in \Nbb ~~\mbox{s.t.}~~ \phi_n^*=0 ~~\forall n \geq
N_0 ~~~ \mbox{and}~~ \Omega :=\cap_{n\geq
N_0}\Omega_n\neq\emptyset.
\end{equation}
Then, $\{\ubf_n\}_{n\in\Nbb}$ is bounded. Moreover, if we
specially use $\forall \mu_n\in[\epsilon_1,2-\epsilon_2]\subset
(0,2)$, we have $\lim_{n\rightarrow\infty}\phi_n(\ubf_n)=0$
provided that $\{\phi_n'(\ubf_n)\}_{n\in\Nbb}$ is bounded.

\item[(Strong convergence)] Assume (\ref{eq:AMSM_asymOpt}) and
$\Omega$ has some relative interior w.r.t.  a hyperplane
$\Pi~(\subset\Hc)$; i.e., there exist
$\tilde{\ubf}\in\Pi\cap\Omega$ and $\exists~ \epsilon
>0$ satisfying $\{\ubf \in \Pi ~\mbox{s.t.}~
||{\ubf-\tilde{\ubf}}||\leq \epsilon\}\subset\Omega$.  Then, by
using $\forall \eta_n \in [\epsilon_1,2-\epsilon_2]\subset(0,2)$,
$\{\ubf_n\}_{n\in\Nbb}$ converges strongly to some $\hat{\ubf}\in
K$, i.e., $\lim_{n\rightarrow\infty} ||\ubf_n-\hat{\ubf}||=0$.
Moreover, $\lim_{n\rightarrow\infty}\phi_n(\hat{\ubf})=0$ if
 $\{\phi_n'(\ubf_n)\}_{n\in\Nbb}$ is bounded and if there exists bounded
$\{\phi_n'(\hat{\ubf})\}_{n\in\Nbb}$ where
$\phi_n'(\hat{\ubf})\in\partial\phi_n(\hat{\ubf})$, $\forall
n\in\Nbb$.
\end{itemize}
\end{theorem}

\vspace{0.5em}   Note that the update rule in
\eqref{eq:APSMupdate} is a composite projection composed of the
subgradient-based projection relative to $\phi_n$ and the metric
projection onto $K$. In fact, the first projection is not the
subgradient projection exactly since the step size parameter
$\mu_n$ does not need to be one. This projection can be rewritten
as
\begin{equation}
\hat{T}_{\mu_n}(\ubf_n):=\ubf_n-\mu_n\frac{\phi_n(\ubf_n)}{||\phi_n'(\ubf_n)||^2}\phi_n'(\ubf_n)
= [(1-\mu_n) I + \mu_n T_{sp(\phi_n)}](\ubf_n),
\end{equation}
where $I$ is the identity mapping. When $\mu_n=1$,
$\hat{T}_{\mu_n}=T_{sp(\phi_n)}$. When $\mu_n < 1$, the projection
moves the original point towards the exact subgradient projection
point, but not fully.  When $\mu_n > 1$, on the other hand, the
projection moves the original point beyond the exact subgradient
projection point. For any $\mu_n \in (0,2)$, $\hat{T}_{\mu_n}$ is
called a $\mu_n$-averaged quasi-nonexpansive mapping, and the
properties of this mapping play an important role in the proof of
Theorem \ref{theo:APSM}. The major difference of the subgradient
projection method from the simple gradient method without
normalization is that we know the exact range of the step size
parameter $\mu_n$ for convergence because of the proper
normalization of the subgradient in the subgradient projection.
(See \eqref{eq:TspFormula} and \eqref{eq:APSMupdate}.) The above
result is general and can be applied to many constrained
optimization problems. Consider the simple case with a fixed cost
function (i.e.,  $\phi_n(\ubf) \equiv \phi(\ubf)$ ~$\forall~n$) of
which minimum is achieved in $K$. Let $\phi^* = \min_{\ubf \in K}
\phi(\ubf)$. Then, we can define $\tilde{\phi}(\ubf) := \phi(\ubf)
- \phi^*$. Then, the condition \eqref{eq:AMSM_asymOpt} is
trivially satisfied, and we have $\lim_{n\rightarrow \infty}
\tilde{\phi}(\ubf_n) =0$, i.e., $\lim_{n\rightarrow \infty}
{\phi}(\ubf_n) = \phi^*$. In this case, the result reduces to the
Polyak's projected subgradient method for which he showed the weak
convergence \cite[Theorem 1]{Polyak:69USSR}. (For the proof of
Theorem \ref{theo:APSM}, refer to
\cite{Slavakis&Yamada&Ogura:06NFAO}. For the introduction of
related projections, see
\cite{Theodoridis&Slavakis&Yamada:01SPM}.)

\subsection{Joint Source and Relay Filter Design via the Projected Subgradient Method}
\label{subsec:algorithmISI}

Even under the LTI formulation (\ref{eq:CLTIfd},
\ref{eq:powerConstSpec1}, \ref{eq:powerConstSpec2},
\ref{eq:sourceSpecFactor}),  the search space for source and relay
filters $T(z)$ and $H(z)$ has countably infinite dimensions since
both $T(z)$ and $H(z)$ can be IIR filters. To avoid the
difficulties in searching in an infinite dimensional space and in
imposing the stability condition, we restrict ourselves to the
case that both $T(z)$ and $H(z)$ have FIRs as in most practical
filtering applications. Then, the stability and causality
constraints are automatically satisfied. In this case, the source
and relay filter responses are respectively given by
\begin{eqnarray}
T(z) &=& t_0 + t_1 z^{-1} + \cdots + t_{L_s-1}z^{-L_s+1}, ~~~\mbox{and}\\
H(z) &=& h_0 + h_1 z^{-1} + \cdots + h_{L_r-1}z^{-L_r+1},
\label{eq:ISIH(z)}
\end{eqnarray}
 where $L_s$ and $L_r$ are the orders of source and relay
filters, respectively. Although the channel responses $H_{sr}(z)$,
$H_{rd}(z)$ and $H_{sd}(z)$ do not need to have finite durations,
we also assume that these channel responses have finite duration
$L$ for simplicity.  For notational convenience, we define the
following:
\begin{eqnarray}
    \wbf_m(\omega) &:=& [1, e^{j\omega}, e^{j2\omega}, \cdots, e^{j(m-1)\omega}]^T,\\
    \tbf &:=& [t_0, t_1, \cdots, t_{L_s-1}]^T ~~~\in \Rbb^{L_s\times 1}, \\
    \hbf &:=& [h_0, h_1, \cdots, h_{L_r-1}]^T ~~~\in \Rbb^{L_r\times 1}, \\
    \ubf &:=& [\tbf^T, \hbf^T]^T  ~~~\in \Rbb^{(L_s+L_r)\times 1}, \\
    \hbf_{ab} &:=& [h_{ab}[0], h_{ab}[1], \cdots, h_{ab}[L-1]]^T
    ~~\mbox{for}~~ (a,b)= (s,r), (r,d) ~\mbox{and}~ (s,d).
\end{eqnarray}
Then, we have
\begin{equation}
T(e^{j\omega})= \wbf_{L_s}^H(\omega) \tbf, ~~H(e^{j\omega})=
\wbf_{L_r}^H(\omega) \hbf, ~~ H_{ab}(e^{j\omega})=
\wbf_L^H(\omega) \hbf_{ab}.
\end{equation}
Here, we assume that $\tbf$ and $\hbf$ are real vectors for
simplicity, but the extension to the complex case is
straightforward. Now from (\ref{eq:CLTIfd}) we define the cost
function as
\begin{equation} \label{eq:ISIrelayCostFunc}
\phi(\ubf) = -  \frac{1}{2\pi} \int_{-\pi}^\pi
\frac{1}{2}\log_2\left( 1 +
\frac{|H_{sd}(e^{j\omega})+H_{sr}(e^{j\omega}) H(e^{j\omega};\hbf)
H_{rd}(e^{j\omega})|^2}{\sigma^2(|H_{rd}(e^{j\omega})
H(e^{j\omega};\hbf)|^2 + 1)}  |T(e^{j\omega};\tbf)|^2 \right)
d\omega,
\end{equation}
where the respective dependence of $H(e^{j\omega})$ and
$T(e^{j\omega})$ on $\hbf$ and $\tbf$ is explicitly shown. The
gradient $\phi^\prime (\ubf)$ at $\ubf$ can be obtained after some
computation as
\begin{equation}  \label{eq:ISIrelayCostGradient}
    \phi^\prime(\ubf) =\frac{1}{2\pi}
        \int_{-\pi}^{\pi}
        \frac{1}{2\ln2} \cdot \frac{1}{1+A(\omega)} \left(\begin{array}{c}B(\omega)\\C_1(\omega) (C_2(\omega)-C_3(\omega) C_4(\omega))\end{array}\right) d\omega,
\end{equation}
where $A(\omega) = \mbox{CNR}(e^{j\omega};
\hbf)\cdot\tbf^T\wbf_{L_s}\wbf_{L_s}^H\tbf$,
        $B(\omega) = \mbox{CNR}(e^{j\omega}; \hbf)\cdot (\wbf_{L_s}\wbf_{L_s}^H+\wbf_{L_s}^*\wbf_{L_s}^T) \tbf$,
        $C_1(\omega) = {\tbf^T\wbf_{L_s}\wbf_{L_s}^H\tbf}$ $/{D(e^{j\omega};\hbf)}$,
        $C_2(\omega) = \nabla_{\hbf}N(e^{j\omega};\hbf)$,
        $C_3(\omega) = \nabla_{\hbf}D(e^{j\omega};\hbf)$ and
        $C_4(\omega) =\mbox{CNR}(e^{j\omega};\hbf)$;
and $N(e^{j\omega};\hbf)$ and $D(e^{j\omega};\hbf)$ are the
numerator and denominator of $\mbox{CNR}(e^{j\omega};\hbf)$,
respectively.  The source power constraint
\eqref{eq:powerConstSpec1}  is given in terms of $\tbf$ and $\hbf$
by
\begin{equation} \label{eq:ISIrelayPowerConstSource}
 \frac{1}{2\pi} \int_{-\pi}^\pi |T(e^{j\omega})|^2 d\omega = \sum_{l=0}^{L_s-1} t_l^2 =||\tbf||^2 \leq P_s
\end{equation}
by Parseval's theorem, and the constraint set for $\tbf$
determined by \eqref{eq:ISIrelayPowerConstSource} is simply a ball
in $\tbf$, denoted by $B_{\tbf}(P_s)$, with no constraint on
$\hbf$. Next, consider the power constraint at the relay. The
relay power constraint \eqref{eq:powerConstSpec2} is expressed in
terms of $\tbf$ and $\hbf$ as
\begin{eqnarray}
    & & \frac{1}{2\pi} \int_{-\pi}^\pi | H(e^{j\omega})|^2(|H_{sr}(e^{j\omega})|^2 \cdot |T(e^{j\omega})|^2+\sigma^2)d\omega \label{eq:ISIrelayPowerConstBicon1}\\
    &=& \frac{1}{2\pi} \int_{-\pi}^\pi \hbf^T\wbf_{L_r}(\omega)\wbf_{L_r}^H(\omega)\hbf \cdot \underbrace{(|H_{sr}(e^{j\omega})|^2
    \tbf^T \wbf_{L_s}(\omega) \wbf_{L_s}^H(\omega) \tbf + \sigma^2)}_{=:g(\omega;\tbf)}d\omega, \nonumber\\
    &=& \hbf^T \left(\frac{1}{2\pi} \int_{-\pi}^\pi g(\omega;\tbf)\wbf_{L_r}(\omega)\wbf_{L_r}^H(\omega)    d\omega   \right) \hbf ~~\le~~
    P_r. \label{eq:ISIrelayPowerConstBicon}
\end{eqnarray}
Note that the constraint set for $(\tbf,\hbf)$ determined by
\eqref{eq:ISIrelayPowerConstBicon} is bi-convex, i.e., convex in
each of $\tbf$ and $\hbf$ but not jointly convex. The constraint
set for $\hbf$ for a given $\tbf$ is an ellipsoid, as shown in
\eqref{eq:ISIrelayPowerConstBicon}. Let us denote this ellipsoid
by $\xi_\hbf (\tbf)$. The above inequality can also be written as
\begin{eqnarray}
&&\tbf^T \left(\frac{1}{2\pi} \int_{-\pi}^\pi
(|H_{sr}(e^{j\omega})|^2
\hbf^T\wbf_{L_r}(\omega)\wbf_{L_r}^H(\omega)\hbf)
\wbf_{L_s}(\omega) \wbf_{L_s}^H(\omega)   d\omega \right) \tbf
\nonumber\\
&& ~~~~~~~~~~~~~~~~~~~ + \frac{\sigma^2}{2\pi} \int_{-\pi}^\pi
\hbf^T\wbf_{L_r}(\omega)\wbf_{L_r}^H(\omega)\hbf d\omega ~~~\le~~~
P_r, \label{eq:ISIrelayPowerConstBicon3}
\end{eqnarray}
and, therefore, the constraint set for $\tbf$ for a given $\hbf$
is also an ellipsoid.

\begin{figure}[htbp] \centerline{
    \begin{psfrags}
    \psfrag{c1}[c]{{ $C_2$}}  %
    \psfrag{c2}[c]{{ $C_1$}}  %
    \psfrag{t}[c]{{ $\tbf$}}  %
    \psfrag{h}[c]{{ $\hbf$}}  %
    \psfrag{phi}[l]{{ $\phi(\tbf,\hbf)$}} %
     \scalefig{0.55}\epsfbox{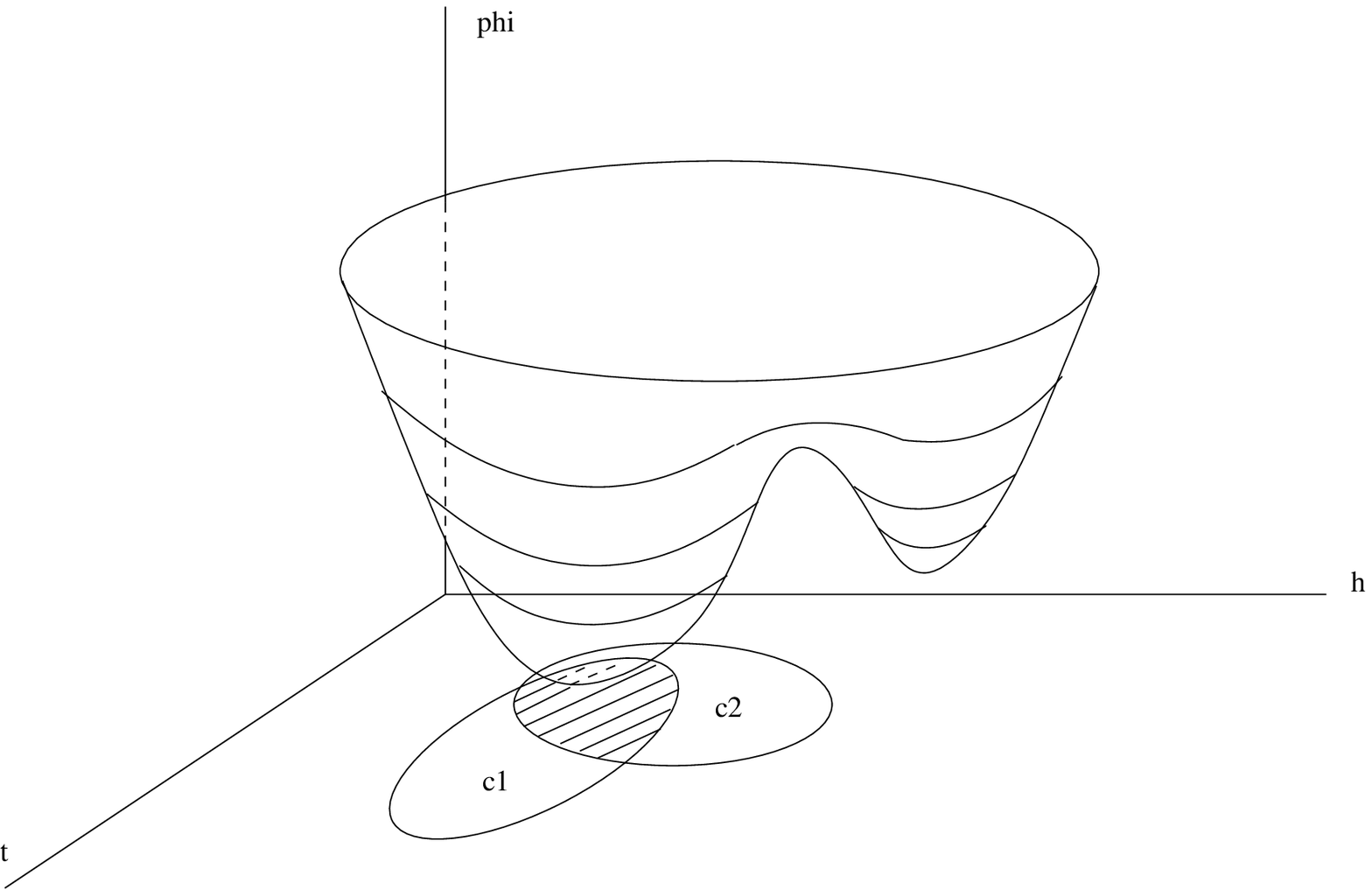}
    \end{psfrags}
} \caption{Cost minimization over the intersection of two
constraint sets} \label{fig:ISIrelaySituation}
\end{figure}

\begin{figure}[htbp]
\centerline{
    \begin{psfrags}
    \psfrag{t0}[c]{{ $t_0$}}  %
    \psfrag{t1}[c]{{ $t_1$}}  %
    \psfrag{h0}[c]{{ $h_0$}}  %
    \psfrag{t0t1h0}[l]{{\small $(t_0^\prime, t_1^\prime, 0)$}}  %
    \psfrag{Eoft0t1}[l]{{ $\xi_\hbf (t_0^\prime, t_1^\prime)$}} %
    \psfrag{fromabove}[c]{{\footnotesize \textsf{View from above}}} %
    \psfrag{c1capc2}[c]{{ $C_1 \cap C_2$}} %
     \scalefig{0.75}\epsfbox{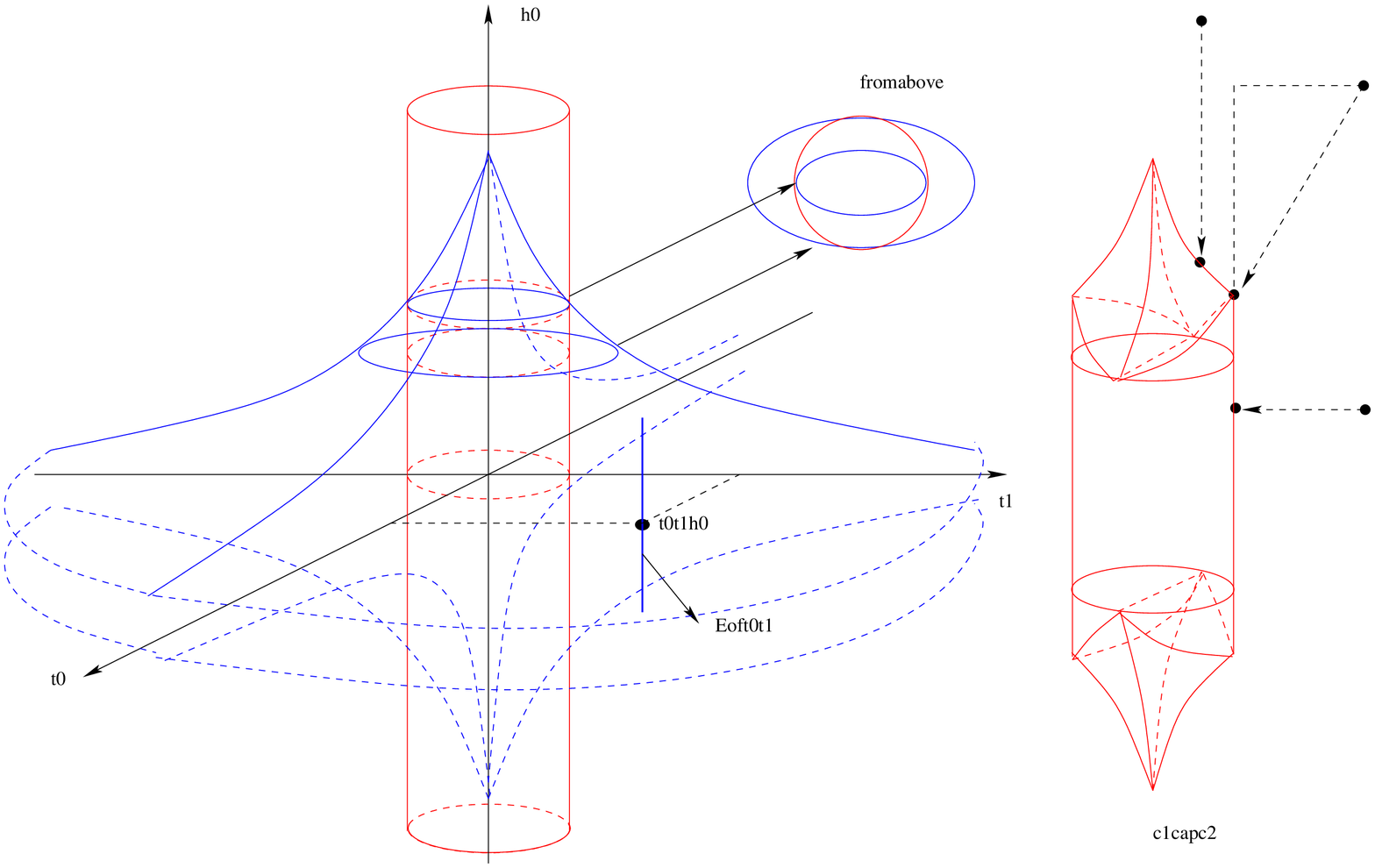}
    \end{psfrags}
} \caption{The structure of two constraint sets in  case of
$\tbf=(t_0,t_1)$ and $\hbf = h_0$}
\label{fig:TwoConstraitSetExample}
\end{figure}

  Now, the problem under the LTI
FIR formulation is  given by
\begin{equation}
\min_{\tbf,\hbf} \phi(\tbf,\hbf)
\end{equation}
such that
\begin{equation}
(\tbf,\hbf) \in C_1 ~~\mbox{and}~~ (\tbf,\hbf) \in C_2,
\end{equation}
where $\phi(\tbf,\hbf)$ is given by \eqref{eq:ISIrelayCostFunc},
and $C_1$ and $C_2$ are the constraint sets determined by
\eqref{eq:ISIrelayPowerConstSource} and
\eqref{eq:ISIrelayPowerConstBicon}, respectively.  The situation
is depicted in Fig. \ref{fig:ISIrelaySituation}. Even though
$\phi(\tbf,\hbf)$ is not jointly convex in $\tbf$ and $\hbf$, we
can still apply the projected subgradient method in Theorem
\ref{theo:APSM}. Suppose now that $K:= C_1 \cap C_2$ is convex.
Then, applying the projected subgradient method
\eqref{eq:APSMupdate} with $\ubf_0 \in K$ leads to the convergence
to a local minimum at least. Due to the nonconvexity of $C_2$,
however, $K$ is not a convex set. To circumvent this issue, let us
investigate the structure of the two constraint sets further. Fig.
\ref{fig:TwoConstraitSetExample} shows the two constraint sets,
$C_1$ and $C_2$, and their intersection $K$ in the case of
$\tbf=(t_0,t_1)$ and $\hbf = h_0$. $C_1$ and $C_2$ are the red
cylinder and the blue shape that looks like a mountain on one side
and is symmetric about the $h_0=0$ plane in the left side of Fig.
\ref{fig:TwoConstraitSetExample}, respectively.
In this case, the ellipsoid $\xi_{\hbf}(\tbf^\prime)$ is a line
segment passing through $(\tbf^\prime,0)$ ending at the upper and
the lower surfaces of the blue shape, as shown in Fig.
\ref{fig:TwoConstraitSetExample}. As mentioned already, for a
given $\hbf$ the constraint set for $\tbf$ is an ellipsoid. (See
\eqref{eq:ISIrelayPowerConstBicon3}.) So, we have an ellipsoidal
cut of the blue shape by a plane perpendicular to the $h_0$-axis.
Thus, $C_1 \cap C_2$ is given by the shape in the right side of
Fig. \ref{fig:TwoConstraitSetExample}, which looks like a cylinder
with bulges on the top and the bottom. If we are willing to
sacrifice the bulging regions from the feasibility set, we can
construct a convex set $\tilde{K} \subset K$ as the Cartesian
product of $B_{\tbf}(P_s)$ and $\Xi_\hbf(P_s,P_r)$, i.e.,
\begin{equation}  \label{eq:FeasSetCartProd}
\tilde{K}:= B_{\tbf}(P_s) \times \Xi_\hbf(P_s,P_r),
\end{equation}
where
\begin{equation} \label{eq:ISIrelayXi}
\Xi_\hbf(P_s,P_r) = \bigcap_{\tbf \in B_{\tbf}(P_s)}
\xi_\hbf(\tbf) = \bigcap_{\tbf \in \mbox{surface}(B_{\tbf}(P_s))}
\xi_\hbf(\tbf).
\end{equation}
The second equality in \eqref{eq:ISIrelayXi} is because $\xi_\hbf
(\tbf^\prime) \subset \xi_\hbf (\gamma \tbf^\prime)$ for $\gamma
\in [0,~1]$. This can easily be verified from
\eqref{eq:ISIrelayPowerConstBicon1}. For $\gamma \tbf^\prime$,
$|T(e^{j\omega})|^2$ will simply scale down to
$\gamma^2|T(e^{j\omega})|^2$, and $|H(e^{j\omega})|^2$ satisfying
the inequality with $|T(e^{j\omega})|^2$ will satisfy the
inequality with $\gamma^2|T(e^{j\omega})|^2$. Here,
$\Xi_\hbf(P_s,P_r)$ is convex since it is the intersection of
ellipsoids. Also, $\tilde{K}$ is convex due to its Cartesian
product structure.  In the considered case of $(t_0,t_1,h_0)$,
$\tilde{K}$ is exactly the red cylinder in the right side of Fig.
\ref{fig:TwoConstraitSetExample}.  Based on the above discussion,
we now provide our first algorithm for the joint source and relay
filter design.

\vspace{0.5em}
\begin{algorithm} \label{algo:algorithm1}
\begin{itemize}
\item Initialization: Set $\ubf_0$ properly, e.g.,
\begin{equation} \label{eq:initialpointofalgo1}
\ubf_0= [\tbf_0^T,
 {\mathbf {0}}^T]^T
\end{equation} with $\tbf_0=[\sqrt{\frac{P_s}{L_s}}[1,1,\cdots,1]^T$. (It is easy to verify that this $\ubf_0 \in \tilde{K} \subset K$.)  %
\item Update:
\begin{equation} \label{eq:LinearRelayUpdate1}
\ubf_{n+1}=\left\{
\begin{array}{ll}
P_{\tilde{K}}\left(\ubf_n-\mu_n\frac{\phi_n(\ubf_n)}{||\phi_n'(\ubf_n)||^2}\phi_n'(\ubf_n)\right), & \mbox{if}~\phi_n'(\ubf_n)\neq 0, \\
\ubf_n, & \mbox{otherwise}, \\
\end{array}
\right.
\end{equation}
where  $\phi(\ubf)$ and $\phi^\prime(\ubf)$ are given by
 \eqref{eq:ISIrelayCostFunc} and
  \eqref{eq:ISIrelayCostGradient}, respectively, $0< \mu_n < 2$ and $\ubf_n=[\tbf_n^T, \hbf_n^T]^T$. %
\item Stopping rule: Stop the update  either when $\ubf_n$ does
not change further or when the number of iterations exceeds a
preset maximum.
\end{itemize}
\end{algorithm}
\vspace{0.5em}

\noindent  Due to the Cartesian product structure of $\tilde{K}$
in \eqref{eq:FeasSetCartProd}, the metric projection
$P_{\tilde{K}}$ can be implemented by the composition of two
separate projections, one in $\tbf$ projecting onto $B_\tbf(P_s)$
and then the other in $\hbf$ projecting onto $\Xi_\hbf(P_s,P_t)$,
i.e.,
\begin{equation}
P_{\tilde{K}}= P_{\Xi_\hbf(P_s,P_t)} \circ  P_{B_\tbf(P_s)},
\end{equation}
where the projection onto a ball is simply given by
\begin{equation}
P_{B_\tbf(P_s)}(\tbf) = \left\{ \begin{array}{cl} \tbf, &
\mbox{if}~~ ||\tbf||^2 \le P_s,\\
\frac{\sqrt{P_s} \tbf}{||\tbf||}, & \mbox{otherwise},
\end{array}
\right.
\end{equation}
and $P_{\Xi_\hbf(P_s,P_t)}$ can be approximated by using the
successive projection method \cite{Censor&Zenios:book}. That is,
we uniformly partition the surface of the $L_s$-dimensional ball
with radius $\sqrt{P_s}$ into $M$ subsets, and select the center
$\tbf_m$ of  subset $m$. Then, $\Xi_\hbf(P_s,P_t) \approx
\cap_{m=1}^M \xi_\hbf(\tbf_m)$ and
\begin{equation}
P_{\Xi_\hbf(P_s,P_t)}(\hbf) \approx P_{\xi_\hbf(\tbf_M)} \circ
P_{\xi_\hbf(\tbf_{M-1})}\circ \cdots \circ
P_{\xi_\hbf(\tbf_{1})}(\hbf),
\end{equation}
where the projection onto an ellipsoid can easily be implemented
by a known method like the one in \cite{Kiseliov:94LMJ}. If
$\phi^\prime(\ubf) \ne 0$ for all  $\ubf \in \tilde{K}$, then
Algorithm \ref{algo:algorithm1} will find the point that yields
the maximum rate within $\tilde{K}$. Otherwise, Algorithm
\ref{algo:algorithm1} finds a local optimum attracting $\ubf_0$,
and  convergence is guaranteed.  However, the complexity of the
projection $P_{\Xi_\hbf(P_s,P_t)}(\hbf)$ is prohibitive even for
small values of $L_s$. Thus, we propose a simplified algorithm to
eliminate this difficulty in the following.

\vspace{0.5em}
\begin{algorithm} \label{algo:algorithm2}
\begin{itemize}
\item Initialization of $\ubf_0$.  %
\item Update:
\begin{equation} \label{eq:LinearRelayUpdate2}
\ubf_{n+1}=\left\{
\begin{array}{ll}
 P_{\xi_\hbf(P_{B_\tbf(P_s)}(\tbf_n))}(\hbf_n) \circ P_{B_\tbf(P_s)}(\tbf_n) \circ \left(\ubf_n-\mu_n\frac{\phi_n(\ubf_n)}{||\phi_n'(\ubf_n)||^2}\phi_n'(\ubf_n)\right), & \mbox{if}~\phi_n'(\ubf_n)\neq 0, \\
\ubf_n, & \mbox{otherwise}. \\
\end{array}
\right.
\end{equation}
 \item Stopping rule: The same
as that in Algorithm \ref{algo:algorithm1}.
\end{itemize}
\end{algorithm}
\vspace{0.5em}

\noindent In the initialization step, we can consider other
initial points than the example in Algorithm
\ref{algo:algorithm1}. For example,
\begin{equation} \label{eq:initialpointofalgo2}
\tbf_0=[\sqrt{P_s}, 0,\cdots, 0]^T ~~\mbox{and}~~ \hbf_0 =
P_{\xi_\hbf(\tbf_0)}([1,\cdots,1]^T).
\end{equation}
 With such an initial point,
we can start the algorithm from the full power use at the relay.
In Algorithm \ref{algo:algorithm2}, the subgradient projection is
the same as that in Algorithm \ref{algo:algorithm1}, but the
projection to $K$ is now different. Here, we first project only
the $\tbf_n$ coordinates to the ball $B_{\tbf}(P_s)$, and then
project the $\hbf_n$ coordinates onto the $\hbf$-ellipsoid
$\xi_{\hbf}(P_{B_{\tbf}(P_s)}(\tbf_n))$ corresponding to
$P_{B_{\tbf}(P_s)}(\tbf_n)$ given by
\eqref{eq:ISIrelayPowerConstBicon}; several projection examples in
the case of $L_s=2$ and $L_r=1$ are shown in the right side of
Fig. \ref{fig:TwoConstraitSetExample}. In this way, the metric
projection to $K$ is approximated and highly simplified, and
$\ubf_n \in K$ for all $n$. The convergence of Algorithm
\ref{algo:algorithm2} is not guaranteed theoretically, but there
is no loss in the feasibility set in this case. It is observed
numerically that Algorithm \ref{algo:algorithm2} is stable and
works well. This is because $K$ seems almost convex as seen in the
example in Fig. \ref{fig:TwoConstraitSetExample}, and the proposed
two-step projection onto $K$ approximates the metric projection
onto $K$ well due to the almost cylindrical structure of $K$
except the top and bottom. Under our formulation, it is
straightforward to impose the strict causality constraint on the
relay filter $H(z)$ that captures the possible delay at the relay
for digital processing; simply remove $h_0$ in \eqref{eq:ISIH(z)}
and follow the same procedure as the causal case.

\section{Numerical Results}
\label{sec:numerical}

To evaluate the performance of the proposed joint filter design
method presented in Section \ref{sec:JointSourceRelayISIDesign},
we ran extensive simulations under various channel conditions. The
channel order $L$ of all the propagation channels $H_{sr}(z)$,
$H_{rd}(z)$ and $H_{sd}(z)$ was selected as $L=5$, and each
channel tap coefficient  was generated independently and
identically-distributedly (i.i.d.) according to a Rayleigh
distribution with a different variance for a different channel,
i.e.,
\begin{equation}
h_{sd}[l] \stackrel{i.i.d.}{\sim} \Nc(0,\sigma_{sd}^2),~~~
h_{sr}[l] \stackrel{i.i.d.}{\sim} \Nc(0, \sigma_{sr}^2)~~~
\mbox{and} ~~~h_{rd}[l] \stackrel{i.i.d.}{\sim} \Nc(0,
\sigma_{rd}^2)
\end{equation}
for $l=0,1,\cdots, L-1$.  Throughout the simulations, we fixed
$\sigma^2=1$ and used $L_s=30$ and $L_r=20$ for the orders of the
source and relay filters, respectively, to allow enough freedom
for the two filters considering $L=5$. We used Algorithm
\ref{algo:algorithm2} with the initial point\footnote{We observed
that Algorithm \ref{algo:algorithm2} with the initialization
\eqref{eq:initialpointofalgo1} did not make a noticeable
difference.}  \eqref{eq:initialpointofalgo2} for the simulations,
and the step size for the algorithm was adaptively changed
according to $\mu_n=1/\sqrt{n}$ as the number of iterations
increases. The stopping condition for the algorithm was either
that the square of the normalized difference of two successive
updates is less than $10^{-5}$ (i.e.,
$||\ubf_{n+1}-\ubf_n||^2/||\ubf_n||^2 \le 10^{-5}$) or that the
number of iterations exceeds 1000. For the numerical integration
to compute the quantities (\ref{eq:ISIrelayCostFunc},
\ref{eq:ISIrelayCostGradient}, \ref{eq:ISIrelayPowerConstBicon}),
we used a 512-point Gaussian quadrature method over $[-\pi,\pi]$.
 Since it is not clear how to optimally
apply the known coding strategies such as DF and CF to the case of
ISI channels, we do not consider these schemes in this section,
and thus we use the AF scheme as the performance reference.

\begin{figure}[htbp] \centerline{
    \begin{psfrags}
      \scalefig{0.6}\epsfbox{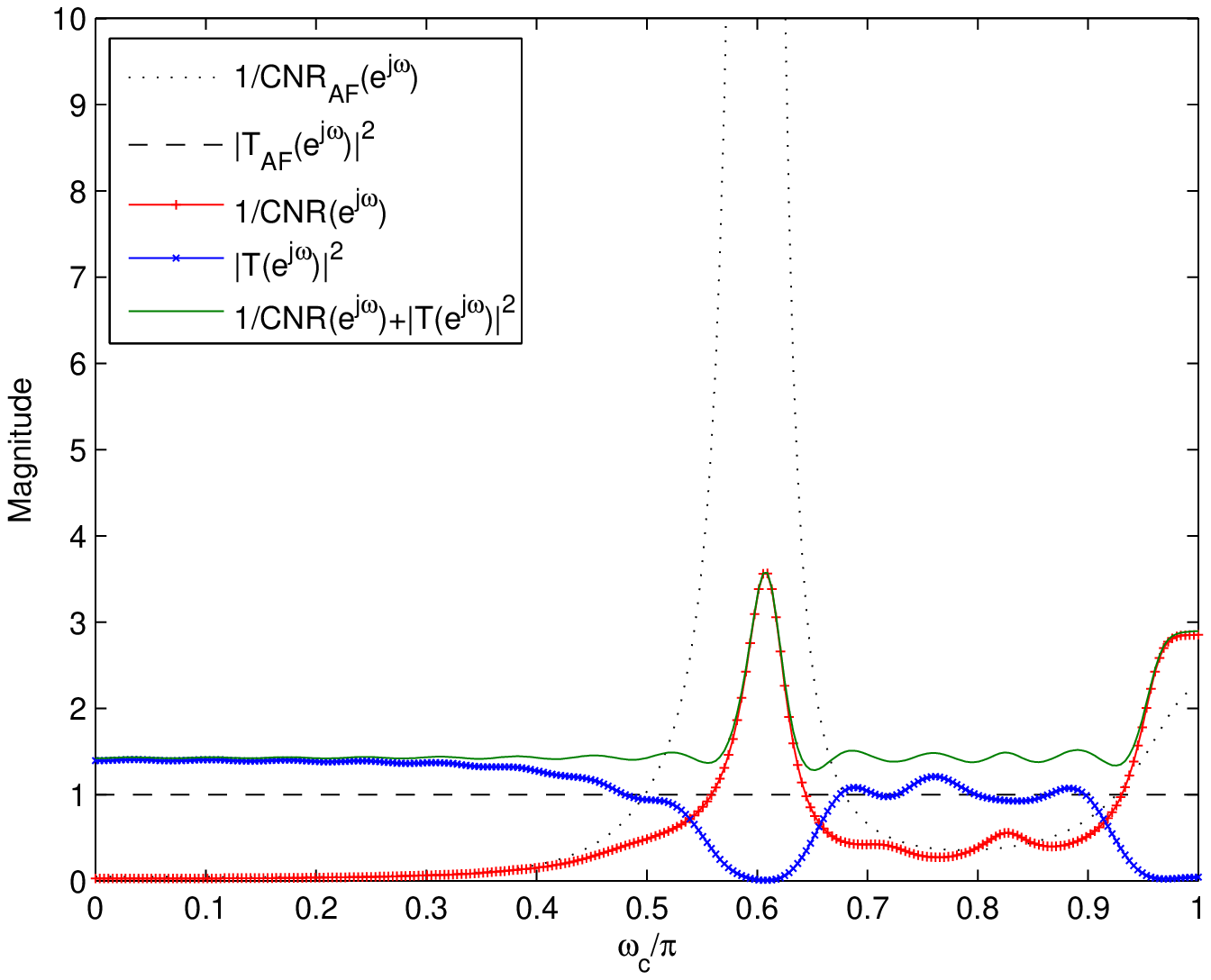}
    \end{psfrags}
} \caption{The frequency responses related to the proposed method}
\label{fig:numerical_spectrum}
\end{figure}

First, Fig. \ref{fig:numerical_spectrum} shows the frequency
responses related to the proposed method. The channel coefficients
for this figure were generated randomly according to
$\sigma_{sr}^2=\sigma_{rd}^2=\sigma_{sd}^2=1$ and given by
$H_{sr}=1.8833 +
0.3254z^{-1}-0.0952z^{-2}+0.0312z^{3}-0.6138z^{-4}$,
$H_{rd}=-0.0728 +
1.3148z^{-1}+0.9783z^{-2}+1.7221z^{3}-0.4123z^{-4}$ and
$H_{sd}=-0.8864 -
1.8402z^{-1}-1.6282z^{-2}-1.1738z^{3}-0.4154z^{-4}$. $P_s=P_r=1$.
The flat\footnote{The reason why the flat input spectrum is used
for the conventional AF scheme is that even for a given relay
filter, e.g., the AF filter, the optimal power allocation to
maximize the transmission rate \eqref{eq:CLTIfd} not only with the
source power constraint \eqref{eq:powerConstSpec1} but also with
the relay power constraint \eqref{eq:powerConstSpec2} in ISI
channels is not a trivial problem; this is not a simple and known
water-filling problem with a single total power constraint because
of the term $|H_{sr}(e^{j\omega})|^2$ in
\eqref{eq:powerConstSpec2} unless $H_{sr}(e^{j\omega})$ is
constant over $[-\pi, \pi]$. As far as we know, this problem was
not handled before, and the two-step projection method in this
paper provides one way to accomplish this in general ISI
channels.} dashed line represents the input spectrum of the
conventional AF scheme, and the dotted line is the inverse of the
CNR density, i.e., the effect noise level, in the case that the AF
scheme is used at the relay. It is seen that there are two peaks
in the noise level for the AF scheme around the normalized
frequency values of 0.6 and 1. Note that the peak around 0.6 is
very high and its  width  is also large, whereas the peak around 1
is mild.
Now the red solid line marked with + denotes the effective noise
level generated by the relay filter designed by the proposed
method. We recognize that the designed relay filter is smart. One
could imagine that a reasonable relay filter would suppress the
frequency band having a bad overall response with the AF scheme
and enhance the frequency band having a good overall response with
the AF scheme so that the water-filling by the source filter might
be enhanced.  This is exactly the case with the mild noise peak
around 1; the designed relay filter reinforces the mild noise
peak. However, the designed relay filter suppresses the severe
noise peak with a large width around 0.6 instead of reinforcing
it. This is because the width of the severe peak is large and,
thus, the loss of this frequency band will reduce the transmission
rate. To this optimally shaped noise level, the water-filling-type
source power allocation is performed by the designed source filter
satisfying both the source and relay power constraints jointly
with the designed relay filter; for the frequency bands around the
two noise peaks the power is not allocated, as seen in the figure.
(See the green solid line without any markers.) Similar behaviors
are observed in other settings even though they are not shown in
this paper.

\begin{figure}[htbp]
\centerline{ \SetLabels
\L(0.25*-0.1) (a) \\
\L(0.76*-0.1) (b) \\
\endSetLabels
\leavevmode
\strut\AffixLabels{
\scalefig{0.5}\epsfbox{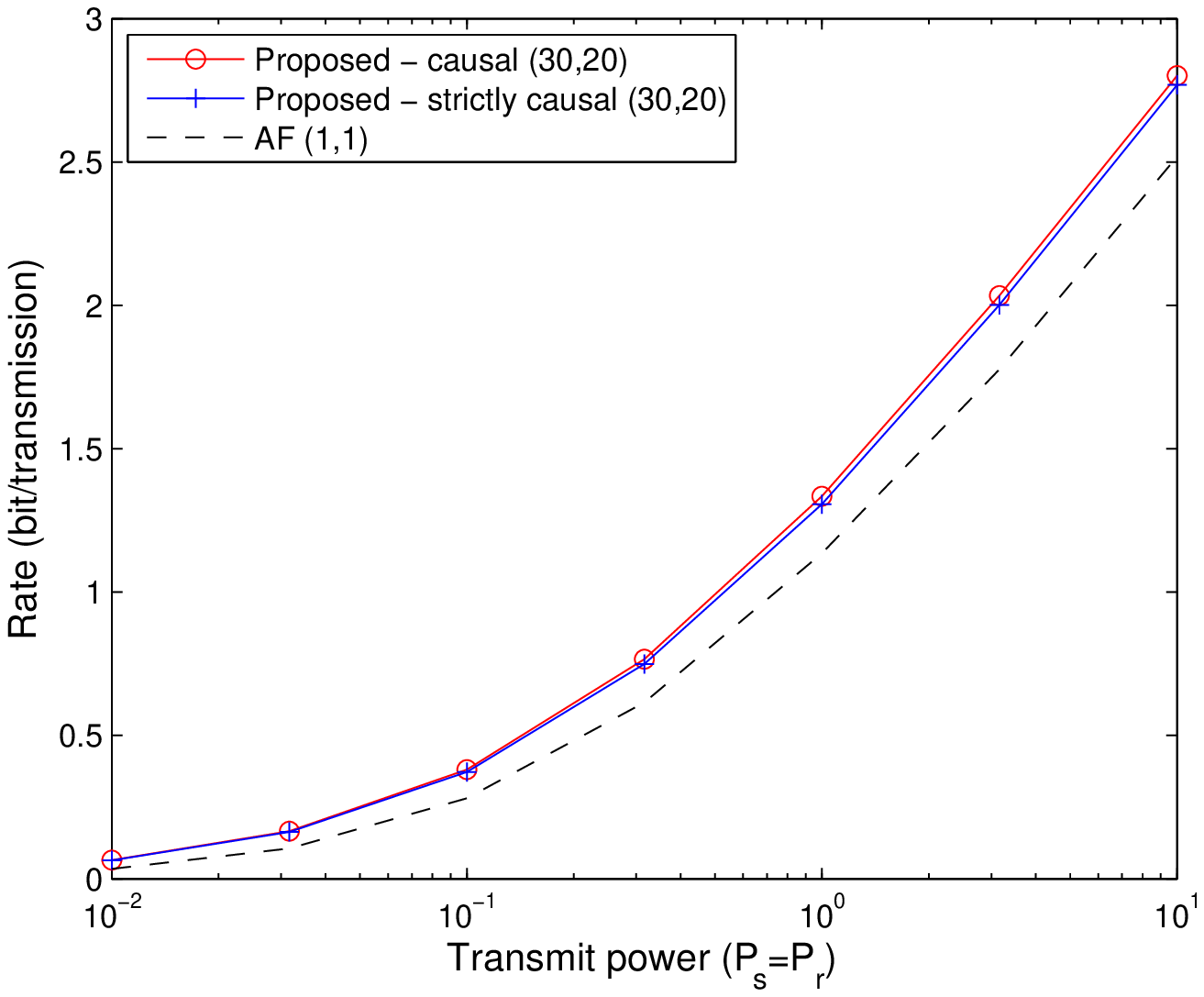}
\hspace{0.5cm}
\scalefig{0.5}\epsfbox{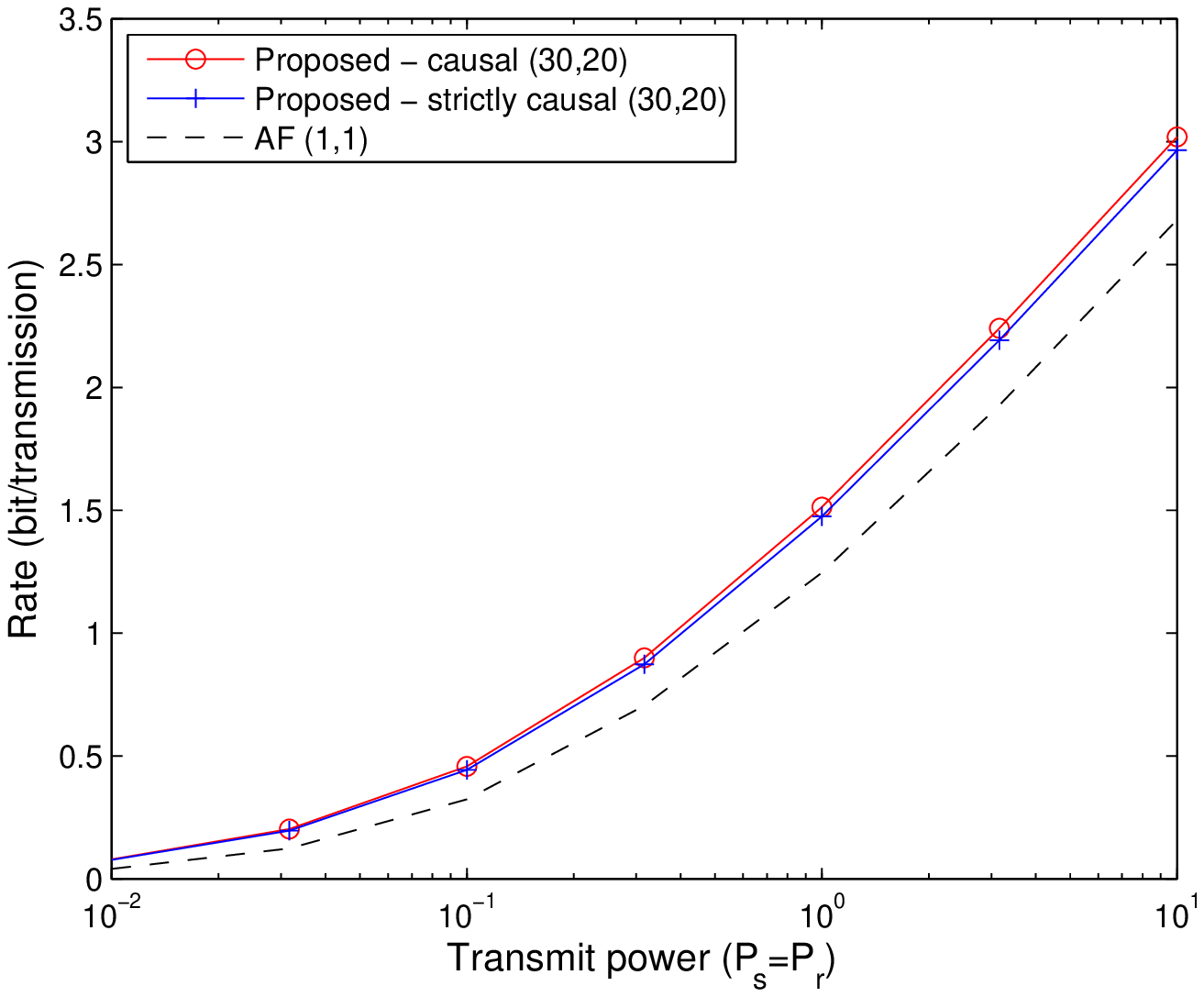} } }
\vspace{0.5cm} \centerline{ \SetLabels
\L(0.25*-0.1) (c) \\
\L(0.76*-0.1) (d) \\
\endSetLabels
\leavevmode
\strut\AffixLabels{
\scalefig{0.5}\epsfbox{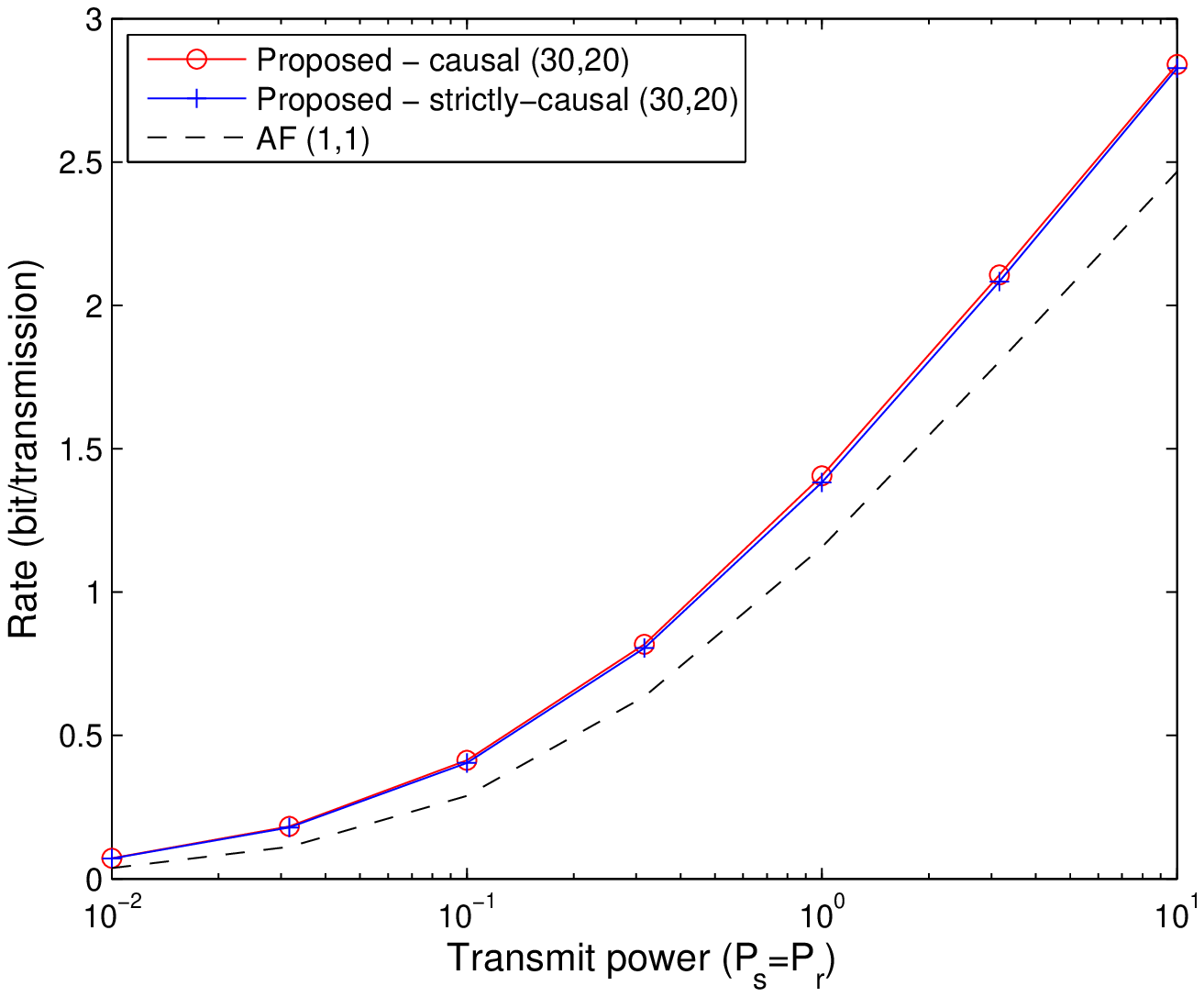}
\hspace{0.5cm}
\scalefig{0.5}\epsfbox{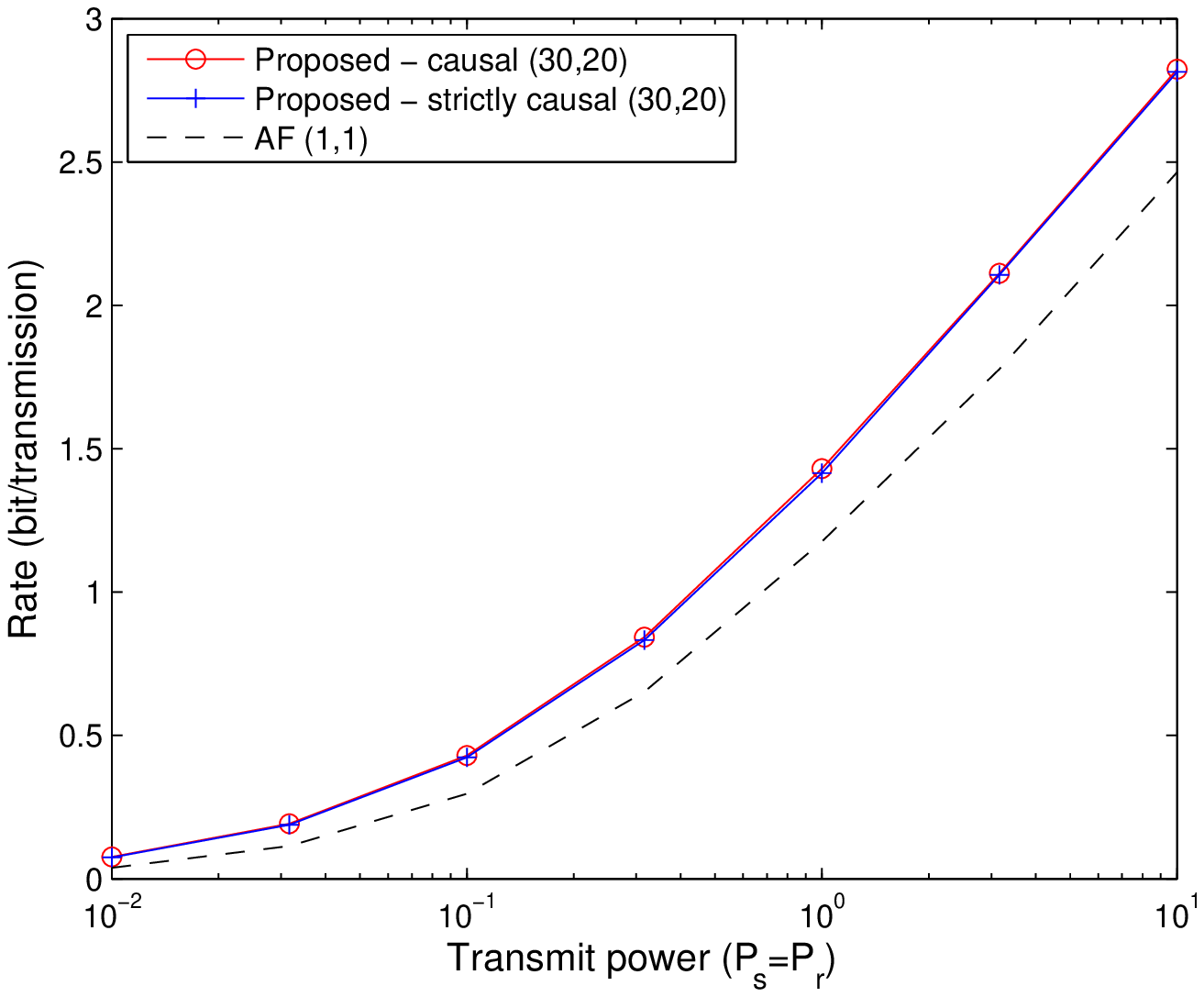} } }
\vspace{0.5cm} \centerline{ \SetLabels
\L(0.25*-0.1) (e) \\
\L(0.76*-0.1) (f) \\
\endSetLabels
\leavevmode
\strut\AffixLabels{
\scalefig{0.5}\epsfbox{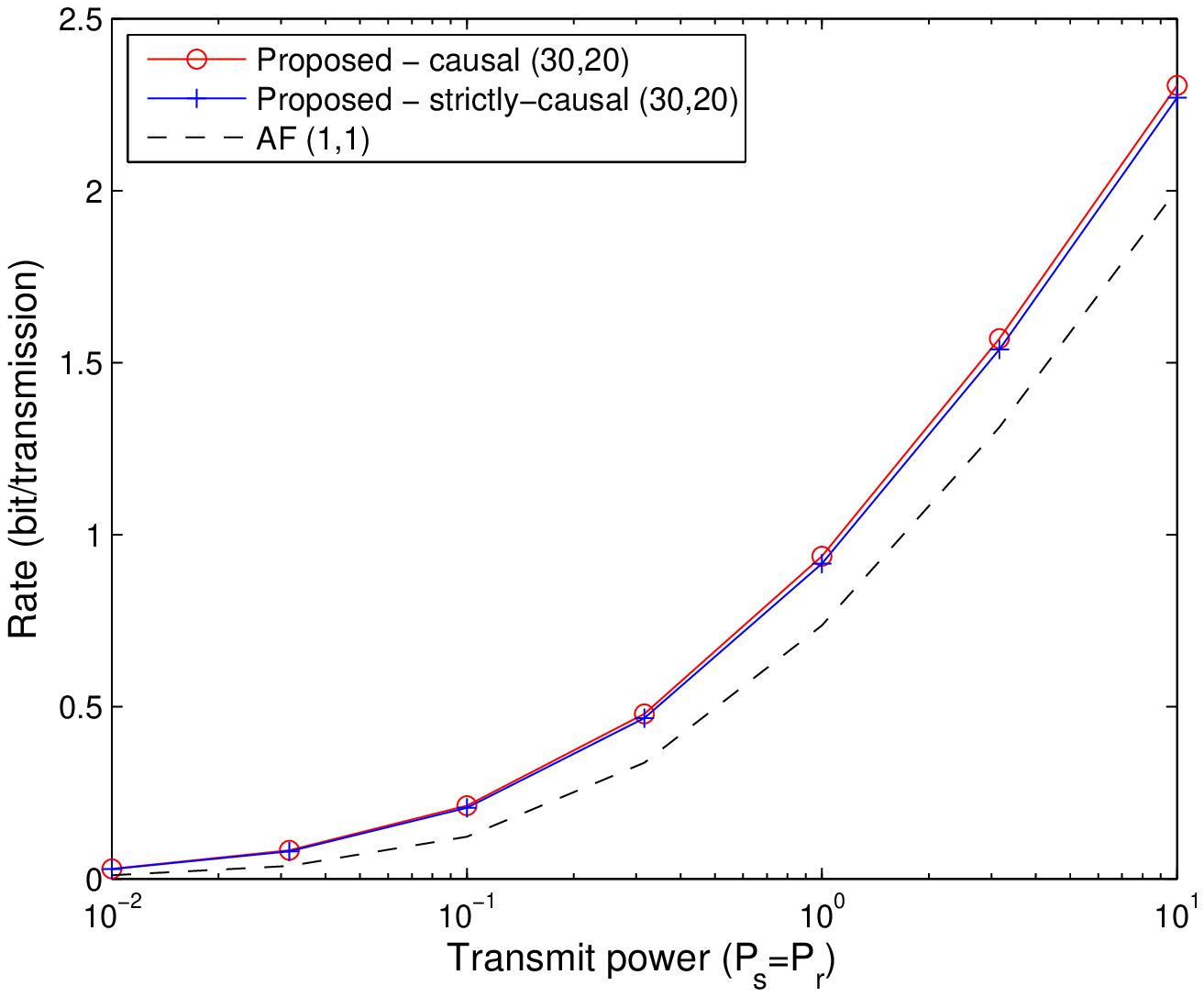}
\hspace{0.5cm}
\scalefig{0.5}\epsfbox{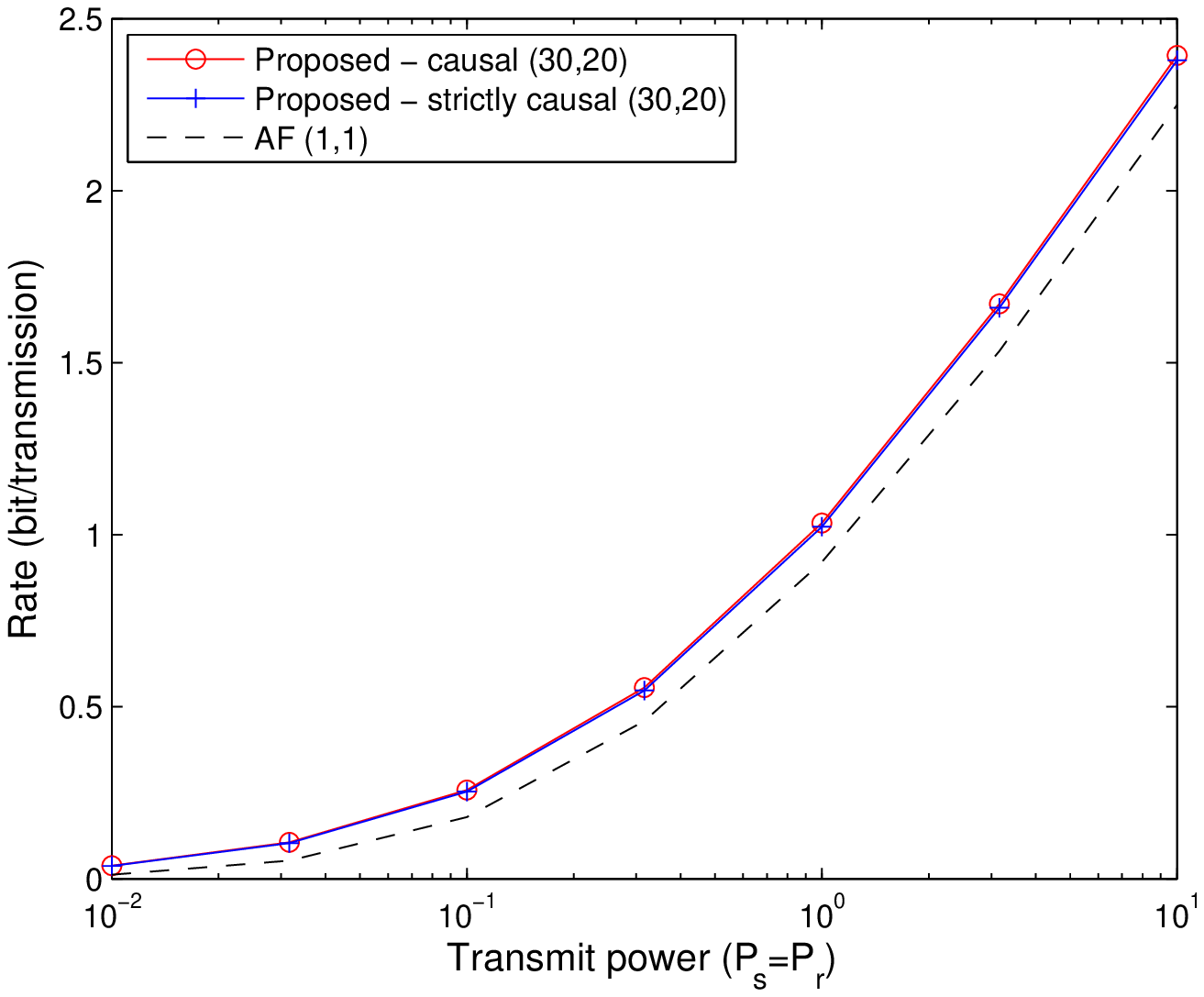} } }
\vspace{0.5cm}  \caption{ Transmission rate averaged over channel
realizations ($P_s=P_r$): (a) $\sigma_{sd}^2=1$,
$\sigma_{sr}^2=1$, $\sigma_{rd}^2=1$, (b) $\sigma_{sd}^2=1$,
$\sigma_{sr}^2=4$, $\sigma_{rd}^2=1$, (c) $\sigma_{sd}^2=1$,
$\sigma_{sr}^2=1$, $\sigma_{rd}^2=4$, (d) $\sigma_{sd}^2=1$,
$\sigma_{sr}^2=1$, $\sigma_{rd}^2=10$, (e) $\sigma_{sd}^2=1/4$,
$\sigma_{sr}^2=1$, $\sigma_{rd}^2=1$, and (f)
$\sigma_{sd}^2=1/10$, $\sigma_{sr}^2=1$, $\sigma_{rd}^2=10$}
\label{fig:averageRate_s1r1}
\end{figure}

Next, let us see the performance of the proposed method in various
channel settings. Fig. \ref{fig:averageRate_s1r1} shows the
transmission rate achieved by the proposed method averaged over
100 independent channel realizations for each of the different
channel gain settings when $P_s = P_r$. We considered both causal
and strictly causal filtering for the proposed method. For the
strictly causal filtering, we eliminated the $t_0$ and $h_0$ terms
in the formulation and kept the same filter orders $L_s$ and $L_r$
 as those in the causal filtering case. We considered two sets of channel conditions; in one set  the direct S-D
link has a reasonable strength compared with
 the S-R and R-D links  (Fig.
\ref{fig:averageRate_s1r1} (a), (b), (c), (d)), and in the other
set the direct link is weak compared with the S-R and R-D links
(Fig. \ref{fig:averageRate_s1r1} (e) and (f)). (We did not
consider the case that the direct link is stronger than the S-R
and R-D links since it is unnecessary to use the relay in this
case.) It is seen in Fig. \ref{fig:averageRate_s1r1} that the gain
obtained by the proposed joint source and relay filtering over the
AF scheme is noticeable. In particular, Fig.
\ref{fig:averageRate_s1r1} (c) shows the performance when
$P_s=P_r$, $\sigma^2=1$, $\sigma_{sd}^2=1$, $\sigma_{sr}^2=1$ and
$\sigma_{rd}^2=4$, which is equivalent to the flat-fading case of
$a=1$ and $b=2$ in  Fig. \ref{fig:flatFadingPlot} (a) and (b) and
Fig. \ref{fig:LTIF_IAF}. Compared with the flat-fading case shown
in Fig. \ref{fig:LTIF_IAF} (c), the gain by the proposed joint
design over the AF scheme in ISI channels is significant. It is
also seen that the loss caused by the strict causality of
filtering is not significant in the proposed scheme. This is
because removing one tap out of 30 or 20 taps does not reduce the
degree of freedom of the filters much. Fig.
\ref{fig:averageRate_s1r1} (e) and (f) show the case that the
direct link is weak compared with the S-R and R-D links. Similar
results are seen in the figure.  Fig. \ref{fig:averageRate_s1r1}
(f) shows the smallest gain among the six cases. This is explained
as follows. As seen in (\ref{eq:CLTIfd}, \ref{eq:CNRformula}), the
gain of the joint filtering depends on various factors.  The worst
situation for the performance of the joint filtering is the case
that $H_{sd}(e^{j\omega}) \approx 0$ and $H_{rd}(e^{j\omega}) >>
0$ such that $|H_{rd}(e^{j\omega})H(e^{j\omega})| >> 1$. In this
case, the CNR density in (\ref{eq:CNRformula}) is approximated by
$CNR(e^{j\omega}) \approx
|H_{sr}(e^{j\omega})H(e^{j\omega})H_{rd}(e^{j\omega})|^2/(\sigma^2
|H_{rd}(e^{j\omega})H(e^{j\omega})|^2)=|H_{sr}
(e^{j\omega})|^2/\sigma^2$; and thus the impact of the relay
filter disappears and only the optimal power allocation by the
source filter under the two power constraints is effective. Even
in this case, a non-negligible gain is observed. Although the
results in the cases of $P_s = 2 P_r$ and $2P_s = P_r$ are not
shown in this paper, we observed similar results to the case of
$P_s =P_r$.

\vspace{-0.5em}
\section{Conclusions}
\label{sec:conclusions} \vspace{-0.5em}

We have considered the linear Gaussian relay problem. By adopting
the LTI relay filtering and realizable input spectra, we have
converted the problem to a joint design problem of source and
relay filters. We have investigated the performance of this joint
LTI filtering in flat-fading relay channels, and have shown the
optimality of the AF scheme within the class of one-tap filters.
In
 general ISI relay channels, we have developed a practical method
 for the joint filter design to maximize the transmission rate
 based on the projected subgradient method under the LTI FIR filtering
 framework, and have shown numerically that the gain of the
 proposed design is noticeable compared with the AF scheme in ISI
 relay channels.

\vspace{-1em}
\section*{Acknowledgement}

The authors of this paper wish to thank Gilwon Lee, Masahiro
Yukawa   and Isao Yamada  for their introduction to the (adaptive)
projected subgradient method.


\end{document}

\begin{figure}[htbp]
\centerline{ \SetLabels
\L(0.25*-0.1) (a) \\
\L(0.76*-0.1) (b) \\
\endSetLabels
\leavevmode
\strut\AffixLabels{ \scalefig{0.4}\epsfbox{figures/apsm.eps}
\scalefig{0.4}\epsfbox{figures/apsm.eps} } } \vspace{0.3cm}
\centerline{ \SetLabels
\L(0.25*-0.1) (c) \\
\L(0.76*-0.1) (d) \\
\endSetLabels
\leavevmode
\strut\AffixLabels{ \scalefig{0.4}\epsfbox{figures/apsm.eps}
\scalefig{0.4}\epsfbox{figures/apsm.eps} } } \vspace{0.5cm}
\caption{$K_v$ vs. $\Delta_1$ ($M=2$, $\Delta=0.02$, SNR = 10 dB):
(a) $A=1$ (b) $A=8$ (c) $A=15$ (d) $A=100$}
\label{fig:periodpatternM2highSNR}
\end{figure}

\begin{figure}[htbp]
\centerline{ \SetLabels
\L(0.17*-0.1) (a) \\
\L(0.50*-0.1) (b) \\
\endSetLabels
\leavevmode
\strut\AffixLabels{ \scalefig{0.45}\epsfbox{figures/apsm.eps}
\hspace{1cm} \scalefig{0.45}\epsfbox{figures/feasibleset.eps}
 } }
 \vspace{0.1cm} \caption{$K_v$ vs. $\Delta_1$ ($M=2$,
$\Delta=0.02$, SNR = -3 dB): (a) $A=1$ (b) $A=100$ (c) $A=1000$}
\label{fig:periodpatternM2lowSNRanal}
\end{figure}